\documentclass[preprint,floatfix]{revtex4}

\pdfoutput=1

\usepackage{pdfpages}
\usepackage{graphicx} 
\usepackage{dcolumn} 
\usepackage{amsmath}
\usepackage{amssymb}
\usepackage{bm}
\usepackage{times}
\usepackage{ulem}
\usepackage[justification=raggedright]{caption}

\usepackage{subfigure}
\usepackage{tabularx}
\usepackage{appendix}
\usepackage{amstext}
\usepackage{array}

\usepackage{bbm}

\definecolor{wine-stain}{rgb}{0.5,0,0} 
\definecolor{bblue}{rgb}{0,0.0,0.5} 

\usepackage[colorlinks=true
,urlcolor=blue
,anchorcolor=blue
,citecolor=blue %red
,filecolor=blue
,linkcolor=wine-stain
,menucolor=blue
,pagecolor=blue
,linktocpage=true
,pdfproducer=medialab
,linktoc=all %section
]{hyperref} 
\usepackage{booktabs}  % for \toprule, \midrule, and \bottomrule macros

\allowdisplaybreaks

\newcommand{\ncmd}{\newcommand}

\ncmd{\lt}{\left}
\ncmd{\rt}{\right}
\ncmd{\tr}[1]{\mbox{Tr}\lt({#1}\rt)}
\ncmd{\half}{\frac{1}{2}}
\ncmd{\eps}{\epsilon}
\ncmd{\veps}{\varepsilon}
\ncmd{\dgr}{\dagger}
\ncmd{\sig}{\sigma}
\ncmd{\gam}{\gamma}
\ncmd{\rtarw}{\rightarrow}
\ncmd{\Rt}{\Rightarrow}
\ncmd{\abs}[1]{\lt\cb{#1}\rt\cb}
\ncmd{\avg}[1]{\lt\lb{#1}\rt\rb}
\ncmd{\dl}{\delta}
\ncmd{\Dl}{\delta}
\ncmd{\sgn}[1]{\mbox{sgn}\lt(#1\rt)}
\ncmd{\kap}{\kappa}
\ncmd{\wtil}[1]{\widetilde{#1}}
\ncmd{\thrfr}{\therefore}
\ncmd{\eq}[1]{Eq. \eqref{#1}}
\ncmd{\fig}[1]{Fig. \ref{#1}}
\ncmd{\Lam}{\Lambda}
\ncmd{\lam}{\lambda}
\ncmd{\dow}{\partial}
\ncmd{\ordr}[1]{\mathcal{O}\lt(#1\rt)}
\ncmd{\dsty}{\displaystyle}
\ncmd{\alert}[1]{\color{red}{#1}}
\ncmd{\mc}{\mathcal}
\ncmd{\mbf}[1]{\mathbf{#1}}
\ncmd{\Deriv}[2]{\frac{d{#1}}{d{#2}}}
\ncmd{\ParDeriv}[2]{\frac{\partial{#1}}{\partial{#2}}}
\ncmd{\step}[1]{\Theta\lt(#1\rt)}
\ncmd{\td}{\tilde} 
\ncmd{\what}{\widehat}
\ncmd{\om}{\omega}
\ncmd{\Om}{\Omega}
\ncmd{\vrho}{\varrho}
\ncmd{\vsig}{\varsigma}
\ncmd{\vkap}{\varkappa}
\ncmd{\bqa}{\begin{eqnarray}} 
\ncmd{\eqa}{\end{eqnarray}}
\ncmd{\nn}{\nonumber \\}
\ncmd{\nnum}{\nonumber}
\ncmd{\comment}[1]{{\color{red}{#1}}}
\definecolor{new_color}{RGB}{50,155,0}

\ncmd{\lb}{\big<}
\ncmd{\rb}{\big>}
\ncmd{\cb}{\big|}
\ncmd{\cH}{{\cal H}}
\ncmd{\lc}{l_c}
\ncmd{\Lc}{l_c}

\ncmd{\sqg}{\sqrt{|g(x)|}}
\ncmd{\gmn}{E_{\mu i}}

\ncmd{\emi}{E_{\mu i}}
\ncmd{\emj}{E_{\mu j}}
\ncmd{\eni}{E_{\nu i}}
\ncmd{\enj}{E_{\nu j}}
\ncmd{\de}{|E(x)|}

\ncmd{\gemn}{g_{E,\mu \nu}}
\ncmd{\gemnp}{g_{E',\mu \nu}}

\ncmd{\bgmn}{\bar E_{\mu i}}
\ncmd{\Fs}{F(\sigma)}
\ncmd{\cJ}{ {\cal J}\left(E',\sigma';E,\sigma \right)  }
\ncmd{\Si}{ \Sigma \left(E',\sigma';E,\sigma \right)  }

\ncmd{\grs}{g_{\alpha \beta}}
\ncmd{\pmn}{\pi^{\mu i}}
\ncmd{\prs}{\pi^{\alpha \beta}}
\ncmd{\fs}{ \frac{3}{\sqrt{2}} \sigma }
\ncmd{\fst}{ 3 \sqrt{2} \sigma }

\ncmd{\ee}{entanglement entropy }

\makeatletter
\newcommand*{\rom}[1]{\expandafter\@slowromancap\romannumeral #1@}
\makeatother

\begin{document}

\title{
Emergent gravity from relatively local Hamiltonians and \\
a possible resolution of the black hole information puzzle
}

\author{Sung-Sik Lee\\
\vspace{0.3cm}
{\normalsize{Department of Physics $\&$ Astronomy, McMaster University,}}\\
{\normalsize{1280 Main St. W., Hamilton ON L8S 4M1, Canada}}
\vspace{0.2cm}\\
{\normalsize{Perimeter Institute for Theoretical Physics,}}\\
{\normalsize{31 Caroline St. N., Waterloo ON N2L 2Y5, 
Canada}}
}

\date{\today}

\begin{abstract}

In this paper, we study a possibility 
where gravity and time emerge from quantum matter.
Within the Hilbert space of matter fields defined on a spatial manifold,
we consider a sub-Hilbert space spanned by 
states which are parameterized by spatial metric.
In those states,
metric is introduced as a collective variable 
that controls local structures of entanglement. 
The underlying matter fields endow the states labeled by metric
with an unambiguous inner product. 
Then we construct a Hamiltonian for the matter fields
that is an endomorphism of the sub-Hilbert space,
thereby inducing a quantum Hamiltonian of the metric.
It is shown that there exists a matter Hamiltonian 
that induces the general relativity 
in the semi-classical field theory limit.
Although the Hamiltonian is not local in the absolute sense,
it has a weaker notion of locality, called {\it relative locality} :
the range of interactions is set by the entanglement 
present in target states on which the Hamiltonian acts.
In general, normalizable states are not invariant 
under the transformations generated by the Hamiltonian.
As a result, a physical state spontaneously breaks the Hamiltonian constraint,
and picks a moment of time.
The subsequent flow of time can be understood 
as a Goldstone mode associated with the broken symmetry. 
The construction allows one to study dynamics of gravity
from the perspective of matter fields.
The Hawking radiation corresponds to
a unitary evolution where entanglement across horizon is gradually transferred
from color degrees of freedom to singlet degrees of freedom.
The underlying quantum states remain pure
as evaporating black holes keep entanglement with early Hawking radiations
in the singlet sector which is not captured by the Bekenstein-Hawking entropy.

\end{abstract}

\maketitle

%\newpage
%\tableofcontents

\newpage

{
\hypersetup{linkcolor=bblue}
\hypersetup{
    colorlinks,
    citecolor=black,
    filecolor=black,
    linkcolor=black,
    urlcolor=black
}
\tableofcontents
}

%\begin{thebibliography}{99}
%\end{thebibliography}

%\newpage
\section{Introduction}

There have been many efforts 
to understand gravity as an emergent phenomenon
from various angles
\cite{
1968SPhD...12.1040S,
WEINBERG198059,
PhysRevLett.75.1260,
1994gr.qc.....4039J,
Volovik1998,
Barcelo:2001tb,
doi:10.1142/S0217751X03014071,
Yang:2006dk,
2007qsst.conf..163S,
Sindoni:2011ej,
1126-6708-2007-12-049,
Verlinde2011,
GU201290,
Padmanabhan:2014jta,
CARLIP2014200,
PhysRevLett.114.031104,
PhysRevD.79.084008,
PhysRevD.81.104033,
2017arXiv170909806G,
2017arXiv170803040S}.
The anti-de Sitter space/conformal field theory (AdS/CFT) 
correspondence\cite{Maldacena:1997re,Witten:1998qj,Gubser:1998bc}
is a concrete example of emergent gravity,
where a gravitational theory in the bulk emerges from a quantum field theory 
defined at the boundary of the anti-de Sitter space.
However, the AdS/CFT correspondence does not directly apply to our universe 
which does not have a spatial boundary.
For our universe, 
it seems more natural that
the bulk spacetime emerges from a theory
defined at a temporal boundary 
in the past or future.
The program of the de Sitter space/conformal field theory (dS/CFT) correspondence aims 
to make this scenario concrete with the guidance 
from the AdS/CFT correspondence\cite{1126-6708-2001-10-034,2011arXiv1108.5735A,2017arXiv171110037A}.

In this paper, we study the possibility
in which time and gravity emerge from quantum matter,
employing a more microscopic perspective built from the quantum renormalization group (RG)\cite{Lee:2013dln,Lee2016}.
Quantum RG provides a prescription to construct holographic duals
for general quantum field theories 
based on the intuition that the emergent space direction in the bulk 
corresponds to a length scale\cite{1998PhLB..442..152A,deBoer:1999xf,Skenderis:2002wp,1126-6708-2004-10-075,Heemskerk:2010hk,2011JHEP...08..051F,PROP:PROP201400007}.
The basic object in quantum RG is wavefunctions defined in the space of couplings.
Instead of specifying a quantum field theory 
in terms of classical values of all couplings allowed by symmetry,
a theory is represented as a wavefunction 
defined in a much smaller subspace of couplings.
The subspace is chosen so that all symmetry-allowed operators can be constructed as composites 
of those operators sourced by the couplings in the subspace.
Then, general theories can be represented as coherent linear superpositions
of theories defined in the subspace.
As a result, the couplings in the subspace are promoted to fluctuating variables.
Metric, which sources the energy-momentum tensor,
also becomes a dynamical variable 
whose fluctuations account for 
composite operators made of the energy-momentum tensor.
While the Wilsonian RG flow is a classical flow defined in the full space of couplings,
the same exact RG flow can be represented as a quantum evolution of the wavefunction defined in the subspace.
The classical flow of the Wilsonian RG is replaced by
a sum of all possible RG paths defined in the subspace of couplings.
The weight for each RG path is determined by an action 
which includes a dynamical gravity\cite{Lee2012}.
%In quantum RG for Euclidean boundary field theories,
%the radial evolution is generated by a non-unitary evolution 
%that projects wavefunctions of couplings
%to an IR fixed point of the theory\cite{Lee2016}.
%This gives rise to a radial direction that is space-like.
%%
%%

In order to realize an emergent time 
in a manner that a space-like direction emerges in quantum RG,
we consider wavefunctions of ordinary quantum matters defined on a space manifold
instead of wavefunctions of couplings defined on a spacetime manifold.
Metric in Lorentzian quantum field theories
%, as a coupling that sources the energy-momentum tensor,
determines connectivity of spacetime 
by setting the strength of derivative terms in local actions.
In quantum states of matter fields we consider here,
metric with the Euclidean signature is introduced as a collective variable
that controls entanglement of matter fields in the space manifold.
Namely, local actions for Lorentzian quantum field theories
are replaced with short-range entangled states of quantum matters.
The metric in quantum states of matter plays the role of a variational parameter
that sets the notion of locality (`short-rangeness')
in how matter fields are entangled in space.
More specifically, we consider a set of wavefunctions of matter fields parameterized by Riemannian metric.
The space spanned by those states forms a sub-Hilbert space
in the full Hilbert space of the matter field.

With wavefunctions of couplings replaced by wavefunctions of matter fields,
we consider a unitary evolution of the quantum states. 
Although an unitary evolution is not same as RG flow,
one may still view the former 
as a coarse graining process 
in which information accessible to local observers
decreases in time through scrambling. 
In particular, we consider an evolution generated by a Hamiltonian
which maps the sub-Hilbert space into the sub-Hilbert space.
%This is analogous to the situation where 
%RG flow is confined to a subset of quantum field theories
%in the full space of quantum field theories.
%%
Since the sub-Hilbert space is parameterized by spatial metric,
the Hamiltonian of the matter fields induces a quantum Hamiltonian
of the metric.
In this way, one can induce quantum theories of metric from matter fields.
The main goal of this paper is to address the following questions :

\begin{enumerate}
	\item Can a matter Hamiltonian induce a quantum theory 
	      that becomes Einstein's general relativity at long distances in the classical limit ?
	\item Is the matter Hamiltonian that gives rise to the general relativity local ? 
	\item What is the nature of time in the emergent gravity ?
	\item How does a quantum state of matter maintain its purity under an evolution 
		that is dual to a black hole evaporation ?
\end{enumerate}
The short answers to these questions are
\begin{enumerate}
\item Yes, one can engineer a matter Hamiltonian 
whose induced dynamics agrees with the general relativity 
in the semi-classical field theory limit.
\item No, the Hamiltonian is not local in the usual sense. 
However, it possesses a relative locality in that the range of interactions depends 
on states on which the Hamiltonian acts.
\item Time arises as a Goldstone mode associated with a spontaneous breaking of the symmetry generated by the Hamiltonian constraint.
\item 
During black hole evaporations,
quantum states stay pure by transferring entanglement from 
color degrees of freedom to singlet sectors.
\end{enumerate}
The rest of the paper gives long answers to the questions.
Here is an outline that may serve as a summary of the paper.

In Sec. \ref{sec:main}, we sketch the main idea that is used
in the explicit examples constructed in the following sections.
This section constitutes a conceptual guide for the rest of the paper.

Sec. \ref{sec:mini} is a warm-up which discusses a toy model 
from which a minisuperspace quantum cosmology emerges.
Although there is no extended space in the toy model,
it still contains the essential idea on how time emerges.
In Sec. \ref{sec:miniA}, we introduce a set of quantum states for $N$ variables.
The states in the set are parameterized by two collective variables.
Those states labeled by the collective variables span a sub-Hilbert space 
in the full Hilbert space of the $N$ fundamental variables.
Throughout the section, we will focus on the sub-Hilbert space.
It becomes the Hilbert space for the induced cosmology
in which the two collective variables become 
the scale factor of a universe and a scalar field, respectively.
The inner product between states in the sub-Hilbert space,
which is inherited from the one defined in the full Hilbert space,
provides a notion of distance between states with different collective variables.
With increasing $N$,
two states with different collective variables 
become increasingly orthogonal.

In Sec. \ref{sec:miniB}, we construct a Hamiltonian for the $N$ variables
that is an endomorphism of the sub-Hilbert space, that is, an operator
that maps the sub-Hilbert space into the sub-Hilbert space.
Through the evolution generated by the Hamiltonian, 
a state with a definite collective variable
evolves into a linear superposition of states 
with different collective variables in the sub-Hilbert space.
The evolution is naturally described as 
a unitary quantum evolution of wavefunctions defined in the space 
of the collective variables.
Therefore, one can identify a Hamiltonian for the collective variables
induced from the matter Hamiltonian.
The key result of this subsection is that 
there exists a Hamiltonian for the $N$ variables
which gives rise to a minisuperspace Wheeler-DeWitt Hamiltonian
for the collective variables.

In Sec. \ref{sec:miniC}, we address the issue of time.
In general relativity which includes minisuperspace cosmology, 
Hamiltonian is a constraint 
which generates time reparameterization transformations.
It is usually assumed that `physical states' are the ones
that are invariant under diffeomorphism
and are annihilated by the Hamiltonian constraint.
This gives rise to the problem of time
because `physical states' are stationary,
and no change is generated under Hamiltonian evolution.
In the present theory of induced quantum cosmology, 
the problem of time is avoided
because there is no normalizable state 
that satisfies the Hamiltonian constraint
in the sub-Hilbert space.
This is shown by diagonalizing the matter Hamiltonian numerically.
Nonetheless, there exist semi-classical states 
which are normalizable. 
They satisfy the Hamiltonian constraint 
to the leading order in the large $N$ limit,
yet break the constraint beyond the leading order.
While the semi-classical states do not satisfy the Hamiltonian constraint exactly,
they are legitimate states as quantum states of matter.
A semi-classical state `picks' a moment of time spontaneously 
because it is forced to have a finite norm.
Non-trivial time evolution of the semi-classical states
can be understood 
as Goldstone modes associated with the 
weak spontaneous symmetry breaking.
In the large $N$ limit,
the time evolution of semi-classical states
coincides with the classical minisuperspace cosmology.

In the following section,
we generalize the discussion on the emergent minisuperspace cosmology
to a fully fledged gravity in $(3+1)$-dimensions. 
The starting point is an $N \times N$ Hermitian matrix field 
defined on a three-dimensional spatial manifold.
In Sec. \ref{sec:gravityA1}, we define a Hilbert space for the induced gravity from the matter field.
The full Hilbert space of the matter field is spanned by eigenstates of the matrix field.
Within the full Hilbert space, 
we focus on a sub-Hilbert space 
spanned by gaussian wavefunctions.
Those gaussian wavefunctions,
which are singlet under a $SU(N)$ internal symmetry,
are parameterized by a Riemannian metric and a scalar field.
Namely, we consider a set of $SU(N)$ invariant wavefunctions 
in which metric and a scalar field enters as collective variables
(equivalently, variational parameters)
that control how the matter field is entangled in space.
General states within the sub-Hilbert space are given by linear superpositions
of states with different collective variables.

Sec. \ref{sec:gravityA2} is devoted to the inner product.
The inner product in the full Hilbert space is 
%%fixed by
%%the inner products between eigenstates of the matrix field.
%%The latter is 
defined in terms of normal modes
of an elliptic differential operator 
associated with a fiducial Riemannian metric.
Although a fiducial metric is introduced to define the inner product in the full Hilbert space,
the fiducial metric decouples in the inner product 
between normalized states in the sub-Hilbert space.
Based on this, 
we show the following properties of the inner product.
First, 
%the inner product in the sub-Hilbert space 
%is invariant under spatial diffeomorphism.
%%
%The inner product in the sub-Hilbert space is inherited 
%from the one defined for the matter field 
%in the full Hilbert space.
the induced inner product in the sub-Hilbert space 
is invariant under diffeomorphisms of the collective variables. 
%%%
Second, we show that two states in the sub-Hilbert space are orthogonal
unless the two have metrics that give same local proper volume.
Third, even for states with same local proper volume,
the overlap decays exponentially 
as the difference in the collective variables increases in the large $N$ limit.
This is explicitly shown for states whose metrics are close to the flat Euclidean metric.

In Sec. \ref{sec:gravityB}, 
we show that the metric, as a variational parameter,
controls the number of degrees of freedom that generate entanglement in space.
As quantum states of matter field defined in continuum,
the number of degrees of freedom per unit coordinate volume is infinite.
Nonetheless, the wavefunctions in the sub-Hilbert space
have a short-distance cut-off scale 
which regularizes the amount of entanglement. 
It is shown that the von Neumann entanglement entropy of states 
in the sub-Hilbert space has two contributions.
One is the color entanglement entropy 
which is generated by the matter field 
in a classical configuration of the collective variables.
The color entanglement entropy of a region in space obeys the area law,
where the area is measured with the metric associated with the state 
in the unit of the short-distance cut-off.
The amount of color entanglement a region in the manifold
can support is not fixed.
Rather it is a dynamical quantity that is determined by the metric.
With increasing proper volume, 
the entanglement entropy increases accordingly.
In the presence of fluctuations of the collective variables,
correlations in the fluctuations give rise to
an additional contribution to the von Neumann entanglement entropy, 
called singlet entanglement entropy.
These two contributions can be approximately separated in semi-classical states
where fluctuations of the collective variables are small.

In Sec. \ref{sec:gravityC}, 
we consider endomorphisms
of the sub-Hilbert space.
We show that there exist
Hermitian operators for the matter field
which induce the momentum density operator
and a regularized Wheeler-DeWitt Hamiltonian density operator
for the collective variables.
Those operators for the matter fields generate 
an evolution of the collective variables 
once the lapse and the shift are fixed.
At long distances and in the large $N$ limit,
the evolution coincides with the time evolution 
of the classical Einstein's gravity
in a fixed gauge. 
The matter Hamiltonian that gives rise to the general relativity
has no absolute locality
because the Hamiltonian, as a quantum operator, contains operators 
with arbitrarily long ranges.
Nonetheless, it is relatively local 
in that the range of interactions
that survive when applied to a state
is limited by the amount of entanglement 
present in the state.
The notion of locality in the Hamiltonian is determined 
relative to states 
to which the Hamiltonian is applied.

The fact that the general relativity can emerge from matter fields 
provides an opportunity to examine the black hole information puzzle 
from the perspective of the underlying matter field.
In Sec. \ref{sec:Hawking}, we consider a formation and evaporation of a black hole
in the induced theory of gravity.
By construction, the time evolution is unitary.
The discussion is centered on how purity of a quantum state
can be in principle maintained during an evolution.
As a black hole evaporates,
the color \ee  across the horizon,
which is identified as the Bekenstein-Hawking entropy\cite{PhysRevD.7.2333,Hawking1975},
decreases in time. 
On the other hand, the singlet \ee increases because Hawking radiation
is emitted in the singlet sector.
As a result, entanglement is gradually transferred from the color sector
to the singlet sector.
This leads to a `neutralization' of entanglement entropy.
The full quantum states remain pure as
black holes keep the entanglement with early Hawking radiation
in the singlet sector
which is not captured by the Bekenstein-Hawking entropy. 
The failure of the Bekenstein-Hawking entropy to account for 
all available states in a black hole 
is attributed to a lack of equilibrium.
It is argued that 
localization can arise dynamically
because both states and Hamiltonian 
effectively flow under the time evolution
generated by the relatively local Hamiltonian.
In Sec. \ref{sec:summary}, we conclude with a summary and discussions on open questions.

\section{The main idea}
\label{sec:main}

\begin{figure}[ht]
\begin{center}
\centering
\includegraphics[scale=0.6]{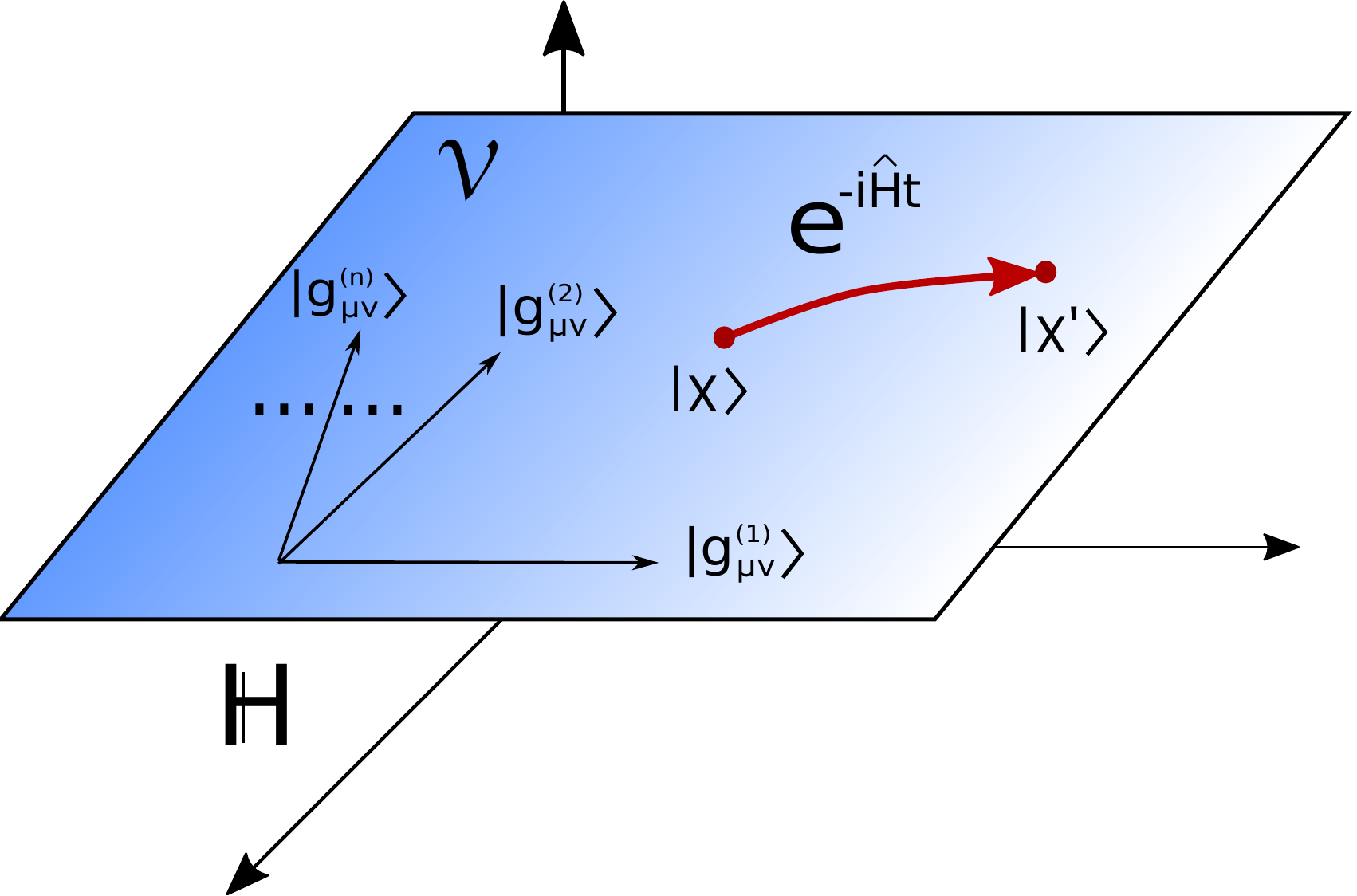}
\end{center}
\caption{
In the full Hilbert space ${\mathbbm H}$ of matter fields,
a sub-Hilbert space ${\cal V}$ is spanned by a set of basis vectors,
where each basis vector is associated with a spatial metric.
A general state in ${\cal V}$ can be written as a linear superposition
of the basis vectors, 
$\cb \chi \rb = \int Dg_{\mu \nu} ~  \cb g_{\mu \nu} \rb ~ \chi( g_{\mu \nu} )$,
where $\chi( g_{\mu \nu} )$ is a wavefunction defined in the space of spatial metric.
An endomorphic Hamiltonian $\hat H$ generates a map from ${\cal V}$ into ${\cal V}$.
Therefore, $\cb \chi' \rb = e^{-i \hat H t} \cb \chi \rb$ can be also written as
$\cb \chi' \rb = \int Dg_{\mu \nu} ~  \cb g_{\mu \nu} \rb ~ \chi'( g_{\mu \nu} )$.
The linear map between $\chi( g_{\mu \nu} )$ and $\chi'( g_{\mu \nu} )$
can be written as
$\chi'( g_{\mu \nu} ) = exp \left[ -i ~t ~{\cal H} \left( g_{\mu \nu}, \frac{\partial}{\partial g_{\mu \nu}} \right) \right] \chi( g_{\mu \nu} )$,
where ${\cal H} \left( g_{\mu \nu}, \frac{\partial}{\partial g_{\mu \nu}} \right) $ is 
identified as an induced Hamiltonian for the metric.
}
\label{fig:endomorphism}
\end{figure}

\begin{figure}[ht]
\begin{center}
\centering
\includegraphics[scale=0.4]{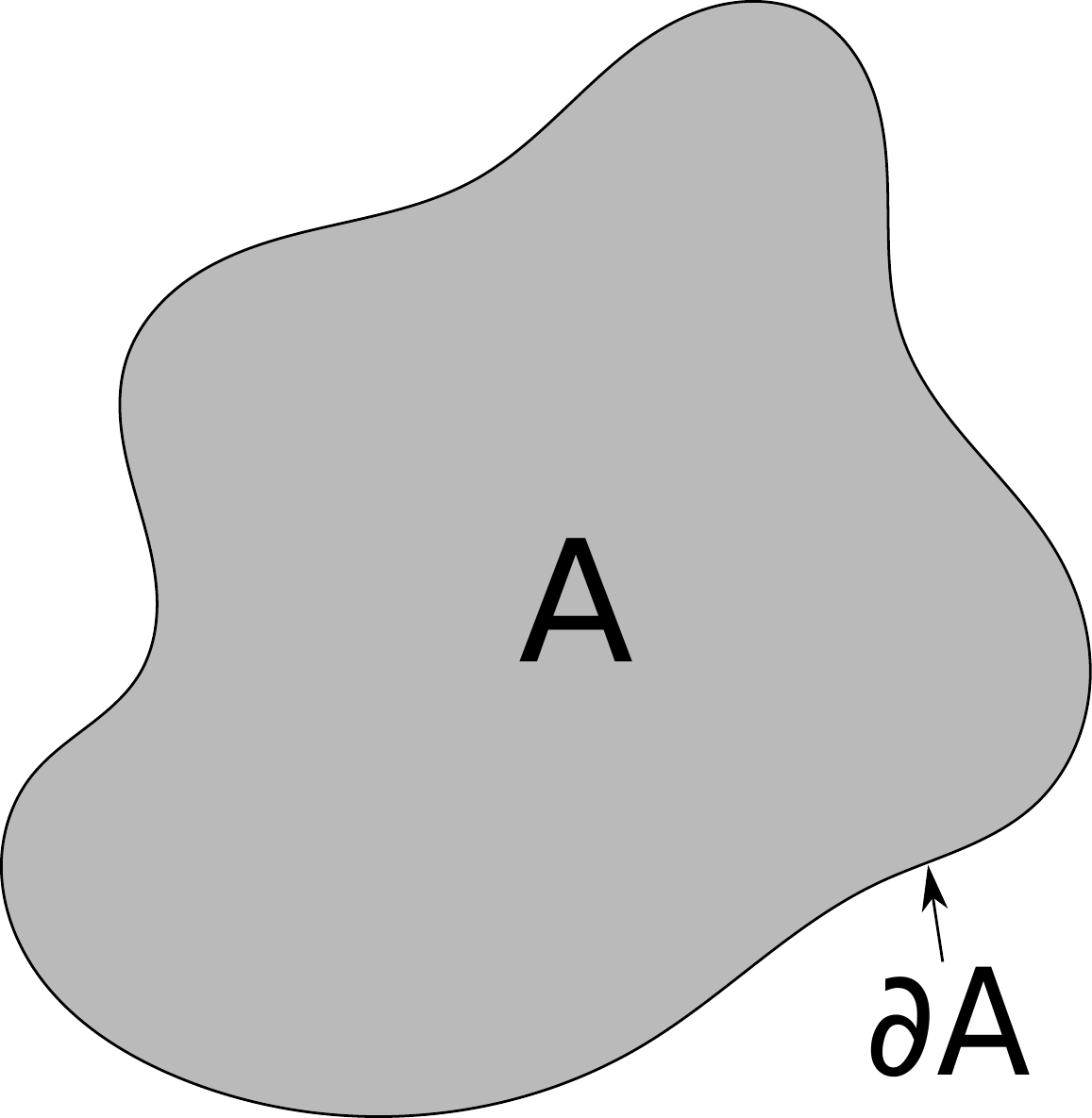}
\end{center}
\caption{
A metric is associated with each basis state in ${\cal V}$ 
such that the von Neumann entanglement entropy of a region $A$ 
is proportional to the proper area of $\partial A$ 
measured with respect to the metric.
}
\label{fig:perimeter}
\end{figure}

In this section, we sketch the main idea of the paper
that is summarized in \fig{fig:endomorphism}.
Our starting point is a Hilbert space of matter fields 
defined on a spatial manifold with a fixed topology.
Within the full Hilbert space (${\mathbbm H}$) of the matter fields,
we consider a sub-Hilbert space (${\cal V}$)
that is spanned by a set of short-range entangled states. 
${\mathbbm H}$ is equipped with an inner product,
which gives a unique inner product in ${\cal V}$.
Each basis state of ${\cal V}$ is associated with a set of  collective variables.
Among the collective variables is spatial metric.
In this discussion of the conceptual idea, 
we focus on the metric, 
ignoring other collective variables.
A spatial metric  
%%$g_{\mu \nu}(x)$
is assigned to each basis state 
such that the von Neumann entanglement entropy of a basis state
for any region in space
is proportional to the proper area of the boundary of the region
measured with 
the metric for the basis state
%$g_{\mu \nu}(x)$
(see \fig{fig:perimeter}). 
Two states which support different amounts of entanglement in a region
are assigned to have different metrics 
so that they give different proper sizes of the region
in proportion to the entanglement.
In this sense, the metric can be viewed as a collective variable
which controls the amount of entanglement in quantum states of the matter fields.
A general state in ${\cal V}$ can be expressed as a linear superposition of the basis states,
$\cb \chi \rb = \int Dg_{\mu \nu} ~  \cb g_{\mu \nu} \rb ~ \chi( g_{\mu \nu} )$.
Here $\cb g_{\mu \nu} \rb $ is the basis state associated with $g_{\mu \nu}(x) $,  
$\chi( g_{\mu \nu} )$ is a wavefunction defined in the space of spatial metric,
and $D g_{\mu \nu}$ is a measure that is defined based on the inner product
in ${\cal V}$.
In the limit that the number of matter fields is large, 
two states with different metrics become orthogonal.
This way, the sub-Hilbert space of the matter field is identified
as a Hilbert space for spatial metric.

Next we study dynamics of the matter field within the sub-Hilbert space
by considering a Hamiltonian $\hat H$ 
that maps ${\cal V}$ into ${\cal V}$.
If an initial state $\cb \chi \rb$ is prepared to be in ${\cal V}$,
$e^{-i t \hat H} ~ \cb \chi \rb$ can be written as a linear superposition
of  $\Bigl\{ \cb g_{\mu \nu}(x) \rb \Bigr\}$.
Because the unitary time evolution is a linear map acting on the sub-Hilbert space,
one can identify a differential operator 
${\cal H}\left( g_{\mu \nu}, \frac{\partial}{\partial g_{\mu \nu}}  \right)$
that acts on the wavefunction of metric such that 
$e^{-i \hat H t} \cb \chi \rb = 
\int D g_{\mu \nu} ~  \cb g_{\mu \nu} \rb ~ 
e^{-i {\cal H}\left( g_{\mu \nu}, \frac{\partial}{\partial g_{\mu \nu}}  \right) t}
\chi (g_{\mu \nu}) $. 
%This is illustrated in \fig{fig:endomorphism}.
We identify 
${\cal H}\left( g_{\mu \nu}, \frac{\partial}{\partial g_{\mu \nu}}  \right)$
as an induced Hamiltonian of the metric.
By requiring that the induced Hamiltonian 
becomes the Wheeler-DeWitt Hamiltonian
in the classical limit,
we construct a matter Hamiltonian 
that induces the general relativity.
%in the semi-classical limit.
%%The main goal of this paper is to study properties of the Hamiltonian.
%%This paper is concerned about understanding properties of the Hamiltonian
%%that induces the general relativity from a simple choice of the sub-Hilbert space.
%%We also discuss potential consequences of this scenario of emergent gravity
%%for the problem of time and the black hole information puzzle.

The matter Hamiltonian that induces the general relativity  
turns out to be a non-local Hamiltonian.
Yet, it has a weaker notion of locality called relative locality.
While the Hamiltonian is non-local as a quantum operator,
the range of interaction that survives when applied to a state  
is determined by the entanglement present in the target state.
%on which the Hamiltonian acts.
There is an intuitive way to understand this.
The Hamiltonian that induces the general relativity can not be local 
%%without being ultra-local
because one can not have a local gradient term in the Hamiltonian 
without introducing a fixed background\footnote{
One alternative possibility for a ultra-local Hamiltonian
has been considered in Ref. \cite{Lee2016}.
}.
Therefore, any background independent theory can not
have an absolute notion of locality.
On the other hand, the general relativity is reduced to a local 
effective field theory when fluctuations of the metric are weak.
Small fluctuations of metric propagate 
on top of a saddle point configuration
that is dynamically determined,
and the notion of locality in the effective field theory
is determined by the saddle point metric.
In the present construction, the metric is determined by the entanglement
present in quantum matter.
Therefore, the notion of locality should be set by the amount of entanglement
present in states of matter fields.

In the following two sections, 
we work out  examples 
which elucidate the idea outlined in this section.
In Sec. \ref{sec:mini}, 
we provide a toy example of quantum mechanical system in zero space dimension.
In this model, the spatial metric is reduced to one scale factor,
and a minisuperspace quantum cosmology emerges.
In Sec. \ref{sec:gravity}, 
we generalize the construction to 
a fully fledged gravity in three space dimension.
In Sec. \ref{sec:Hawking},
we discuss possible implications 
of the induced gravity for the black hole information puzzle.

\section{Emergent minisuperspace cosmology}
\label{sec:mini}
%%NEW%%

Based on the general idea outlined in the previous section,
in this section we consider a quantum mechanical system 
of $N$ variables from which a minisuperspace quantum cosmology emerges.
We start by defining a sub-Hilbert space of the $N$ variables
which becomes the Hilbert space
for two collective variables : a scale factor and a scalar.
The sub-Hilbert space is spanned by a set of basis vectors 
labeled by the two collective variables.
After examining the kinematic structure of the sub-Hilbert space,
we explicitly construct a Hamiltonian of the $N$ variables
that induces the Wheeler-DeWitt Hamiltonian of the minisuperspace cosmology
for the scale factor and the scalar.
Finally, we address the problem of time in quantum cosmology.
By numerically diagonalizing the matter Hamiltonian,
we show that there is no normalizable state that satisfies the Hamiltonian constraint
 within the sub-Hilbert space.
From this, we conclude that the requirement that a physical state should have a finite norm
forces quantum states spontaneously break the Hamiltonian constraint,
and the subsequent time evolution arises as a Goldstone mode associated with the broken symmetry.

%%NEW%%

\subsection{Hilbert space}
\label{sec:miniA}

\begin{figure}[ht]
\begin{center}
\centering
\includegraphics[scale=0.6]{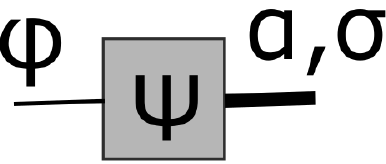}
%\includegraphics[scale=0.2]{Psi1.pdf}
%\includegraphics[scale=0.2,trim={0 -1cm 0 0}]{Psi1.pdf}
%\includegraphics[width=1\linewidth]{Psi1.pdf}
%\scalebox{0.5}{\includegraphics[scale=1]{Psi1.pdf}}
%\includegraphics[scale=1]{Psi1.pdf}
\end{center}
\caption{
A tensor representation of the wavefunction $\Psi(\phi;\alpha,\sigma)$.
Once the collective variables $\alpha, \sigma$ are fixed, 
the tensor defines a wavefunction for $\phi_a$. 
}
\label{fig:Psi}
\end{figure}

We consider a system of $N$ compact variables
whose Hilbert space is spanned by
$\Bigl\{ \cb \phi \rb ~ | ~ 0 \leq \phi_a < 2 \pi,
a=1,2,..,N
\Bigr\}$
with the inner product 
$\lb \phi' \cb \phi \rb = \prod_{a=1}^N \delta( \phi'_a - \phi_a )$.
Within the full Hilbert space, 
we consider states which are invariant 
under permutations of the $N$ flavors.
Furthermore, we focus on wavefunctions 
that depend only on the first harmonics of $\phi_a$
through 
$O_c =  \sum_{a=1}^{N} \cos \phi_a$ and
$O_s =  \sum_{a=1}^{N} \sin \phi_a$.
Wavefunctions that depend on $O_c$ and $O_s$
can be spanned by two-parameter family of 
`plane waves', $e^{i ( k_c O_c +k_s O_s)}$.
We denote the sources for $O_c$ and $O_s$ 
as $k_c =  e^{3 \alpha} \cosh \fs$
and $k_s = e^{3 \alpha} \sinh \fs$ to label
the basis states in terms of two non-compact variables $(\alpha, \sigma)$ as 
\bqa
\cb \alpha, \sigma \rb =  
\int_0^{2 \pi} \prod_{a=1}^N d\phi_a ~
\cb \phi \rb
\Psi( \phi; \alpha,\sigma),
\label{eq:as}
\eqa
where the wavefunction is written as
\bqa
\Psi(\phi; \alpha, \sigma) =
%\frac{3 \sqrt{N}}{ 2^{1/4}  \sqrt{4 \pi} (2\pi)^{N/2}} e^{3 \alpha}
\frac{1}{ (2\pi)^{N/2}} 
e^{ i e^{3 \alpha} \left[
 \cosh \fs ~O_c
 + \sinh \fs ~ O_s \right]}.
\label{Psi}
\eqa 
Here 
%%$\alpha, \sigma$ are two non-compact variables, and
the normalization is chosen such that
$\lb \alpha, \sigma \cb \alpha, \sigma \rb =1$.
%% are variational parameters in the wavefunction for the $N$ fundamental variables,
%%
${\cal V}$ denotes the sub-Hilbert space
spanned by 
\begin{equation}
\Bigl\{ \cb \alpha, \sigma \rb ~ \cb ~ -\infty < \alpha < \infty, -\infty < \sigma < \infty \Bigr\}.
\label{eq:V}
\end{equation}
%%where states in ${\cal V}$ are written as
%%and obeys $\Psi(-\phi; \alpha, -\sigma) = \Psi(\phi;\alpha,\sigma)$.
The wavefunction $\Psi(\phi;\alpha,\sigma)$ can be viewed as a tensor
which depends on $\phi, \alpha, \sigma$
as is shown in \fig{fig:Psi}.
If we considered more general wavefunctions that include higher harmonics,
we would have to introduce more collective variables
to span the extended Hilbert space.
However, we focus on the two-parameter family of basis states in our discussion
to keep the form of wavefunction simple.
We are mainly interested in constructing a simple example 
of emergent cosmology to demonstrate the proof of principle
discussed in Sec. \ref{sec:main}.

%Although ${\cal V}$ is parameterized by two non-compact variables,
%states with different $\alpha,\sigma$ are not orthogonal in general.
The overlap between states in ${\cal V}$ is given by
\bqa
\lb \alpha', \sigma' \cb \alpha, \sigma \rb & = & 
%%\frac{9 N e^{3( \alpha' + \alpha) } }{4 \pi \sqrt{2}} 
\left[
\int_0^{2\pi} \frac{ d \phi}{2 \pi}  
e^{ i  ( e^{3 \alpha} \cosh \fs - e^{3 \alpha'} \cosh \fs' ) \cos \phi 
+
i ( e^{3 \alpha} \sinh \fs - e^{3 \alpha'} \sinh \fs' ) \sin \phi }
\right]^N \nn 
& = & 
%%\frac{9 N e^{3( \alpha' + \alpha) } }{4 \pi \sqrt{2}} 
\left[
I_0( i R_{\alpha',\sigma';\alpha,\sigma} )
\right]^N,
\label{innera}
\eqa
where $I_0(x)$ is the modified Bessel function, 
and
\bqa
R_{\alpha',\sigma';\alpha,\sigma} 
= \sqrt{ 
\left( e^{3 \alpha} \cosh \fs - e^{3 \alpha'} \cosh \fs' \right)^2 +
\left( e^{3 \alpha} \sinh \fs - e^{3 \alpha'} \sinh \fs' \right)^2 
}
\label{R}
\eqa 
is a measure of distance between two states.
For $N \gg 1$, 
the Bessel function decays exponentially in $R$ 
as
$\left[ I_0(iR) \right]^N  \approx e^{-N \frac{R^2}{4} }$.
Roughly speaking, two states with 
$R_{\alpha',\sigma';\alpha,\sigma} $ 
greater than $N^{-1/2}$ 
are orthogonal.

In the large $N$ limit, the overlap is proportional to the delta function
upto a multiplicative factor,
\bqa
\lim_{N \rightarrow \infty} 
\lb \alpha', \sigma' \cb \alpha, \sigma \rb & = & 
\mu(\alpha,\sigma)^{-1} \delta( \alpha' - \alpha) \delta( \sigma' - \sigma),
\label{ortho}
\eqa
where
\bqa
\mu(\alpha,\sigma) = \frac{9 N e^{6 \alpha } }{4 \pi \sqrt{2}}.
\eqa
The overlap defines a natural measure
in the space of $\alpha, \sigma$,
\bqa
D\alpha D\sigma \equiv 
\mu(\alpha,\sigma) 
d\alpha d\sigma,
\label{eq:measure}
\eqa
which guarantees that 
\bqa
\lim_{N \rightarrow \infty} 
\int D\alpha D\sigma ~ \lb \alpha', \sigma' \cb \alpha, \sigma \rb & = & 1
\label{norm0}
\eqa
for any $\alpha'$ and $\sigma'$. 
%%Physically,  \eq{eq:measure} assigns 
%%the unit volume to every region in the space of collective variables
%%occupied by a linearly independent state.

\subsection{Hamiltonian for induced quantum cosmology}
\label{sec:miniB}

We emphasize that $\phi_a$'s, which we call `matter fields', 
are the only fundamental degrees of freedom.
$\{ \alpha, \sigma \}$ parameterizes collective modes of the matter fields.
If a Hamiltonian for the matter fields generates a dynamical flow within $\cal V$,
the dynamical flow can be understood as an evolution
generated by an induced Hamiltonian for the collective variables.
In the following, we  construct a Hamiltonian for the matter fields
which induces a minisuperspace quantum cosmology
for the collective variables, 
where $\alpha$ and $\sigma$ become
the scale factor of a flat universe and a scalar field, respectively.

We first look for a Hamiltonian
$\hat \cH(\alpha,\sigma)$
whose action on $\cb \alpha, \sigma \rb$ induces 
\bqa
\hat \cH(\alpha,\sigma) \cb \alpha, \sigma \rb =  h_{\alpha,\sigma} \cb \alpha, \sigma \rb,
\label{eq:12}
\eqa
where $h_{\alpha,\sigma}$ is a Wheeler-DeWitt differential operator
for the minisuperspace cosmology
of a flat three-dimensional universe,
\bqa
h_{\alpha,\sigma} & = & \frac{1}{2} \left[  
\kappa^2 
e^{-3 \alpha}
\left( 
  \frac{\partial^2}{\partial \alpha^2} 
-  \frac{\partial^2}{\partial \sigma^2} \right)
+ \frac{ e^{3 \alpha}}{\kappa^2} V(\sigma)
\right]
\label{Has2}
\eqa
with a potential $V(\sigma)$.
It is not difficult to construct a Hamiltonian
that does the job for a given state $\cb \alpha, \sigma \rb$.
We try the standard quadratic kinetic term with a potential term,
\bqa
\hat \cH(\alpha,\sigma) &= & 
\frac{ 1}{2 \sqrt{N}} \left[
\frac{ e^{-3 \alpha} }{2} \sum_a  \hat \pi_a^2  
+e^{3 \alpha}  
U(\hat \phi, \alpha,\sigma)
\right],
\label{eq:quadp}
\eqa
where  $\hat \pi_a$ is the conjugate momentum for $\hat \phi_a$
with the commutation relation $[ \hat \pi_a, \hat \phi_b ] = -i \delta_{a,b}$.
%Here the volume factor $e^{3 \alpha}$ is chosen 
%for the minisuperspace cosmology
%deduced from a three dimensional theory. 
Requiring that $ \hat \cH(\alpha,\sigma) \cb \alpha, \sigma \rb$ is in ${\cal V}$
fixes $U(\hat \phi, \alpha,\sigma)$ to be 
\bqa
U(\hat \phi, \alpha,\sigma)&=& 
 -\frac{ 1}{2} \sum_a \left( \cos 2 \hat \phi_a +1 \right)
- \frac{\sinh \fst}{2} \sum_{a \neq b} \cos \hat \phi_a \sin \hat \phi_b \nn
&& 
-\frac{ \cosh \fst + 3}{4} \sum_{a\neq b} \cos \hat \phi_a \cos \hat \phi_b 
-\frac{ \cosh \fst - 3}{4} \sum_{a\neq b} \sin \hat \phi_a \sin \hat \phi_b.
\eqa
Applying $ \hat \cH(\alpha,\sigma)$ to $\cb \alpha, \sigma \rb $,
one indeed obtains \eq{eq:12} with
$V(\sigma) =  \frac{1}{18} \left( \cosh 3\sqrt{2}  \sigma -1 \right) $
%\eqa
and $\kappa^2 = \frac{1}{9\sqrt{N}}$.
\eq{eq:12} implies that the action of $\hat \cH(\alpha,\sigma)$ 
on $\cb \alpha, \sigma \rb$
is equivalent to a differential operator acting
on the collective variables. 
This proves that  $ \hat \cH(\alpha,\sigma) \cb \alpha, \sigma \rb $
is in ${\cal V}$.
The induced differential operator $h_{\alpha,\sigma}$ is $O(\sqrt{N})$
because
$\frac{\partial}{\partial \alpha} \sim \frac{\partial}{\partial \sigma} \sim \sqrt{N}$
to the leading order in the large $N$ limit
\footnote{
This can be understood from 
$ \int \prod_a d \phi_a  
\Psi^*(\phi;\alpha,\sigma)
\frac{\partial^2}{\partial \alpha^2}
\Psi(\phi;\alpha,\sigma)$
$\sim
\int \prod_a d \phi_a  
\Psi^*(\phi;\alpha,\sigma)
\frac{\partial^2}{\partial \sigma^2}
\Psi(\phi;\alpha,\sigma)
\sim N$.
}.

$\hat \cH(\alpha,\sigma)$, as a quantum operator of the matter fields, depends on $\alpha,\sigma$.
This means that $\hat \cH(\alpha,\sigma)$ can not generate the desired dynamics for general states in ${\cal V}$, that is,
$\hat \cH(\alpha,\sigma) \cb \alpha', \sigma' \rb \neq  h_{\alpha',\sigma'} \cb \alpha', \sigma' \rb$
if  $(\alpha', \sigma') \neq (\alpha, \sigma)$.
In order for the Hamiltonian flow to stay within ${\cal V}$ for arbitrary initial states in ${\cal V}$,
one effectively has to choose different Hamiltonians for states with different collective variables.
%%Of course, such an operator is not a linear map if
%%states with different collective variables are not orthogonal.
%%However, 
Such a `state-dependent' operator
can be realized through a linear map in the large $N$ limit
because states with different collective variables are orthogonal
in the large $N$ limit as is shown in \eq{ortho}. 
Based on this intuition, we consider the following Hamiltonian,
\bqa
\hat H = \frac{1}{2} 
\int D\alpha D \sigma ~ 
\Bigl[
\hat \cH(\alpha,\sigma) \cb \alpha,\sigma \rb \lb \alpha, \sigma \cb
+ \cb \alpha,\sigma \rb \lb \alpha, \sigma \cb \hat \cH^\dagger(\alpha,\sigma) 
\Bigr].
\label{Has}
\eqa
It is noted that
$\hat P_{\alpha,\sigma} \equiv  \cb \alpha,\sigma \rb \lb \alpha, \sigma \cb$
becomes orthogonal projection operators in the large $N$ limit. 
%For a finite $N$, $\hat P_{\alpha,\sigma} \hat P_{\alpha',\sigma'} \neq 0$
%for $\alpha \neq \alpha'$, $\sigma \neq \sigma'$.
$\hat H$ is made of the projection operator and $\hat \cH(\alpha,\sigma)$.
In the large $N$ limit, 
the projection operator first picks 
a state with a definite $\{ \alpha,\sigma \}$ 
before $\hat \cH(\alpha,\sigma)$ is applied
\footnote{
For a finite $N$, 
there is a smearing
of $\delta \alpha, \delta \sigma \sim \frac{1}{\sqrt{N}}$.
}.
This way, the operator tailored for 
each set of collective variables is applied
to the state with the corresponding collective variables.
Because $\hat \cH(\alpha,\sigma)$
depends on $\alpha, \sigma$, 
one may regard $\hat H$ as a state dependent operator
whose action on the Hilbert space depends on states it acts on\cite{Papadodimas2013}.
However, it is still a linear operator\footnote{
For example, 
$\hat H = \hat \tau_x \frac{ 1 + \hat \tau_z}{2}
+ \hat \tau_y \frac{1 - \hat \tau_z}{2}$
acts as $\hat \tau_x$ and $\hat \tau_y$
on $\cb \uparrow \rb$ and$\cb \downarrow \rb$, respectively.
}.

\eq{eq:12} allows us to write $\hat H$ as
\bqa
\hat H 
%&=& \frac{1}{2} 
%\int D\alpha D \sigma ~ 
%\Bigl[
%\left( h_{\alpha,\sigma} \cb \alpha,\sigma \rb \right) \lb \alpha, \sigma \cb
%+ \cb \alpha,\sigma \rb \left( h_{\alpha,\sigma}^* \lb \alpha, \sigma \cb \right) 
%\Bigr] \nn
&=& 
\int D\alpha D \sigma ~ 
\left( \tilde H_{\alpha,\sigma} \cb \alpha,\sigma \rb \right) \lb \alpha, \sigma \cb,
\eqa
where
$
\tilde H_{\alpha,\sigma}= \frac{1}{2} 
\left(
h_{\alpha,\sigma}
+ h^\dagger_{\alpha,\sigma}
\right)
$ and
$h^\dagger_{\alpha,\sigma}$ is the Hermitian conjugate of $h_{\alpha,\sigma}$
defined from 
\bqa
\int D\alpha D\sigma ~f^*(\alpha,\sigma) \left[ h_{\alpha,\sigma} g(\alpha,\sigma) \right]
= 
\int D\alpha D\sigma ~\left[ h_{\alpha,\sigma}^\dagger f(\alpha,\sigma) \right]^*
g(\alpha,\sigma).
\label{Herm1}
\eqa
It is noted that $h_{\alpha,\sigma}^\dagger$ differs from $h_{\alpha,\sigma}$ 
only by terms that are at most $O(1)$.

\begin{figure}[ht]
\begin{center}
\includegraphics[scale=0.6]{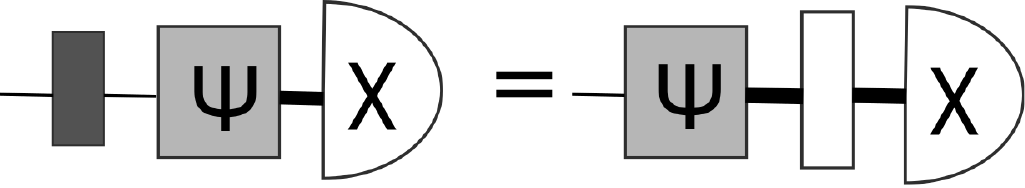}
\end{center}
\caption{
The filled box represents an operator 
that acts on a quantum state of the matter field,
$\cb \chi \rb = \int d\alpha d\sigma ~ \cb \alpha, \sigma \rb \chi(\alpha,\sigma)$.
If the operator is an endomorphism of ${\cal V}$,
it can be represented as an operator (represented by the empty box) 
that acts on the wavefunction $\chi(\alpha,\sigma)$ for the collective variables.
}
\label{fig:precess}
\end{figure}

For general states constructed from  linear superpositions of $\cb \alpha, \sigma \rb$, 
\bqa
\cb \chi \rb = \int D \alpha D \sigma ~ 
\cb \alpha, \sigma \rb
\chi( \alpha, \sigma ),
\eqa
$\hat H$ acts as
\bqa
\hat H \cb \chi \rb 
%&=&  \int D\alpha D\sigma D \alpha' D\sigma' ~ 
%\left( \tilde H_{\alpha,\sigma} \cb \alpha, \sigma \rb \right)  \lb \alpha,\sigma \cb \alpha', \sigma' \rb
%\chi(\alpha',\sigma') 
%\nn
&=& \int D\alpha D\sigma ~ 
\cb \alpha, \sigma \rb
\left( H_{\alpha,\sigma} \chi(\alpha,\sigma) \right),
\label{trans}
\eqa
where
\bqa
H_{\alpha,\sigma} ~ \chi(\alpha,\sigma)  
\equiv  \tilde H_{\alpha,\sigma} 
\int D \alpha' D\sigma' ~ 
\lb \alpha,\sigma \cb \alpha', \sigma' \rb
\chi(\alpha',\sigma').
\label{conv}
\eqa
Therefore, the Hamiltonian for the matter fields translates to a linear operator
acting on the wavefunction of the collective variables. 
This is illustrated in \fig{fig:precess}.
The induced Hamiltonian for the collective variables 
consists of two parts.
The first is the convolution of the wavefunction with 
$\lb \alpha,\sigma \cb \alpha', \sigma' \rb \approx e^{-\frac{N}{4} R_{\alpha,\sigma;\alpha',\sigma'}^2}$.
The convolution smears out sharp features 
which vary at scales shorter than 
$\Delta R_{\alpha,\sigma; \alpha', \sigma'} \sim N^{-1/2}$
%%( e^{3\alpha \pm \frac{3}{\sqrt{2}} \sigma } )   
in the space of $\alpha, \sigma$.
For slowly varying $\chi(\alpha,\sigma)$ with 
$\frac{1}{ \sqrt{N}  } \frac{ \partial \ln \chi(\alpha,\sigma)}{ \partial \alpha} ,
\frac{1}{ \sqrt{N} } \frac{ \partial \ln \chi(\alpha,\sigma)}{ \partial \sigma} \ll 1$,
$\int D \alpha' D\sigma' ~ 
\lb \alpha,\sigma \cb \alpha', \sigma' \rb
\chi(\alpha',\sigma') \propto \chi(\alpha,\sigma)$,
and the convolution merely changes
the normalization of $\chi(\alpha,\sigma)$.
The second is the minisuperspace Wheeler-DeWitt Hamiltonian, $\tilde H_{\alpha, \sigma}$.
Combined, $H_{\alpha,\sigma}$ can be understood as a 
regularized Wheeler-DeWitt Hamiltonian.

\subsection{Emergent time as a Goldstone mode}

\label{sec:miniC}

\eq{trans} implies that $\hat H$ 
induces the Wheeler-DeWitt Hamiltonian of a minisuperspace cosmology.
In gravity, time evolution is a part of diffeomorphism,
and states that are invariant under diffeomorphism
are annihilated by the Hamiltonian.
A state that satisfies the Hamiltonian constraint
represents a whole history rather than a moment of time.
Recovering time from a stationary state 
is the problem of time in gravity\cite{1992grra.conf..211K,1992gr.qc....10011I}.
States that satisfy the Hamiltonian constraint correspond to 
quantum states with zero energy.
For a finite $N$, however, 
there is no guarantee that there exists a state with zero energy in $\cal V$.
This is because the configuration space is compact,
and the energy level is discrete.
In order to check this explicitly,
we diagonalize $\hat H$ in  \eq{Has}.
The matrix elements of the Hamiltonian is written as
\bqa
&& \lb \phi' \cb \hat H \cb \phi \rb 
= \frac{1}{4 \sqrt{N}} \int D \alpha D \sigma ~ 
\Biggl[ \left( - \frac{e^{-3\alpha}}{2} \sum_a \frac{\partial^2}{\partial \phi_a'^2} + e^{3 \alpha} U(\phi',\alpha,\sigma) \right)
 \Psi(\phi', \alpha, \sigma) \Biggr] \Psi^*(\phi, \alpha, \sigma ) \nn
&& + \frac{1}{4 \sqrt{N}} \int D \alpha D \sigma ~  \Psi(\phi', \alpha, \sigma ) 
\Biggl[ \left( - \frac{e^{-3\alpha}}{2} \sum_a \frac{\partial^2}{\partial \phi_a^2} + e^{3 \alpha} U(\phi,\alpha,\sigma) \right)
 \Psi^*(\phi, \alpha, \sigma) \Biggr].
\eqa
Explicit integrations over $\alpha, \sigma$ result in
\bqa
&& \lb \phi' \cb \hat H \cb \phi \rb 
=
 i N \text{sgn}(\Delta_c)  \frac{   \Theta(\Delta_c^2-\Delta_s^2)  }{  2^{N+5} \pi^{N}  (\Delta_c^2-\Delta_s^2)^{5/2}} 
\Biggl\{
  \frac{ 4 }{9 \kappa^2 }   \left(\Delta_c^2+2 \Delta_s^2\right)
 \nn
&& + 9 \kappa^2    \Bigl[
6  \Delta_s  \Delta_c  (O_c' O_s'+O_c O_s)
+\Delta_c^4     -   3 \Delta_c^2 ( O_c'^2 + O_c^2) 
 - 2 \Delta_s^2  \left( \Delta_c^2 + O_s'^2 + O_s' O_s + O_s^2 \right)
   \Bigr]
   \Biggr\}   \nn
 &&   
 + N \frac{    \Theta(\Delta_s^2-\Delta_c^2)  }{  2^{N+4} \pi^{N}   (\Delta_s^2-\Delta_c^2)^{5/2}} 
  \Biggl\{ 
   -\frac{2}{9 \kappa^2 } \left(\Delta_c^2+2 \Delta_s^2\right) 
  + 9 \kappa^2  \Bigl[ 
     (O_c^2-O_s^2 )^2
  + ( O_c'^2 - O_s'^2)^2   \nn
  &&
    + O_c' O_c \Delta_s^2 
    + ( O_s' O_s - O_c' O_c ) ( O_c'^2 + O_c^2 )
    + O_s'^2 O_c^2 + O_s^2 O_c'^2
    -O_s O_s' ( O_s'^2 + O_s^2 )
   \Bigr]
      \Biggr\},
 \label{OH}
\eqa
where
 $\Delta_c = O_c' - O_c$ and $\Delta_s = O_s' - O_s$
 with
$O_c =  \sum_{a=1}^{N} \cos \phi_a$,
$O_s =  \sum_{a=1}^{N} \sin \phi_a$,
$O_c' =  \sum_{a=1}^{N} \cos \phi_a'$,
$O_s' =  \sum_{a=1}^{N} \sin \phi_a'$.
We note that the matter fields
are subject to strong all-to-all interactions.

\begin{figure}[ht]
 \begin{center}
 \subfigure[]{
\includegraphics[scale=0.85]{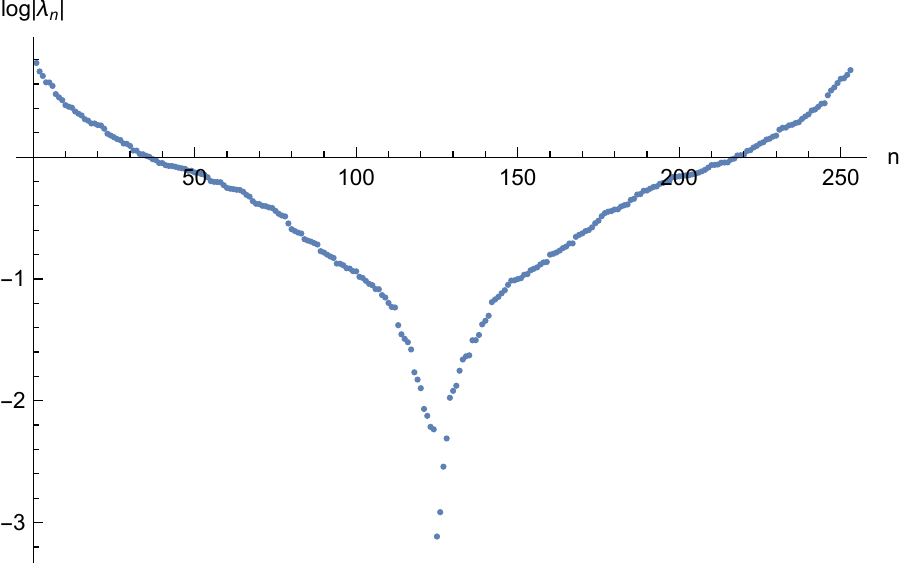} 
 \label{fig:12}
 } 
\hfill
 \subfigure[]{\includegraphics[scale=0.85]{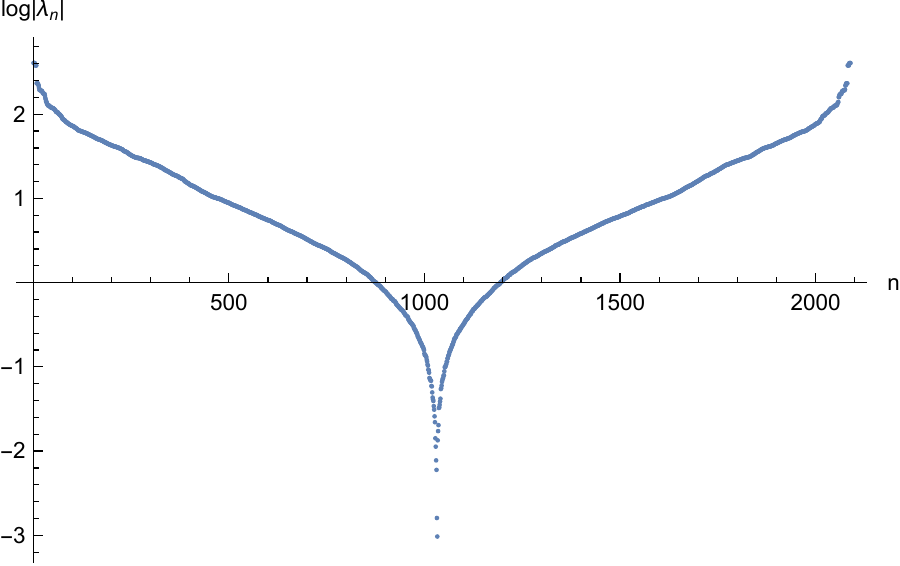} 
 \label{fig:24}
 }
\\
 \subfigure[]{\includegraphics[scale=0.85]{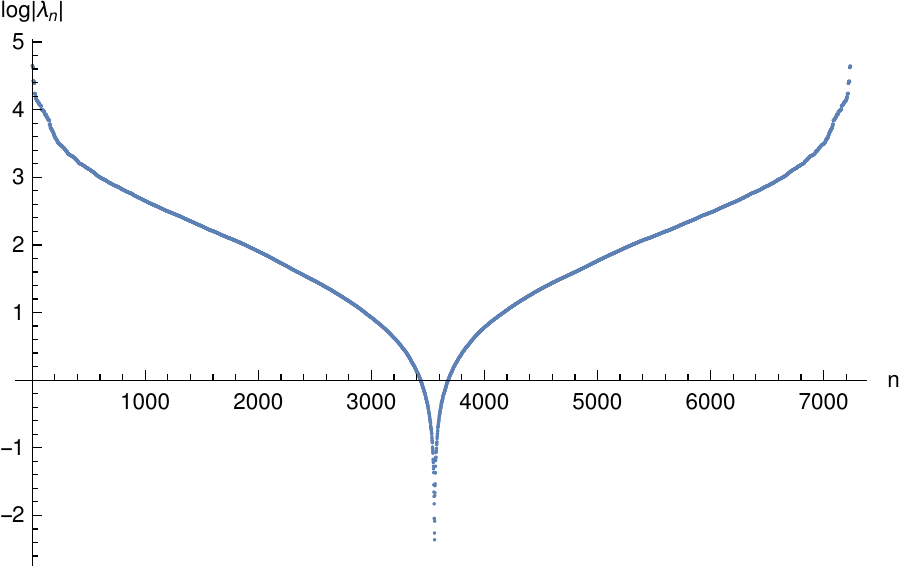} 
 \label{fig:36}
 }
\hfill
 \subfigure[]{\includegraphics[scale=0.85]{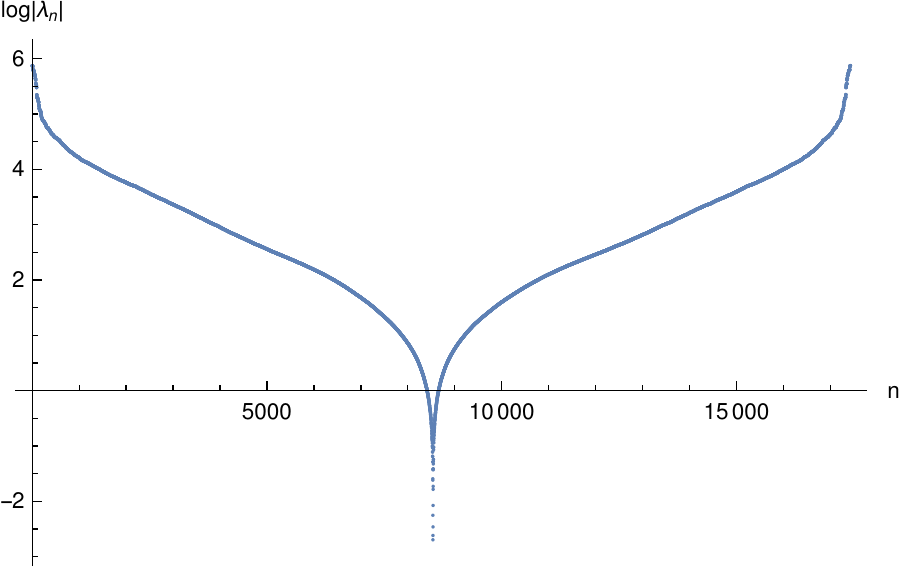} 
 \label{fig:48}
 }
 \end{center}
 \caption{
The eigenvalues of $\hat H$ whose eigenstates
have nonzero projection to ${\cal V}$.
The $x$-axis is a label of the eigenvalues sorted in the ascending order, 
and the $y$-axis denotes logarithm of the absolute magnitude of eigenvalues.
The eigenvalues on the left (right) of the dip are negative (positive).
The Hamiltonian has been numerically diagonalized for $N=3$
by discretizing the compact space of each $\phi_a$ into $L$ segments, where 
(a) $L=12$,
(b) $L=24$,
(c) $L=36$,
(d) $L=48$.
With increasing $L$, 
the bandwidth of the eigenvalues keeps increasing,
which reflects the fact that the spectrum is unbounded 
both from the above and below in the continuum limit.
On the other hand, the eigenvalue that is smallest in magnitude saturates to a nonzero value.
This suggests that there is no state in ${\cal V}$ with zero energy in the continuum limit.
 }
 \label{fig:Eigen}
 \end{figure}

We numerically diagonalize $\hat H$ for $N=3$.
Indeed, all eigenstates which have nonzero projection in ${\cal V}$ 
have non-zero eigenvalues, as is shown in \fig{fig:Eigen}.
This may seem contradictory because 
$H_{\alpha,\sigma} \chi_0(\alpha,\sigma) = 0$
is a hyperbolic equation, which can be 
solved once a boundary condition is provided.
Solutions to the Wheeler-DeWitt equation formally give zero energy states.
%$\cb 0 \rb =  \int D \alpha D \sigma ~ 
%\cb \alpha, \sigma \rb
%\chi_0( \alpha, \sigma ) 
%$.
The reason why such states do not appear in the spectrum is because
they are not normalizable\cite{1992grra.conf..211K,1992gr.qc....10011I}. 

The fact that there is no normalizable state which 
satisfies the Hamiltonian constraint means that 
a physical state in ${\cal V}$ inevitably breaks the time translational symmetry
by the virtue of having a finite norm.
This is analogous to the fact that 
states that are invariant under spatial translations 
in the Euclidean space are not normalizable, 
and physical states (such as wave packets) necessarily break the translational symmetry.  
However, there exist normalizable semi-classical states 
which are annihilated by $\hat H$ to the leading order in $1/\sqrt{N}$
and break the symmetry only weakly,
\bqa
\chi(\alpha,\sigma) =  e^{ -\frac{(\alpha-\bar \alpha)^2 + (\sigma-\bar \sigma)^2}{2 \Delta^2} }
e^{  \frac{i}{\kappa^2} \left( \bar \pi \alpha + \bar \pi_\sigma \sigma \right) },
\label{semi}
\eqa
where $\bar \alpha, \bar \sigma, \bar \pi, \bar \pi_\sigma$ are classical 
coordinates and momenta which satisfy
\bqa
 e^{-3 \bar \alpha} ( -\bar \pi^2 + \bar \pi_\sigma^2 ) + e^{3 \bar \alpha} V(\bar \sigma) = 0.
\eqa
$\Delta$ determines the uncertainty of $\alpha$ and $\sigma$.
With $\kappa^2 \ll \Delta \ll 1$,
\eq{semi} represents a semi-classical state in 
which both coordinates and momenta are well defined with 
$\delta \alpha ~ \delta \pi  \sim \delta \sigma ~ \delta \pi_\sigma 
\sim \kappa^2 \sim \frac{1}{ \sqrt{N} }$,
and the Hamiltonian constraint is satisfied to the leading order in $1/\sqrt{N}$.

 \begin{figure}[ht]
 \begin{center}
 \subfigure[]{\includegraphics[scale=0.4]{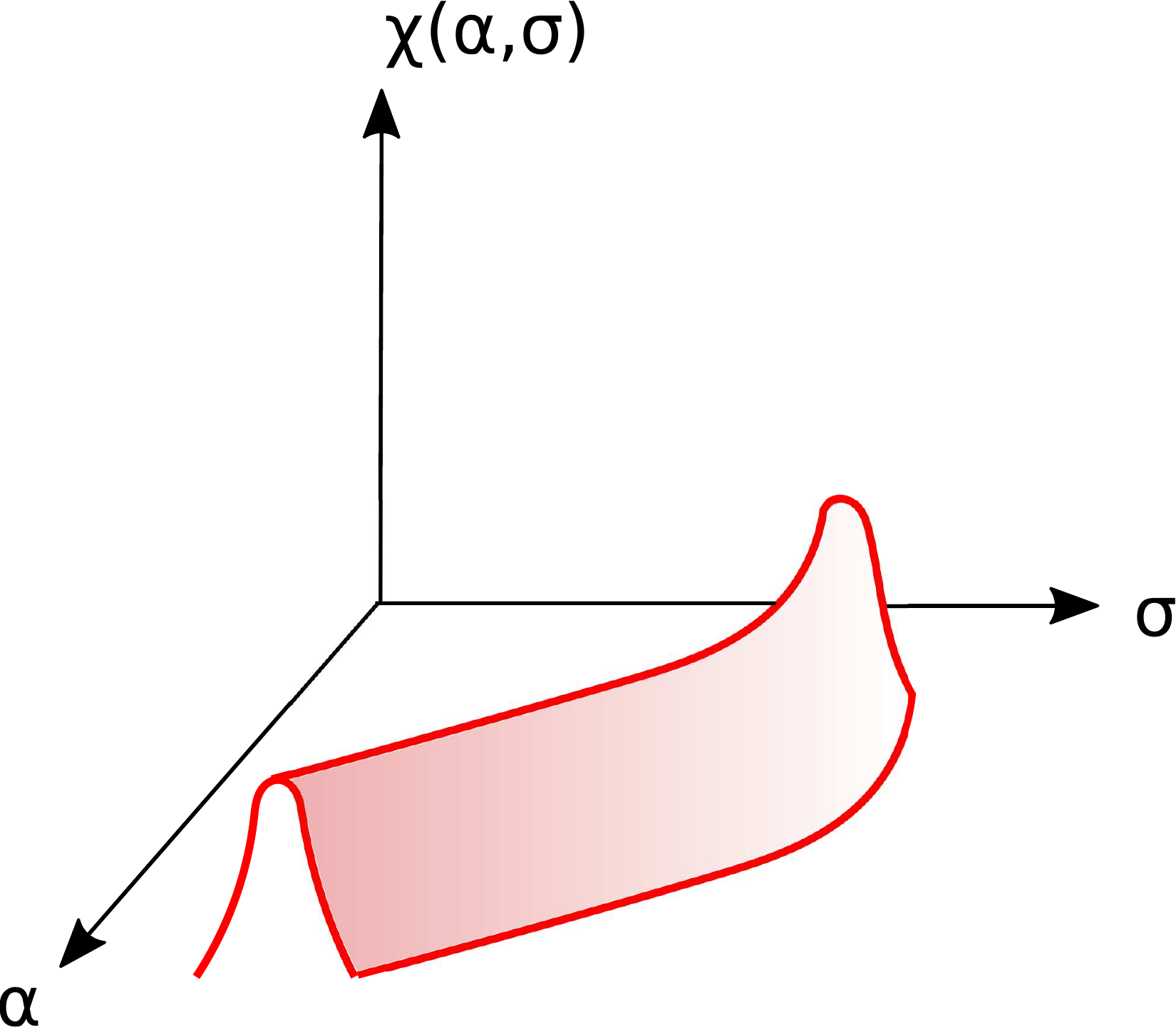} 
 \label{fig:wf1}
 } \hfill
 \subfigure[]{\includegraphics[scale=0.4]{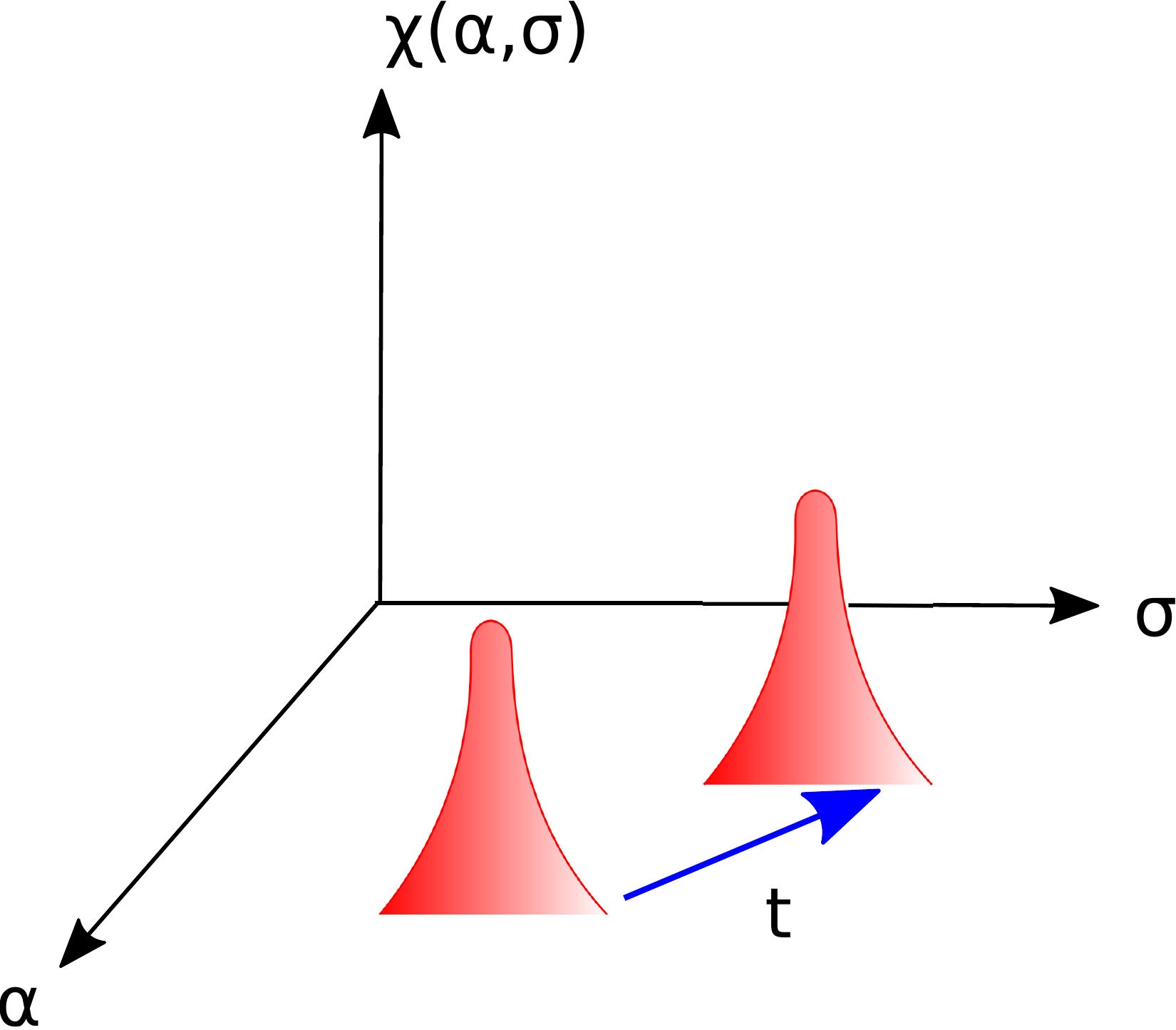} 
 \label{fig:wf2}
 }
 \end{center}
 \caption{
(a) States that satisfy the Hamiltonian constraint 
are extended in the space of the collective variables,
and are not normalizable.
(b) Physical states with finite norms spontaneously break the symmetry generated by the Hamiltonian.
The Goldstone mode associated with the spontaneously broken symmetry gives rise to a non-trivial time evolution.
 }
 \label{fig:emergentT}
 \end{figure}

Because semi-classical states are not exactly annihilated by $\hat H$,
they spontaneously break the symmetry generated by the Hamiltonian.
The spontaneous symmetry breaking amounts to picking a moment of time in a history.
The following evolution of the state generated by $\hat H$ creates one-parameter family of states.
The evolution can be viewed as 
a Goldstone mode associated with the 
spontaneously broken symmetry.
This is illustrated in  \fig{fig:emergentT}.
We call the parameter along the orbit $t$.
However, $t$ itself is not a physical observable 
because there is no independent way of measuring $t$
in a closed quantum system.
It is merely a parameter that labels a sequence of states generated by the Hamiltonian evolution. 
What is physical is relation between physical observables,
e.g., the value of $\sigma$ when $\alpha$ takes a certain value.

%%%%%%%
%%%%NEW
%%%%%%%
Now we examine how  semi-classical states evolve under $\hat H$.
Since \eq{semi} has a fast oscillating phase factor,
the convolution integration in \eq{conv} gives rise to a 
suppression in the norm of the wavefunction,
\bqa
\int D \alpha' D\sigma' ~ 
\lb \alpha,\sigma \cb \alpha', \sigma' \rb
\chi(\alpha',\sigma') \approx A_\chi \chi(\alpha,\sigma),
\label{suppression}
\eqa
where
$A_\chi = e^{
-  18 e^{-6 \bar \alpha}
\left(
\frac{ \cosh 3 \sqrt{2} \bar \sigma}{2} ~ \bar \pi^2
- \sqrt{2} \sinh 3 \sqrt{2} \bar \sigma  ~ \bar \pi \bar \pi_\sigma
+ \cos 3 \sqrt{2} \bar \sigma ~ \bar  \pi_\sigma^2
\right)
} < 1$,
and it is used that $\chi(\alpha,\sigma)$ is sharply peaked
at $(\bar \alpha, \bar \sigma)$.
As a result, $H_{\alpha, \sigma}$ becomes the Wheeler-DeWitt Hamiltonian
upto a multiplicative factor that depends on the wavefunction,
\bqa
H_{\alpha,\sigma} ~ \chi(\alpha,\sigma)  
\approx A_\chi \tilde H_{\alpha,\sigma} 
\chi(\alpha,\sigma).
\label{conv12}
\eqa
By choosing a lapse that absorbs $A_\chi$,
the state after an infinitesimal step of the parameter time can be written as
\bqa
\cb \chi; dt \rb & = & e^{-i n^{(1)} A_\chi^{-1} \hat H dt } \cb \chi \rb  \nn
%& = & \int D \alpha^{(0)} D  \sigma^{(0)}  ~ \cb  \alpha^{(0)} , \sigma^{(0)}  \rb 
%e^{-i n^{(1)}  \tilde H_{\alpha^{(0)} ,\sigma^{(0)} } dt } \chi(\alpha^{(0)} ,\sigma^{(0)} ) \nn
& = & \int  D \alpha^{(0)} D  \sigma^{(0)}  D \alpha^{(1)} D \sigma^{(1)} D \pi^{(1)} D \pi_\sigma^{(1)}
~ 
\cb  \alpha^{(1)}, \sigma^{(1)} \rb 
e^{\frac{i}{\kappa^2} \left[ \pi^{(1)} ( \alpha^{(1)}  -\alpha^{(0)}) + \pi_\sigma^{(1)} ( \sigma^{(1)} - \sigma^{(0)} ) 
\right] } \nn
&&
e^{-in^{(1)} dt  \frac{e^{-3 \alpha^{(0)}}}{2 \kappa^2} 
 \left[  
   -\pi^{(1)2} + \pi_\sigma^{(1)2} 
+  e^{6 \alpha} V(\sigma^{(0)})
+O(\kappa^2) 
\right]
 } \chi(\alpha^{(0)},\sigma^{(0)}).
 \label{chidt}
\eqa
Here $n^{(1)}$ determines the speed of the flow along the orbit.
$O(\kappa^2)$ represents sub-leading terms 
that are generated from the measure and the smearing. 
The measure for the conjugate momenta has been defined as
$D\pi D\pi_\sigma \equiv \mu(\alpha,\sigma)^{-1} d\pi d \pi_\sigma$.
In the large $N$ limit, \eq{chidt} remains a semi-classical state 
centered at a different classical configuration.
In the next step, we choose the lapse $n^{(2)} A_{\chi(dt)}^{-1}$.
Repeating these steps, one obtains a state at parameter time $t$, 
\bqa
\cb \chi; t \rb &=& e^{-i  \int_0^t d\tau ~ n(\tau) \hat H } \cb \chi \rb \nn
%\label{TE}
%\eqa
%\bqa
%\cb \chi; t \rb 
& = & 
\int  D \alpha(\tau) D \sigma(\tau) D \pi(\tau) D \pi_\sigma(\tau)
~~
\cb \alpha(t), \sigma(t)  \rb
~ e^{iS} ~ \chi(\alpha(0),\sigma(0)),
\eqa
where $n(\tau)$ is a time-dependent speed of time evolution
which can be chosen at one's will, 
and
\bqa
S = \frac{1}{\kappa^2} \int_0^t d \tau 
\left\{
\pi \partial_\tau \alpha + \pi_\sigma \partial_\tau \sigma 
- n(\tau)   \frac{e^{-3 \alpha}}{2} 
 \left[  
    -\pi^{2} + \pi_\sigma^{2} 
+  e^{6 \alpha} V(\sigma)
+O(\kappa^2) 
\right]
\right\}.
\eqa
This is a minisuperspace quantum cosmology
for the three-dimensional flat universe with one scalar field.
In the large $N$ limit, the classical path dominates the path integration.

There is a sense in which the emergent time in the present theory
resembles an internal time generated by relative motions of a subsystem
in stationary states\cite{PhysRevD.27.2885}.
To make the connection, one views $\Psi(\phi;\alpha,\sigma)$ as a wavefunction
of an enlarged system that includes not only the matter fields but also the collective variables 
as independent dynamical degrees of freedom.
In this case, \eq{eq:12} is understood as the Wheeler-DeWitt equation for the whole system
(with a wrong sign in the kinetic term for the matter field).
Although the full state is stationary, 
one defines a time flow in terms of the evolution of the matter fields relative to the collective variables.
What is different in the present construction are two-fold.
First, the collective variables are not independent dynamical degrees of freedom.
Instead, they describe collective excitations of the matter fields. 
Accordingly, the quantization of the collective variables follow from that of the matter fields.
Second, the inability to find normalizable states in the Hilbert space of the matter fields
provides a dynamical mechanism to pick a moment of time 
in the induced theory of cosmology.

\section{Emergent gravity }
\label{sec:gravity}

In this section, we extend the discussion on 
the emergent minisuperspace cosmology 
to gravity in $(3+1)$ dimensions.
%% the main idea sketched  in Sec. \ref{sec:main}.
The biggest difference from the previous section
is that we are now dealing with an infinite dimensional Hilbert space.
To be concrete, we consider an $N \times N$ matrix field 
defined on a three dimensional manifold.
Within the full Hilbert space, we define a sub-Hilbert space of the matter field
that becomes a Hilbert space for two collective fields 
: a spatial metric and a scalar field\footnote{
The collective variables are fields in this case.}.
The sub-Hilbert space is spanned by a set of basis vectors
each of which is labeled by the metric and the scalar field.
As variational parameters of wavefunctions of the matter field,
the spatial metric sets the notion of locality in 
how matter fields are entangled in space,
while the scalar field determines the range of mutual information
in each basis state.
After we discuss the covariant regularization of the wavefunctions
and the inner product within the sub-Hilbert space,
we explain the connection between the collective variables and 
entanglement in details.
Building on the intuitions we learned from the previous two sections,
we then construct a matter Hamiltonian that 
induces the general relativity at long distances 
in the large $N$ limit.

\subsection{Construction of a Hilbert space for metric from matter fields}
\label{sec:gravityA}

In this subsection, 
we define a sub-Hilbert space of a matrix field
and an inner product 
that is invariant under spatial diffeomorphisms.

\subsubsection{Hilbert space}
\label{sec:gravityA1}

We consider an $N \times N$ Hermitian matrix field $\Phi(x)$ 
defined on a compact three dimensional manifold.
The full Hilbert space of the matrix field is spanned by
the eigenstates of the field operator, 
$ \hat \Phi_{ab}(x) \cb \Phi \rb =  \Phi_{ab}(x) \cb \Phi \rb$.
In order to define an inner product 
in the infinite dimensional Hilbert space,
%% the normalization of $\cb \Phi \rb$,
we need to introduce a discrete basis that spans
the space of $\Phi_{ab}(x)$.
For this, we  choose an elliptic differential operator
whose eigenvectors form a complete basis,
%Here we consider the following eigenvalue equation,
%%defined with a metric, $g_{E,\mu \nu}(x)$ and a scalar field, $\sigma(x)$,
\bqa
%% \left[ - g^{\mu \nu} \nabla^g_\mu \nabla^g_\nu  + M^2 \right] 
K_{(E,\sigma)} f^{(E,\sigma)}_n (x) =  \lambda^{(E,\sigma)}_n  f^{(E,\sigma)}_n(x)
\label{Kf}
\eqa
with  
\bqa 
K_{(E,\sigma)} = 
\left[ - g_E^{\mu \nu} \nabla^E_\mu  \nabla^E_\nu   + \frac{e^{2 \sigma}}{\lc^2}  \right].
\label{KES}
\eqa
Here $\nabla^E_\mu$ is the covariant derivative 
defined with respect to a Riemannian metric, $g_{E,\mu \nu}(x)$,
which is parameterized by a triad,
\bqa
g_{E,\mu \nu}(x)  = E_{\mu i}(x) E_{\nu}^i(x).
\label{ge}
\eqa
In \eq{ge}, the local Euclidean index $i$ is raised or lowered
with $\delta^{ij} = \delta_{ij}$, 
and repeated indices are summed over $i=1,2,3$.
$\sigma(x)$ is a scalar that determines the `mass' in the unit of a fixed length scale, $\lc$. 
$\lambda^{(E,\sigma)}_n$ is the $n$-th eigenvalue 
and $f^{(E,\sigma)}_n(x)$  is the eigenfunction with the normalization condition, 
$\int dx |E| f^{(E,\sigma)*}_n(x) f^{(E,\sigma)}_m(x) = \delta_{n,m}$
with $|E| \equiv | det ~ \emi |$.
For a choice of $( E_{\mu i}, \sigma )$, 
the set of eigenvectors 
$\left\{f^{(E,\sigma)}_n(x) \cb n=1,2,... \right\}$ forms a complete basis.
A general field configuration can be decomposed as
$ \Phi_{ab}(x) =  \sum_n \Phi^{(E,\sigma)}_{ab,n}  f^{(E,\sigma)}_n(x) $,
where $\Phi^{(E,\sigma)}_{ab,n}$
represents the amplitude of the $n$-th normal mode
in the basis of $\left\{ f^{(E,\sigma)}_n(x) \right\}$.
In order to define an inner product, 
we choose a fiducial triad and scalar, $( \hat E_{\mu i}, \hat \sigma )$.
In terms of the normal mode associated with 
$K_{(\hat E, \hat \sigma)}$,
the inner product is defined to be
\bqa
\lb \Phi' \cb \Phi \rb = \prod_{a,b} \prod_n 
\left[
\sqrt{\pi}
\delta 
\Bigl( \Phi^{' (\hat E, \hat \sigma)}_{ab,n} - \Phi^{(\hat E, \hat \sigma)}_{ab,n} \Bigr)
\right].
\label{inner_phi}
\eqa
Two states with different amplitudes in any of the normal modes are orthogonal.
The inner product defines a natural measure  for a functional integration of the matter field 
in terms of the normal modes as 
\bqa
D^{(\hat E,\hat \sigma)} \Phi 
\equiv \prod_{a,b} \prod_n 
\left[
\frac{d \Phi^{(\hat E,\hat \sigma)}_{ab,n}}{\sqrt{\pi}}
\right].
\label{hatD}
\eqa
This guarantees that 
\bqa
\int D^{ (\hat E, \hat \sigma) } \Phi ~ \lb \Phi' \cb \Phi \rb f(\Phi)  = f(\Phi')
\eqa
for general functional $f(\Phi)$.
Obviously, the inner product and the measure 
depends on the choice of the fiducial triad and scalar,
$(\hat E_{\mu i}, \hat \sigma)$.
A measure defined in terms of a different triad and scalar field $( E,  \sigma)$
is related to \eq{hatD} through a Jacobian,
\bqa
D^{ ( \hat E, \hat \sigma) } \Phi =
J^{ (\hat E, \hat \sigma)}_{( E,  \sigma) }
D^{ ( E,  \sigma) } \Phi ,
\label{measureT}
\eqa
where
$J^{ (\hat E, \hat \sigma)}_{( E,  \sigma) }$
is the determinant of the matrix, 
\bqa
a_{mn} = \int dx |\hat E| ~ f^{(\hat E,\hat \sigma)*}_m(x) f^{( E,  \sigma)}_n(x).
\label{matrix_element}
\eqa
In general, the Jacobian is not unity.
However, in special cases with $|E(x)| = |\hat E(x)|$, 
$J^{ (\hat E, \hat \sigma)}_{( E,  \sigma) } = 1$
because $(a^{-1})_{nm} = a_{mn}^*$.

\begin{figure}[ht]
 \begin{center}
 \subfigure[]{
\includegraphics[scale=0.6]{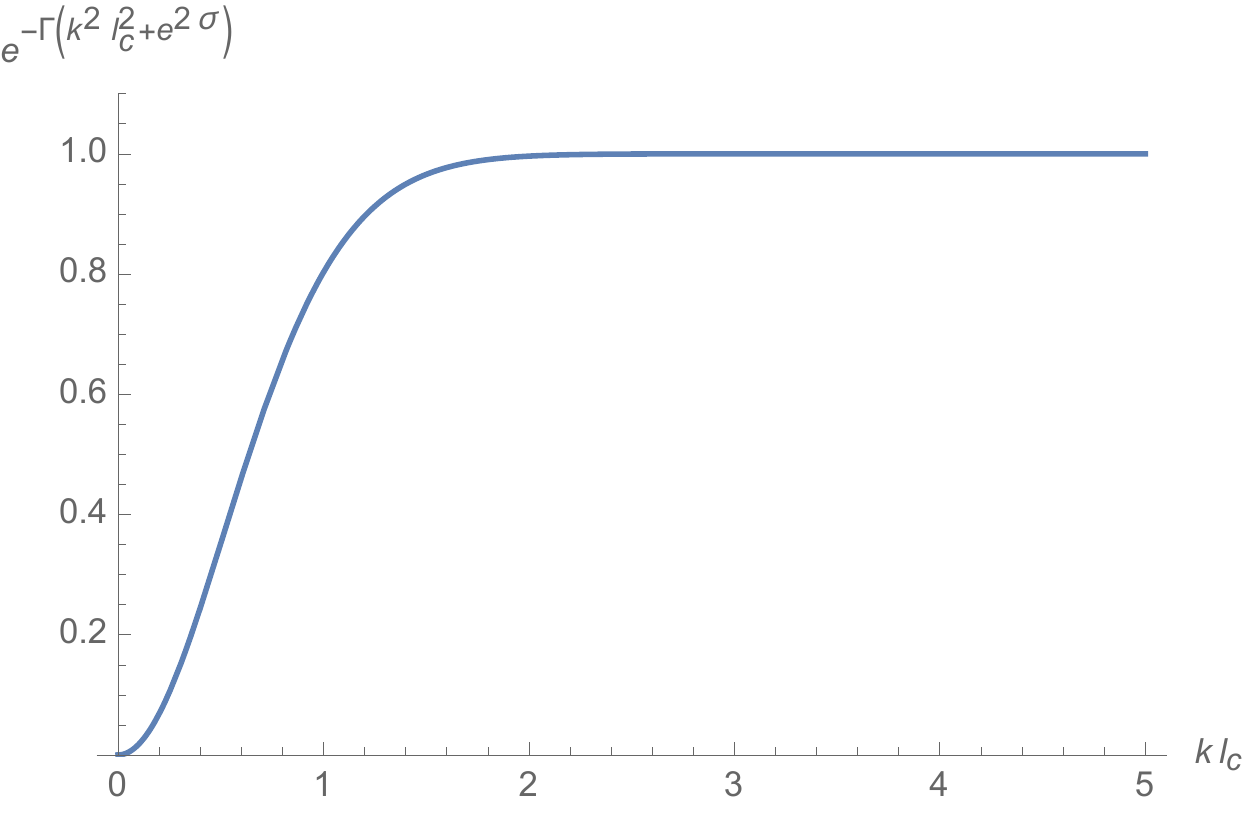} 
\label{fig:Gamma0.01}
 } 
%\hfill
\subfigure[]{
\includegraphics[scale=0.6]{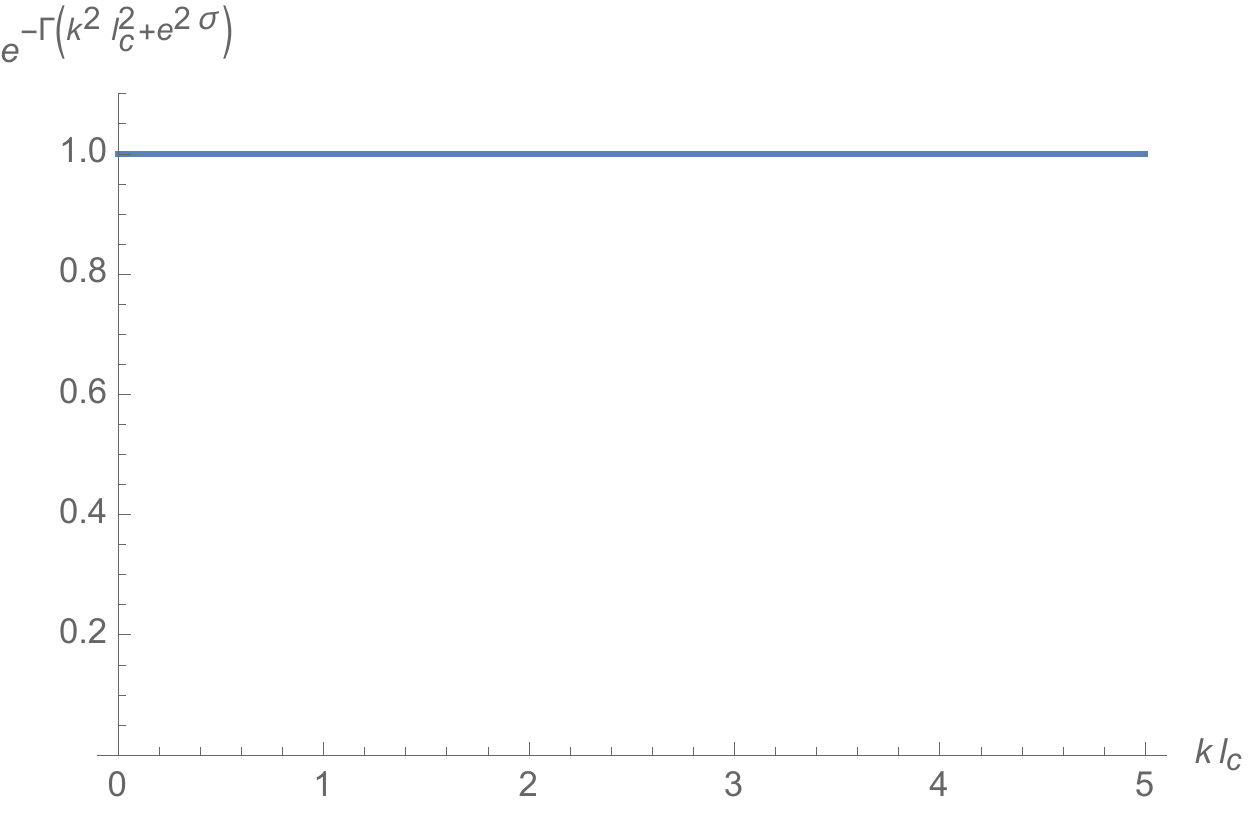} 
\label{fig:Gamma10}
}
\end{center}
\caption{
$e^{
-\Gamma \left[ ( k \lc )^2 + e^{2 \sigma}   \right] 
}$
plotted as a function for $k \lc$ for
(a) $e^\sigma =10^{-2}$
and (b) $e^\sigma =10^2$.
For $e^\sigma \ll 1$, 
the kernel disperses quadratically in $k$ at small momenta
before the dispersion is lost at large momenta with $k \gg \lc^{-1}$.
For $e^\sigma \gg 1$, the momentum dependence is suppressed at all momenta,
and the wavefuction becomes a direct product state in space.
}
\label{fig:Gamma}
\end{figure}

%%
%This defines the full Hilbert space of the matter field and the inner product.
Within the full Hilbert space, 
we focus on singlet states 
that are invariant under 
global $SU(N)$ transformations, 
$\Phi(x) \rightarrow U^\dagger \Phi(x) U$,
where $U$ is SU(N) matrix.
In particular,
we consider a sub-Hilbert space, ${\cal V}$ 
spanned by a set of basis states that are 
labeled by $\{ E_{\mu i}(x), \sigma(x) \}$,
\bqa
\cb E, \sigma \rb 
= \int D^{ (\hat E, \hat \sigma) } \Phi ~
\cb \Phi \rb
~
\Psi( \Phi; E, \sigma ).
%\right\},
\label{g}
\eqa
Here $\Psi( \Phi; E, \sigma )$ is a 
%wavefunction
%of the matter field characterized
%by the metric $g_{E,\mu \nu}$ 
%and a scalar field $\sigma$.
%In particular, we choose 
short-range entangled wavefunction of the matter field
in which the metric ($g_{E,\mu\nu}$) and the scalar field ($\sigma$)
set local structures of entanglement.
%%the metric determines the amount of entanglement
%%a region in space supports, 
%%and the scalar field determines the range of mutual information.
Wavefunctions for such short-range entangled states can be written as 
an exponential of a local functional,
\bqa
  \Psi(\Phi; E, \sigma) = 
    e^{
- \int dx ~ |E(x)| ~ 
{\cal L}[\Phi(x); \emi(x),\sigma(x)]
-  \frac{1}{2} S_0[E,\sigma]
%     S[ \Phi; E, \sigma] 
}
\label{eq:psi}
\eqa 
in which the triad and the scalar
enter as variational parameters.
For simplicity,
we choose ${\cal L}[\Phi(x); \emi(x),\sigma(x)]$ 
to be a gaussian form\footnote{
The gaussian wavefunction has 
the $O(N^2)$ internal symmetry.
},
%${\cal L}$ is a gaussian given by
\bqa
%S[\Phi;E, \sigma] = \frac{ e^{i \epsilon}}{2}  
%S[\Phi;E, \sigma] = 
{\cal L}[\Phi(x); \emi(x),\sigma(x)] =
\frac{1}{2}  
tr\left[ \Phi e^{-  \Gamma \left[ \lc^2 K_{(E,\sigma)} \right] }  \Phi \right].
%+  \frac{1}{2} S_0[E,\sigma].
%%\right).
\label{SG}
\eqa
$tr[ ..]$ denotes the trace over matrix indices.
%%
%%
%%In \eq{SG}, 
%%gradient terms should be regularized. 
%$e^{-  \Gamma \left[ \lc^2 K_{(E,\sigma)} \right] }$ in \eq{SG} is a regularized gradient term,
%where 
$\Gamma[ \lc^2 \lambda ] \equiv \int_{\lc^2}^\infty \frac{dt}{t} e^{ -\lambda t}$
is the incomplete Gamma function
and $\lc$ is the cut-off length scale.
$ \Gamma \left[ \lc^2 K_{(E,\sigma)} \right] $
is a regularized derivative operator 
that creates local entanglement 
at distance scales larger than $\lc$. 
%In order to have finite entanglement for the matter field defined in continuum,
%a short-distance cut-off is introduced as
%
%
It has the following asymptotic behaviors,
\bqa
\Gamma(x) = - \ln x - \gamma_E +O(x) ~~~ \mbox{ for $x \ll 1$}, \nn
\Gamma(x) =  \frac{e^{-x}}{x} \left( 1+ O(x^{-1}) \right)  ~~~ \mbox{ for $x \gg 1$},
\eqa
where $\gamma_E$ is the Euler-Mascheroni constant.
For modes with eigenvalues $\lambda^{(E,\sigma)}_n \ll \lc^{-2}$, 
the kernel becomes the usual two-derivative operator,
$e^{-  \Gamma \left[ \lc^2 K_{(E,\sigma)} \right] }  \sim  K_{(E,\sigma)}$.
At large wavevectors with $\lambda^{(E,\sigma)}_n \gg \lc^{-2}$,
the gradient term is suppressed and one has
$e^{-  \Gamma \left[ \lc^2 K_{(E,\sigma)} \right] }   \approx 1$.
Basically, $e^{-  \Gamma \left[ \lc^2 K_{(E,\sigma)} \right] }$ 
behaves 
as a two-derivative term at long distances
while it becomes a constant a short distances.
Only those modes with wavelengths larger than $\lc$ 
have non-negligible entanglement in space.
A plot of $e^{-  \Gamma \left[ \lc^2 K_{(E,\sigma)} \right] } $ is shown in \fig{fig:Gamma}.
$S_0[E,\sigma]$  is chosen to enforce the normalization condition, $\lb E, \sigma \cb E, \sigma \rb = 1$.
From
\bqa
\lb E, \sigma \cb E, \sigma \rb 
%& = &
%\int D^{(\hat E,\hat \sigma)} \Phi  ~
%e^{  
%- \int d^3 x ~ |E| ~ 
%tr\left[ \Phi e^{-  \Gamma \left[ \lc^2 K_{(E,\sigma)} \right] }  \Phi \right]
%  - S_0[E,\sigma]
% }   \nn
& = &
\int D^{( E, \sigma)} \Phi  ~
  J^{ (\hat E, \hat \sigma)}_{( E,  \sigma) }
~
e^{  
- \int d^3 x ~ |E| ~ 
tr\left[ \Phi e^{-  \Gamma \left[ \lc^2 K_{(E,\sigma)} \right] }  \Phi \right]
  - S_0[E,\sigma]
 },
%\nn & = &
%\exp\left[
%\frac{N^2}{2} 
% ~ \tr{   \Gamma \left[ \lc^2 K_{(E,\sigma)} \right] }
% - S_0[E,\sigma]
% + \ln J^{ (\hat E, \hat \sigma)}_{( E,  \sigma) }
%\right],
\label{muH}
\eqa
we obtain 
\bqa 
S_0[E,\sigma] = 
\frac{N^2}{2}  ~ \tr{   \Gamma \left[ \lc^2 K_{(E,\sigma)} \right] }
 + \ln J^{ (\hat E, \hat \sigma)}_{( E,  \sigma) }.
\label{Sn}
\eqa
Here $\tr{..}$ denotes the trace of differential operators.
In \eq{muH}, 
\eq{measureT} is used.
%$ D^{( \hat E, \hat \sigma)} \Phi  ~
%= D^{( E, \sigma)} \Phi  ~
%  J^{ (\hat E, \hat \sigma)}_{( E,  \sigma) }$.

It is noted that the particular choice in \eq{SG} is not crucial.
In order to include more general wavefunctions,
one needs to introduce more collective variables 
which source different operators in \eq{SG}.
Here we choose the simplest form of wavefunction
to have a tractable example.

\begin{figure}[ht]
\includegraphics[scale=0.50]{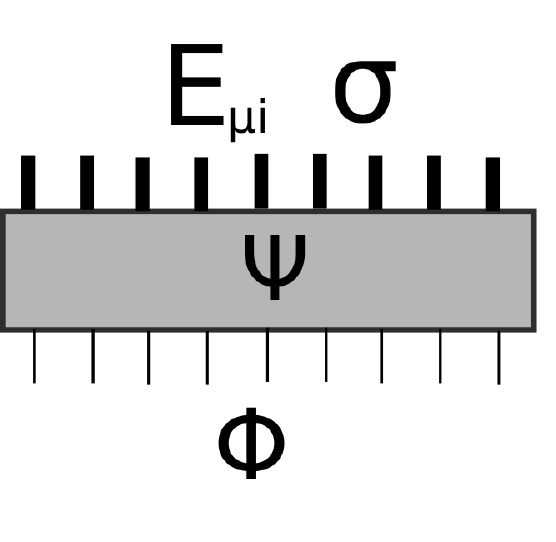}
\caption{A tensor representation
of the wavefunction $\Psi(\Phi; E,\sigma)$,
where the triad and a scalar field are collective variables of the matter field. 
}
\label{fig:Psi2}
\end{figure}

${\cal L}$ can be understood as local tensors that generate short-range entangled states,
$ \Psi(\Phi; E, \sigma) \propto \prod_x e^{  - dx \de ~ {\cal L}[\Phi(x); \emi(x),\sigma(x)] }$.
Here $\emi$ and $\sigma$ play the role of variational parameters (see \fig{fig:Psi2})\cite{Lee2016}.
The metric sets the notion of distance in how matter fields are entangled in \eq{SG}.
Because the proper cut-off length scale below which the matter field 
is unentangled is measured with the metric,
the metric controls the number of degrees of freedom
that participate in entanglement.
On the other hand, $\sigma$ determines the range of mutual information.
In the large $\sigma$ limit, $\Psi(\Phi; E, \sigma) $ becomes a direct product state in real space.
The precise connection between entanglement
and the collective variables
will be established in Sec. \ref{sec:gravityB}.

\begin{figure}[ht]
\begin{center}
\includegraphics[height=6.5cm,width=4cm]{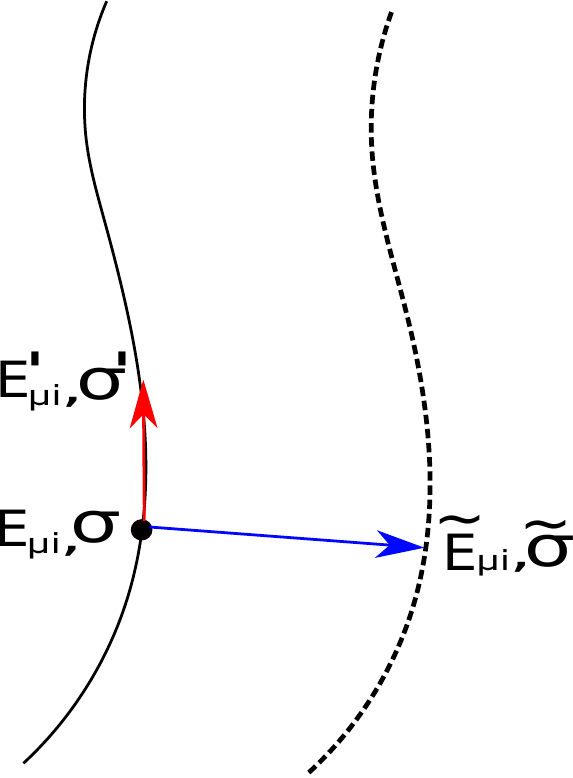} 
\end{center}
\caption{ 
The curves represent gauge orbits 
generated by local SO(3) transformations,
$E_{\mu i}'(x) = O_i^j(x) E_{\mu j}$, 
$\sigma'(x) = \sigma(x)$
in the space of 
$\{ E_{\mu i}(x), \sigma(x) \}$,
where $O_i^j(x)$ is space dependent $SO(3)$ matrices.
Configurations connected by SO(3) transformations
represent a same physical state.
On the other hand, a diffeomorphism,
$\tilde E_{\mu i}(x) = E_{\mu i}(x)  - \nabla_\mu \xi^\nu E_{\nu i}$,
$\tilde \sigma(x) = ( 1 - \xi^\mu \partial_\mu ) \sigma(x)$ 
generally gives a physically distinct state of the matter field. 
}
\label{fig:orbit}
\end{figure}

We note that 
$\Psi(\Phi; E,\sigma)$ 
depends on triad
only through $\gemn$.
Because metric is invariant under local $SO(3)$ transformations, 
$\emi(x) \rightarrow O_i^{~j}(x) E_{\mu j}(x)$,
there is a gauge redundancy in labeling states in ${\cal V}$
in terms of triad.
Each gauge orbit generated by $SO(3)$ transformations
corresponds to one state in ${\cal V}$.
Unlike the $SO(3)$ gauge transformation,
a diffeomorphism of the collective variables generates a different state
of the matter field in general.
In order to see this, we note that
$\Psi(\Phi; E,\sigma)$  is invariant
upto a multiplicative factor 
under diffeomorphisms of the collective variables and the matter field.
%under which the collective variables as well as the matter field
%are shifted as
Under a diffeomorphism generated by an infinitesimal vector field, $\xi^\mu$
\footnote{
Under the diffeomorphism, the triad is transformed as
$\tilde E_{\mu i}(x) = 
\left(
\delta_\mu^\nu - \partial_\mu \xi^\nu
 \right) 
\left( \delta_i^j - \xi^\alpha \omega_{\alpha i}^{~~j} \right) 
( 1 - \xi^\beta \partial_\beta )
E_{\nu j}(x)$, 
where $\omega_{\alpha i}^{~~j}$ is a $SO(3)$ spin connection.
The spin connection is determined from the compatibility condition, 
$ \nabla_\mu E_{\nu i} \equiv 
\partial_\mu E_{\nu i} - \Gamma_{\mu \nu}^\alpha E_{\alpha i} + \omega_{\mu i}^{~~~j} E_{\nu j} 
=0 $,
where $\Gamma_{\mu \nu}^\alpha$ is the torsionless Christoffel symbol
that is compatible with $g_{E,\mu \nu}$.
From $ \nabla_\mu E_{\nu i} =0$,
one can write
$\tilde E_{\mu i}(x) = E_{\mu i}(x)  - \nabla_\mu \xi^\nu E_{\nu i}$,
where the covariant derivative of a space vector is defined as
$\nabla_\mu \xi^{\nu} \equiv \partial_\mu \xi^{\nu} + \Gamma_{\mu \alpha}^\nu \xi^{\alpha}$.
},
\bqa
\tilde E_{\mu i}(x) &=& E_{\mu i}(x)  - \nabla_\mu \xi^\nu E_{\nu i}, \label{diffE} \\
\tilde \sigma(x) &=& ( 1 - \xi^\mu \partial_\mu ) \sigma(x), \label{diffsigma} \\
\tilde \Phi(x) &=& ( 1 - \xi^\mu \partial_\mu ) \Phi(x), \label{diffphi} 
\eqa
%%
%%
%Under the diffeomorphism, 
the wavefunction is transformed as
\bqa
\Psi(\tilde  \Phi; \tilde E, \tilde \sigma ) = 
\left[ J^{ (\tilde E,\tilde \sigma)}_{(E,  \sigma) } \right]^{\frac{1}{2}}
\Psi( \Phi; E, \sigma ).
\label{diffeo}
\eqa
Therefore $\cb \tilde E, \tilde \sigma \rb$ represents a state 
in which the matter field is shifted in space,
and is in general distinct from 
$\cb E, \sigma \rb$ 
as a quantum state of the matrix field (see \fig{fig:orbit}).

\subsubsection{Inner product}

\label{sec:gravityA2}

The inner product between states in ${\cal V}$ is written as
\bqa
&& \lb E', \sigma' \cb E, \sigma \rb =
\int 
D^{ (\hat E, \hat \sigma) } \Phi ~
~ \Psi^*(\Phi;E',\sigma') \Psi(\Phi;E,\sigma).
%%~ e^{ - \int d^3 x  \sqrt{|G|}  ~ tr \left( G^{\mu \nu} \nabla^G_\mu \Phi \nabla^G_\nu \Phi + M^2  \Phi^2 \right) },
\label{innerg0}
\eqa 
While both $D^{(\hat E, \hat \sigma)} \Phi$ and 
$ \Psi^*(\Phi;E',\sigma') \Psi(\Phi;E,\sigma)$ depend on the fiducial metric,
 $\lb E', \sigma' \cb E, \sigma \rb$ does not
because the dependence on the fiducial metric in the measure
is canceled by the normalization factor in \eq{Sn}.
This can be seen
by rewriting the functional integration in \eq{innerg0} in terms of 
the measure associated with $E_{\mu i}$ or $E'_{\mu i}$.
In terms of the measure associated with $(E,\sigma)$,
\eq{innerg0} can be written as
\bqa
&& \lb E', \sigma' \cb E, \sigma \rb =
 \left[
   J^{ (E',\sigma')}_{(E,  \sigma) } 
\right]^{\frac{1}{2}}
%%e^{- \frac{1}{2} \left( S_0[E',\sigma'] + S_0[E,\sigma] \right) } \times \nn
e^{
-\frac{N^2}{4}  
~ \left\{ \tr{
     \Gamma \left[ \lc^2 K_{(E',\sigma')} \right]}
+  \tr{  \Gamma \left[ \lc^2 K_{(E,\sigma)} \right] 
   }   \right\}
} \times \nn
&&
\int 
D^{( E, \sigma)} \Phi ~
~ e^{
- \frac{1}{2} \int d^3 x    ~ tr ~  \Phi \left( 
 |E'| e^{-  \Gamma \left[ \lc^2 K_{(E',\sigma')} \right] }
+ |E| e^{-  \Gamma \left[ \lc^2 K_{(E,\sigma)} \right] }
\right) \Phi
 }.
\label{innerg01}
\eqa 
The fiducial metric drops out in \eq{innerg01}.
This has an important consequence :
the inner product between states in ${\cal V}$ 
is invariant under spatial diffeomorphisms,
%%that is,
\bqa
\lb E', \sigma' \cb E, \sigma \rb 
= \lb \tilde E', \tilde \sigma' \cb \tilde E, \tilde \sigma \rb,
\label{inner_diff_inv}
\eqa
where 
$\{ \tilde E_{\mu i}(x), \tilde \sigma(x) \}$ 
and
$\{ \tilde E_{\mu i}'(x), \tilde \sigma'(x) \}$ 
are respectively related to
$\{ \emi(x), \sigma(x) \}$ 
and
$\{ \emi'(x), \sigma'(x) \}$ 
through a diffeomorphism 
in Eqs. (\ref{diffE})-(\ref{diffsigma}).
See Appendix \ref{diffeo_inner} for the proof
of \eq{inner_diff_inv}.

Once the Gaussian integration is performed in \eq{innerg01},
the overlap can be written as
\bqa
\lb E,\sigma \cb E', \sigma' \rb 
= e^{
-\int dx ~ |E|~
\delta v_a(x)  {\cal M}_{ab}(x)  \delta v_b(x)
}
\label{eq:EsEs}
\eqa
to the quadratic order in 
the difference of the collective variables,
$v_a(x) = \left( h_{E,\mu \nu}(x), \delta \sigma(x) \right)$
with index $a$ running over different collective variables,
where $h_{\mu \nu}=g_{E', \mu \nu} - g_{E, \mu \nu}$ and $\delta \sigma = \sigma' - \sigma$.
${\cal M}_{ab}(x)$ is a positive kernel
which is order of $N^2$.
In the large $N$ limit, 
the cubic and higher order terms in $\delta v_a$ 
are negligible
in \eq{eq:EsEs} 
because $\delta v_a \sim 1/N$.

One can show that 
\eq{innerg01} vanishes identically unless 
$|E(x)| = |E'(x)|$ at all $x$.
Therefore ${\cal M}(x) = \infty$ if $|E(x)| \neq |E'(x)|$.
This is because metrics with different local proper volumes
support eigenmodes with different normalizations.
The mismatch in the normalization of modes with arbitrarily large momenta
gives rise to zero overlap if there is any region in space
with $|E(x)| \neq |E'(x)|$.
The proof is given in Appendix \ref{zero_proper}.
It automatically follows that two states with metrics 
which give different global proper volumes are orthogonal.

Two states with $|E(x)| = |E'(x)|$ are not orthogonal in general.
Nonetheless, $\lb E', \sigma' \cb E, \sigma \rb$ decays exponentially in  
$h_{\mu \nu}$ and $\delta \sigma$
in the large $N$ limit.
This is because each of the $N^2$ components of the matrix field
contributes an overlap which is less than $1$ 
when the collective variables do not match.
%%Each component is independent, and 
%%the net overlap is exponentially small in the large $N$ limit.
Since $\lc$ is the only scale, 
${\cal M}(x) \sim \frac{N^2}{\lc^3} \left( 1 + O(\lc \nabla) \right)$
in \eq{eq:EsEs}.
This form of ${\cal M}(x)$ is 
confirmed through an explicit computation
of the overlap between states with 
metrics close to the Euclidean metric 
in  Appendix \ref{sec:overlap}.
Two states whose collective variables differ by 
\bqa
h_{\mu}^{\nu}(x) \sim \frac{1}{N}, ~~~~
|\delta \sigma(x)| \sim \frac{e^{-3/2 \sigma} }{N}
\label{hds}
\eqa
%$h_{\mu}^{\nu}(x) \sim 1/N$,
%$e^{3/2 \sigma} |\delta \sigma(x)| \sim 1/N$
or more over a proper volume larger than $\lc^3$ 
are nearly orthogonal 
even when $|E(x)| = |E'(x)|$ (See Appendix \ref{sec:overlap}).
With increasing $N$, 
the overlap approaches the delta function
upto a normalization factor.
In the large $N$ limit,
the overlap can be formally written as
\bqa
\lim_{N \rightarrow \infty}
\lb E', \sigma' \cb E, \sigma \rb 
& = & \tilde \mu^{-1}(E, \sigma)
\prod_{x} \left[
\delta \Bigl( \sigma'(x) - \sigma(x) \Bigr)
\prod_{(\mu,\nu)}  
\delta \Bigl( \gemnp(x) - \gemn(x) \Bigr) 
\right], \nn
\label{innerg}
\eqa 
where $\tilde \mu^{-1}(E, \sigma)$ is a measure 
determined from the determinant of \eq{Mp}.
%
%
%replaces
%$e^{ \cJ } \prod_{p} \left[ \det {\cal M}(p) \right]^{-1/2} $  in \eq{innerg1}
%for non-flat metrics.
The full expression for $\tilde \mu(E, \sigma)$ 
can be in principle computed from \eq{innerg01}.
Here we don't need an explicit form of the measure.

The overlap provides 
%%a sense of distance
%%between $SO(3)$ gauge orbits in the space of $\{ \emi(x), \sigma(x) \}$.
%%Accordingly, it defines 
the natural measure for the functional integration over the collective variables.
We define the measure from the condition that
\bqa
\int D \emi D \sigma ~
\lb E', \sigma' \cb E, \sigma \rb = 1
\label{measure}
\eqa
for any $E_{\mu i}'$ and $\sigma'$.
Formally, the measure is written as
 $DE D\sigma \equiv \mu(E, \sigma) \prod_{x} \left[ d\emi(x) d\sigma(x) \right]$ 
 with $\mu(E,\sigma) = \tilde \mu(E, \sigma) \prod_x \left[
 \int dE_{\mu i}'(x) \delta\left( g_{E',\mu\nu}(x)  -\gemn(x)  \right)
 \right]^{-1}$,
where the last factor divides out the $SO(3)$ gauge volume.
The measure defined by this condition 
is invariant under diffeomorphism.
This can be checked from a series of identities,
\bqa
&& \int D \emi D \sigma ~
\lb E', \sigma' \cb E, \sigma \rb 
=
\int D \emi D \sigma ~
\lb \tilde{ E'},   \tilde \sigma' \cb  E,  \sigma \rb  \nn
&& = 
\int D \tilde E_{\mu i} D \tilde \sigma ~
\lb \tilde{ E'}, \tilde \sigma' \cb \tilde E, \tilde \sigma \rb 
= 
\int D \tilde E_{\mu i} D \tilde \sigma ~
\lb  E',  \sigma' \cb  E,  \sigma \rb,
\label{diffeo_measure}
\eqa
where $\{ \tilde E_{\mu i}, \tilde \sigma \}$
is related to $\{ \emi, \sigma \}$ through
a diffeomorphism.
For the first equality, we use the fact that 
\eq{measure} holds for any $E_{\mu i}$ and $\sigma$.
The second equality is a simple change of variables.
For the third equality, we use the fact that
the inner product is invariant under diffeomorphism.
\eq{diffeo_measure} implies that $ D \emi D \sigma = D \tilde E_{\mu i} D \tilde \sigma$.

 \begin{figure}[ht]
\includegraphics[scale=0.50]{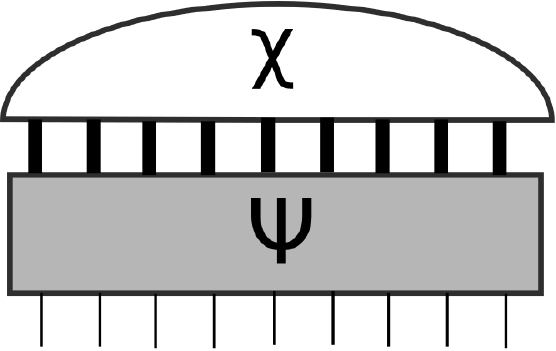}
\caption{
A tensor representation of a general state.
The thick lines represent the collective variables which are 
contracted with the wavefunction $\chi(E,\sigma)$.
 }
\label{fig:chi}
\end{figure}

General states in ${\cal V}$ can be expressed as linear superpositions of $\cb E, \sigma  \rb$,
\bqa
\cb \chi \rb = \int DE D\sigma~
\cb E, \sigma \rb
\chi(E,\sigma),
\label{gs-1}
\eqa
where $\chi(E,\sigma)$ is 
invariant under local $SO(3)$ transformations.
Its tensor representation is shown in \fig{fig:chi}.
It is  normalized such that
$
%%\lb \chi \cb \chi \rb = 
\int 
DE' D\sigma'
DE D\sigma 
~
\chi^*(E',\sigma')
\lb E', \sigma' \cb E, \sigma \rb 
\chi(E,\sigma)
 = 1$.
In the large $N$ limit, 
$\lb E', \sigma' \cb E, \sigma \rb$
is sharply peaked  at $g_{E',\mu \nu}=g_{E,\mu \nu}$, $\sigma'=\sigma$,
and the normalization condition reduces to
$\int 
DE D\sigma 
~
|\chi(E,\sigma)|^2
 = 1$.
Similar to \eq{Herm1},
we define the Hermitian conjugate of a differential operator 
acting on the collective variables from
\bqa
\int DE D\sigma ~f^*(E,\sigma) H \left[ g(E,\sigma) \right]
= 
\int DE D\sigma ~\left[ H^\dagger f(E,\sigma) \right]^* g(E,\sigma).
\label{Herm2}
\eqa

%%%
%%%
%%%

%%%
%%%
%%%

 \subsection{Metric as a collective variable for entanglement}

\label{sec:gravityB}

In this subsection, we discuss the physical meaning
of the metric and the scalar field as collective variables
for the matter field.
In particular, we show that the metric controls
the number of degrees of freedom that are entangled in space,
and the scalar field determines the rate at which 
the mutual information decays in space.
Being a wavefunction defined in continuum,
the size of the Hilbert space per unit {\it coordinate volume} is  infinite.
However, the number of degrees of freedom that contribute to entanglement 
is controlled by the proper volume measured in the unit of the short-distance cut-off, $\lc$.
The metric sets the notion of distance in the short range entangled states of the matter.

 \begin{figure}[ht]
\includegraphics[scale=0.55]{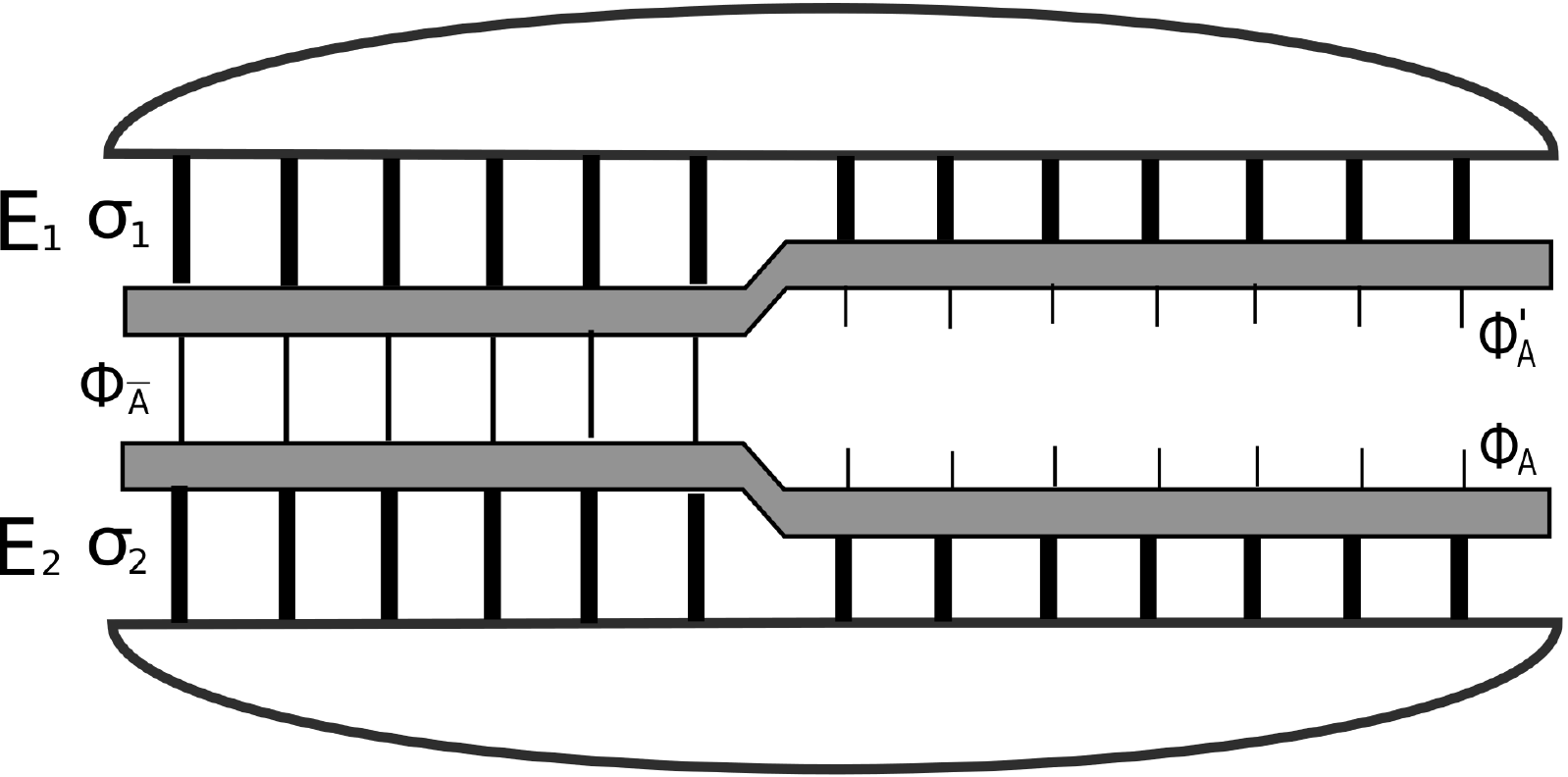}
\caption{
A tensor representation of 
the density matrix 
of region $A$ in space.
 }
\label{fig:density_matrix}
\end{figure}

Let us consider a region $A$ in space.
For general states in \eq{gs-1},
the density matrix of the region is given  by
\bqa
&& \rho_A(\Phi'(x_A), \Phi(x_A) ) = 
 \int D^{(\hat E, \hat \sigma)} \Phi(x_{\bar A}) 
DE_1 D\sigma_1 
DE_2 D\sigma_2 ~ \nn
& & ~~~~
\chi^*(E_1, \sigma_1)
  \Psi^*( \Phi'(x_A),\Phi(x_{\bar A}); E_1, \sigma_1)
   \Psi( \Phi(x_A),\Phi(x_{\bar A}); E_2, \sigma_2)
   \chi(E_2, \sigma_2), 
\eqa
where $\bar A$ is the complement of $A$ (See \fig{fig:density_matrix}).
The replica method allows one to express the
von Neumann \ee as
\bqa
S(A) &=&  
 -\lim_{n \rightarrow 1} \frac{1}{n-1}
 \left( Z_n -1 \right),
\label{eq:SA}
\eqa 
where $Z_n = \tr{ \rho_A^n }$.
The entanglement entropy for general states 
depends both on
$\Psi( \Phi; E, \sigma)$
and $\chi(E, \sigma)$ in 
a complicated way,
where the former represents the wavefunction of the matter field
for a fixed collective variable $(E,\sigma)$
and the latter encodes fluctuations of the collective variables.
Here we focus on $\chi(E, \sigma)$ 
that is peaked at a classical configuration
$(\bar E, \bar \sigma)$ with small fluctuations around it.
For such semi-classical wavefunctions for the collective variables,
the entanglement entropy can be approximately decomposed into two contributions,
\bqa
S(A) \approx S_\Phi(A) + S_{E, \sigma}(A).
\label{eq:decom}
\eqa
Here $S_\Phi(A)$ is the entanglement entropy of the matter degrees of freedom
defined in the classical background collective variables $(\bar E, \bar \sigma)$,
%\cite{PhysRevB.79.115421},
\bqa
S_\Phi(A) = F_A + F_{\bar A} - F,
\eqa
where
\bqa
F & =&  - \ln \int D^{ (\bar E, \bar \sigma) } \Phi ~ e^{ - 2 \int dx  |\bar E|  {\cal L}[\Phi; \bar E, \bar \sigma] }, \nn
F_{A (\bar A)} & =&  
- \ln 
\left. 
\int D^{ (\bar E, \bar \sigma) } \Phi ~  e^{ - 2 \int_{A (\bar A)} dx  |\bar E|  {\cal L}[\Phi; \bar E, \bar \sigma] }
\right
|_{ \Phi(x_{\partial A}) = 0 }.
\label{FF}
\eqa
%Here we drop the regular contribution from $ C(\bar E, \bar \sigma)$,
%and focus on the contribution that is potentially singular in the small $\lc$ limit.
On the other hand, 
$S_{E, \sigma}(A)$ is the entanglement generated by correlations between fluctuations of the collective variables,
\bqa
S_{E, \sigma}(A) = 
 -\lim_{n \rightarrow 1} \frac{1}{n-1}
 \left(
 Z_n^{E, \sigma}
 -1
  \right),
\eqa
where
\bqa
&& Z_n^{E, \sigma}
= 
\int 
\prod_{j=1}^n 
DE^j D\sigma^j  ~
\Bigg\{
~ e^{S_0[E^j,\sigma^j] } ~
\tilde \chi^* \Bigl( E^{j}(x_{\bar A}),\sigma^j(x_{\bar A}); E^{j}(x_{A}), \sigma^{j}(x_A) \Bigr) \times \nn
&& 
~~~~~~~~~~
\left.
\tilde \chi \Bigl( E^{j}(x_{\bar A}),\sigma^j(x_{\bar A}); E^{j-1}(x_A), \sigma^{j-1}(x_A) \Bigr)
 \right|_{
 \begin{array}{l}
{\scriptstyle \sigma^{2,..,n}(x_{\partial A}) = \sigma^1(x_{\partial A})} ,   \\
{\scriptstyle g_{E^{2,..,n},\mu\nu}(x_{\partial A}) = g_{E^1,\mu \nu}(x_{\partial A})}
\end{array}
 }
\Bigg\}
.
 \label{Zn22}
\eqa
with
$ \tilde \chi(E,\sigma) = e^{- \frac{1}{2}S_0[E,\sigma]} ~ \chi(E,\sigma)$.
Here $E^0 = E^n$ and $\sigma^0 = \sigma^n$.

The derivation of Eqs. (\ref{eq:decom})-(\ref{Zn22}) is given in Appendix \ref{EE_decomposition}.
Here we provide an intuitive explanation of the result.
When 
$\chi(E,\sigma) \propto \delta( g_{E,\mu \nu} -g_{\bar E,\mu \nu} ) \delta(\sigma-\bar \sigma)$,
there is no fluctuations in the collective variables.
In this case, the entanglement entropy is given by
that of $\Psi(\Phi;\bar E, \bar \sigma)$. 
Because $\Psi(\Phi;\bar E, \bar \sigma)$ is written as an exponential of a local functional,
the entanglement entropy is related to the `free energy' difference 
caused by a Dirichlet boundary condition
as is shown in \eq{FF}\cite{PhysRevB.79.115421}.
Now, suppose the wavefunction for the collective variables has a small but nonzero width
around the semi-classical configuration.
As a simple example, let us assume that there are only two  
configurations of the collective variables,
$\chi(E,\sigma) = 
A \delta( g_{E,\mu \nu} -g_{ E^1,\mu \nu} ) \delta(\sigma- \sigma^1)
+
B \delta( g_{E,\mu \nu} -g_{ E^2,\mu \nu} ) \delta(\sigma- \sigma^2)$,
where 
$(E^1,\sigma^1)$
and 
$(E^2,\sigma^2)$
are distinct from each other 
but are close to their average, $(\bar E, \bar \sigma)$.
On the one hand, there is an entanglement 
generated by the matter field
whose wavefunction is well approximated by $\Psi(\Phi;\bar E, \bar \sigma)$.
This entanglement is given by $S_\Phi$.
However, $\Psi(\Phi;\bar E, \bar \sigma)$ does not 
capture the entire correlation present in the system.
There is an additional correlation 
generated by fluctuations of the collective variables.
Since 
$\Psi(\Phi;E^1,\sigma^1)$
and 
$\Psi(\Phi;E^2,\sigma^2)$ 
are almost orthogonal when $N$ is large,
these fluctuations of the collective variables give rise to
an additional entanglement which is captured by $S_{E,\sigma}$.

\begin{figure}[ht]
\includegraphics[scale=0.50]{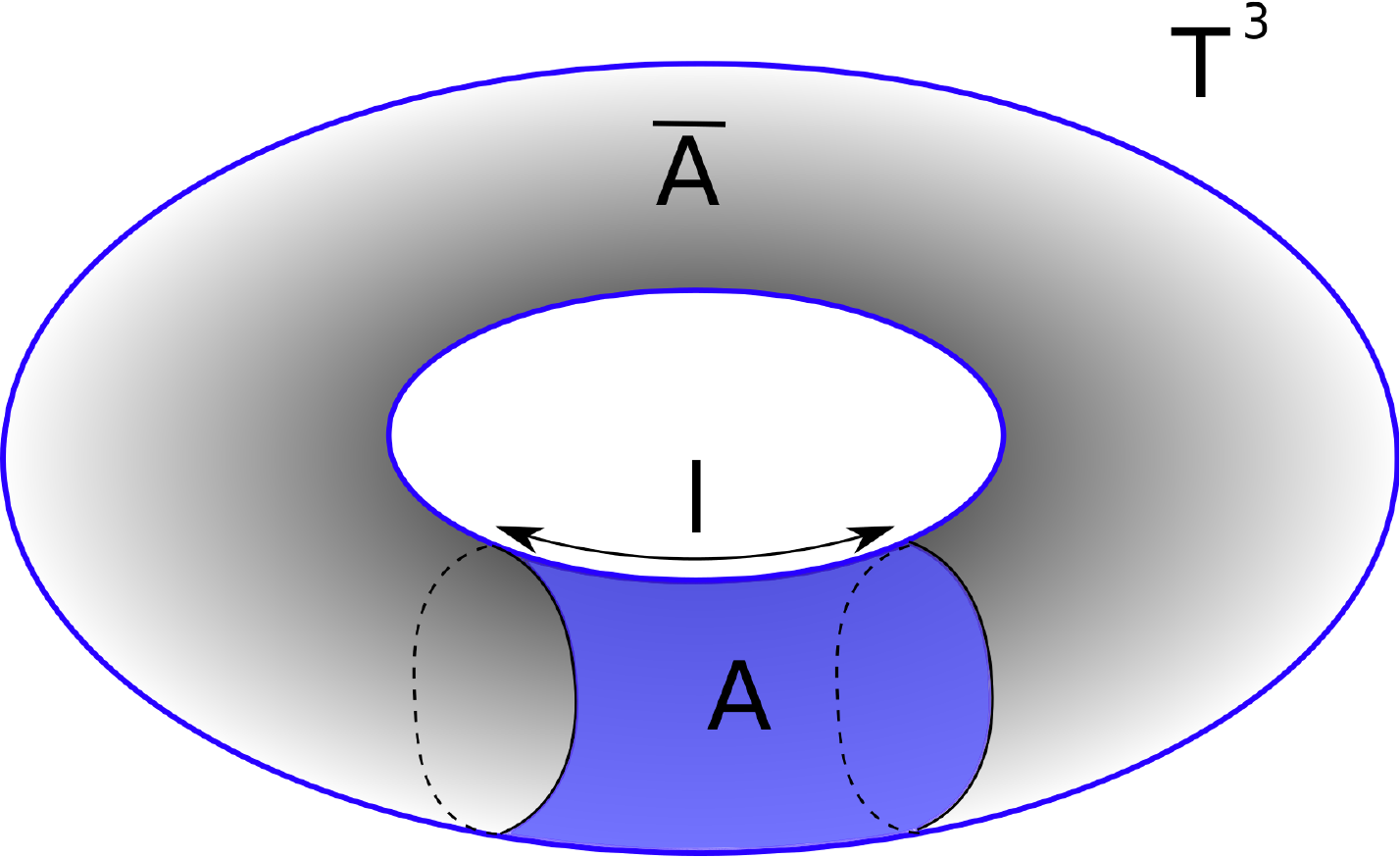}
\caption{
A region with linear size $l$ in the compact space 
which has $T^3$ topology.
}
\label{fig:torus}
\end{figure}

In the limits that 
$e^\sigma << 1$ and 
the linear proper size of $A$ is 
much larger than $\lc$,
$S_\Phi$ is proportional to the area  of $\partial A$
and the number of matter fields.
When the metric is flat and $\sigma$ is constant,
one can compute $S_\Phi$ explicitly.
Consider a region,
$A = \{(
x_1,x_2,x_3) \cb 
0 < x_1 < l, 
0 \leq x_2 < l_c, 
0 \leq x_3 < l_c 
\}$ 
in $T^3$
with the flat metric
$g_{\bar E, \mu \nu} = a^2 \delta_{\mu \nu}$
as is shown in \fig{fig:torus}.
In the small $\lc$ limit with fixed $a \lc$,
the entanglement entropy of region $A$ 
is given by (see Appendix \ref{computation_EE} for derivation)
\bqa
S_\Phi(A) =
%%N^2 \frac{ {\cal A}_{\partial A} }{64 \pi \lc^2},
\frac{ {\cal A}_{\partial A} }{ 4 \kappa^2 },
\label{EE}
\eqa
where ${\cal A}_{\partial A}$ is the area of $\partial A$
measured with the metric $g_{\bar E, \mu \nu}$,
and  
\bqa
\kappa^2 \equiv  \frac{4 \pi l_c^2 }{N^2}.
\label{kappa}
\eqa
The \ee is given by
the proper area of the boundary measured in the unit of $\kappa^2$.
$\kappa$ is much smaller than the cut-off scale $\lc$ in the large $N$ limit.
Although \eq{EE} has been derived in the flat metric,
the same formula is expected to hold for general metrics
to the leading order in the limit that
the curvature is much smaller than $\lc^{-1}$.
This is because the leading order contribution,
which is divergent in the $\lc \rightarrow 0$ limit,
comes from short-wavelength modes
for which geometry can be regarded locally flat
and the WKB approximation is valid.

The entanglement entropy of a fixed region $A$ increases 
as the proper area of $\partial A$ increases.
%In order to understand this point,
%let us consider again the flat geometry with scale factor $a$
%whose metric is given by $\gemn = a^2 \delta_{\mu \nu}$.
This can be understood in terms of mode softening with increasing proper volume.
The eigenvalue for the mode with momentum 
$k_\mu = \frac{2\pi}{\lc}(n_1, n_2, n_3)$ with integer $n_\mu$ is given by 
$\lambda_k = \left( \frac{2 \pi}{a \lc } \right)^2 ( n_1^2 + n_2^2 + n_3^3 ) + \frac{ e^{2 \sigma}}{\lc^2}$.
Because of the cut-off scale, 
only the modes with wave-numbers, $n_i < a$ 
contribute significantly to the entanglement entropy
as is shown in Appendix \ref{computation_EE}.
With increasing $a$, 
more modes become soft and contribute to entanglement. 
Therefore, the number of degrees of freedom
that generate entanglement in space
is a dynamical quantity
rather than a fixed number.
This has an important consequence.
There is no fundamental limit in the amount of information a `finite' region in space can hold 
because the proper size of the region is a dynamical variable which can be as large as it can be.
This may sound unphysical until we think about our universe,
which was once of the Planck size 
yet contained the vast amount of 
information on the current universe.

$S_\Phi(A)$ is the contribution from the degrees of freedom
that carry non-trivial charge under the $SU(N)$ symmetry.
The classical metric controls the entanglement encoded in the non-singlet degrees of freedom.
For this reason, we call $S_\Phi(A)$ `color entanglement entropy'.
On the other hand, $S_{E, \sigma}(A)$ is encoded in the wavefunction 
for the collective variables.
It is the entanglement generated by correlations 
in fluctuations of the singlet collective variables.
We call $S_{E, \sigma}(A)$ `singlet entanglement entropy'.

In terms of $N$ counting, $S_{E, \sigma}(A)$ is $O(1)$
while $S_\Phi(A)$ is $O(N^2)$ for semi-classical states.
However, the singlet \ee can be larger than the color \ee in some states.
This is because the singlet entanglement entropy can scale with the volume of a region 
if there exist long-range correlations in fluctuations of the collective variables
while color entanglement entropy scales with the area of its boundary.
This point will become important in the discussion
on black hole evaporation in Sec. IV.

In the von Neumann entanglement entropy,
$\sigma$ doesn't enter to the leading order 
in the small $\lc$ limit.
However, the scalar field plays a more distinct role
in determining mutual information.
The mutual information between two regions $A$ and $B$ is defined to be
$I(A,B) = S(A) + S(B) - S(A \cup B)$.
For semi-classical states,
the mutual information is again decomposed as
$I(A,B) \approx I_\Phi(A,B) + I_{E, \sigma}(A,B)$,
where $I_\Phi(A,B)$ ($I_{E, \sigma}(A,B)$) is the contribution 
from color degrees of freedom (singlet collective variables).
In order to examine the relation between the color mutual information
and the collective variables, 
it is useful to write the expression for the color entanglement entropy as
\bqa
S_\Phi(A) &=&  
 -\lim_{n \rightarrow 1} \frac{1}{n-1}
 \Bigl(
\int D^{(\hat E, \hat \sigma)} \Phi^j  ~ O_{\partial A} ~ 
\prod_j |\Psi( \Phi^j; \bar E, \bar \sigma)|^2
 -1
  \Bigr)
\label{SPhi}
\eqa
with
$O_{\partial A} \propto 
\prod_{x \in \partial A} 
\prod_{j=2}^{n} 
\int d \lambda_j(x) ~
e^{i \tr{ \lambda_j(x) \left[ \Phi^j(x) - \Phi^1(x) \right] } }$.
$\lambda_j$ is an $N \times N$ Hermitian Lagrangian multiplier
which enforces the Dirichlet boundary condition
at the boundary.
This is only schematic 
because the measure for $\prod_{x \in \partial A}$
hasn't been specified. 
However, it is still useful 
in understanding the connection between the mutual information
and the collective variables.
Suppose $A$ and $B$ represent infinitesimally small balls 
centered at $x$ and $y$, respectively.
In the limit that the proper distance between $x$ and $y$ is large,
the color mutual information is dominated by the connected correlation function 
between the fundamental fields inserted at $x$ and $y$ 
which exhibits the slowest decay in \eq{SPhi}.
A straightforward calculation shows that the color mutual information scales as 
\bqa
I_\Phi(A,B)  \sim  N^2 G[x, y; \bar E,\bar \sigma]^2,
%%I_\Phi(A,B)  \sim  N^2 e^{-  2 \int_x^y \frac{e^\sigma}{\lc} ds },
\eqa
where $G[x,y; \bar E,\bar \sigma]$
is the correlation function of the fundamental field.
In the small $e^\sigma$ limit, 
$G[x,y; \bar E,\bar \sigma] \sim \frac{1}{d_{x,y}}$,
where $d_{x,y}$ is the proper distance between $x$ and $y$ measured
with the metric $g_{\bar E,\mu \nu}$.
For fixed $x$ and $y$ in the manifold,
the proper distance between the points
is controlled by the metric,
and so does the mutual information.
For example, states that support small (large) color mutual information between two points
give large (small) proper distance between the points. 
When $e^\sigma$ is not negligible,
the Green's function decays exponentially at large distances,
$G[x,y; \bar E,\bar \sigma] \sim  e^{- \int_x^y \frac{e^\sigma}{\lc} ds }$,
where $ds$ is the infinitesimal proper distance 
along the geodesic that connects $x$ and $y$.
This shows that $\sigma$ determines 
the range of entanglement, 
while the metric sets the notion of locality 
in how matter fields are entangled in space.
In this construction, 
the connection between entanglement
and geometry\cite{PhysRevLett.96.181602,1126-6708-2007-07-062,Casini2011,Lewkowycz2013,PhysRevD.86.065007,2013arXiv1309.6282Q}
has been encoded as a kinematic building block of the theory.

\subsection{Relatively local Hamiltonian}

\label{sec:gravityC}

Having understood the kinematic structure of the sub-Hilbert space,
now we construct a Hamiltonian of matter field which induces
the Wheeler-DeWitt Hamiltonian of the general relativity
in the sub-Hilbert space.

\begin{figure}[ht]
\includegraphics[scale=0.50]{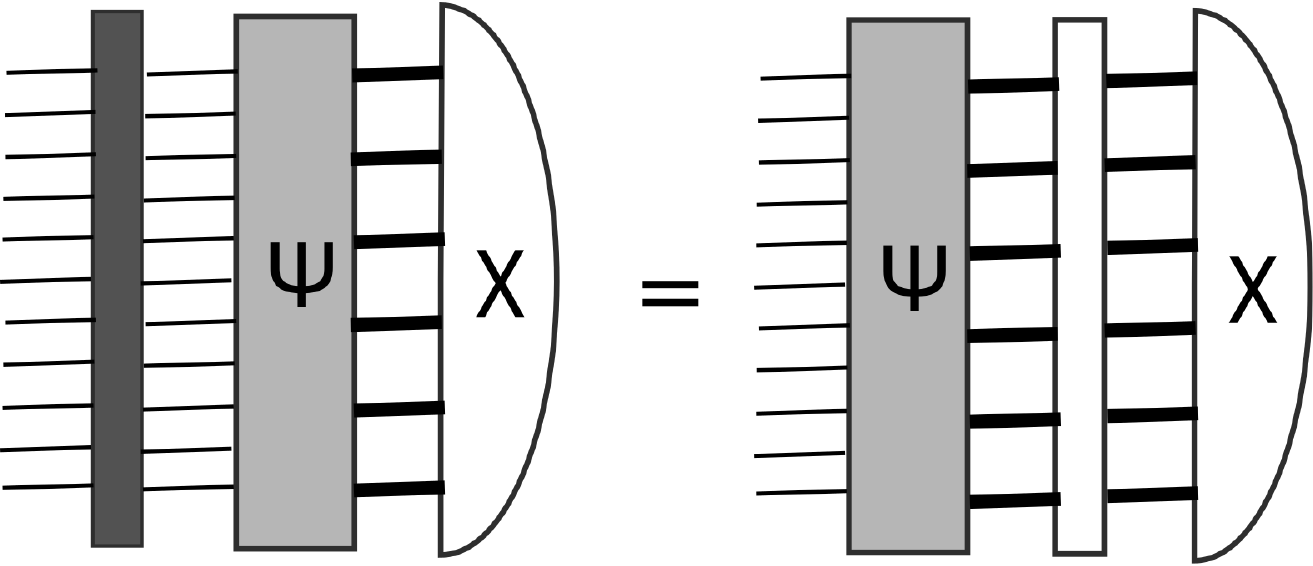}
\caption{
An endomorphism of ${\cal V}$ represented by the dark filled box
induces a map for the collective variable represented 
by the empty box.
}
\label{fig:precess2}
\end{figure}

Hermitian operators
that map ${\cal V}$ to ${\cal V}$
generate unitary evolutions of the collective variables.
%In describing such evolutions, 
%it is enough to focus on states within ${\cal V}$.
One example of such endomorphisms
is the momentum density operator for the matter fields,
\bqa
\hat \cH_\mu(x) & = & 
- \frac{1}{2} 
\left[
\left( \nabla_\mu \hat \Phi_{ab}(x) \right)
\hat \pi_{ba}(x)
+
\hat \pi_{ba}(x)
\left( \nabla_\mu \hat \Phi_{ab}(x) \right)
\right],
\label{mC} 
\eqa
where $\hat \pi_{ba}(x)$ is the conjugate momentum of $\hat \Phi_{ab}(x)$
with the commutator $[ \hat \pi_{ab}(x), \hat \Phi_{cd}(y) ] = -i \delta_{ad} \delta_{bc} \delta(x-y)$.
Due to \eq{diffeo}, 
the action of $\hat \cH_\mu(x)$ on $\cb E, \sigma \rb$
is equivalent to a differential operator
that induces a diffeomorphism of the collective variables, 
\bqa
\int dx ~n^\mu(x) \hat \cH_\mu(x) \cb E, \sigma  \rb = 
- \int dx ~ n^\mu(x) \cH^{E, \sigma}_{\mu}(x) \cb E, \sigma  \rb,
\label{mCR}
\eqa
where 
\bqa
\cH^{E, \sigma}_{\mu}(x)  = 
- i  \emi \nabla_\nu \frac{\delta}{\delta E_{\nu i} }
+ i  ( \nabla_\mu \sigma ) \frac{\delta}{\delta \sigma(x) }.
\label{inducedM}
\eqa
 \eq{mCR} is proven in  Appendix \ref{diffeomorphism}.
It implies that a shift of the matter field
%%with a fixed configuration of the collective variables 
is equivalent to the inverse shift of the collective fields.
%with the fixed configuration of the matter field.
This follows from the fact that 
$\cb E, \sigma  \rb$
is invariant under
the simultaneous shift
of the matter field and the collective variables.
Only relative shifts between the matter field
and the collective variables matter.
 \eq{inducedM} can be viewed as the induced momentum 
 density operator for the collective variables.
For general states in \eq{gs-1}, 
the operation of \eq{mC} results in
a shift in wavefunctions of the collective variables
as is illustrated in \fig{fig:precess2},
\bqa
&& \int dx~ n^\mu(x)  \hat \cH_\mu(x) 
\left[  \int DE  D\sigma ~ \cb E, \sigma  \rb  \chi(E, \sigma) \right] \nn
&  = &  \int DE  D\sigma ~
\cb E, \sigma  \rb ~
\left[
\int dx ~
n^\mu(x)  \cH^{E, \sigma}_{\mu}(x)  \chi(E,\sigma)  
\right].
\label{eq:Hmu}
\eqa
Here it is used that $DE D\sigma$ is invariant
under diffeomorphism\footnote{
The derivation goes as follows.
From \eq{mCR}, we have
$e^{ -i  \int dx~ n^\mu(x)  \hat \cH_\mu(x) }
\left[  \int DE  D\sigma ~ \cb E, \sigma  \rb  \chi(E, \sigma) \right] 
= \int DE  D\sigma ~ \cb \tilde E, \tilde \sigma  \rb  \chi(E, \sigma)$,
where 
$\tilde E_{\mu i}(x) = E_{\mu i}(x)  -  \nabla_\mu n^\nu E_{\nu i}$,
$\tilde \sigma(x) = ( 1 - n^\mu \nabla_\mu ) \sigma(x)$.
By shifting 	
$ E_{\mu i}(x) \rightarrow E_{\mu i}'(x) = E_{\mu i}(x)  +  \nabla_\mu n^\nu E_{\nu i}$,
$ \sigma(x) \rightarrow \sigma'(x) = ( 1 + n^\mu \partial_\mu ) \sigma(x)$,
and using the fact that the measure is invariant under diffeomorphism,
one obtains
$e^{- i \int dx~ n^\mu(x)  \hat \cH_\mu(x) }
\left[  \int DE  D\sigma ~ \cb E, \sigma  \rb  \chi(E, \sigma) \right] 
= \int DE  D\sigma ~ \cb E, \sigma  \rb  \chi( E', \sigma' )
= \int DE  D\sigma ~ \cb E, \sigma  \rb  
e^{- i  \int dx~ n^\mu(x)  \cH^{E,\sigma}_\mu(x) } \chi( E, \sigma)
$.
The linear terms in $n^\mu$ give \eq{eq:Hmu}.
}.

Similarly, a Hamiltonian for the matter field 
whose trajectories stay within $\cal V$
induces a quantum Hamiltonian for $\emi$ and $\sigma$. 
Our goal is to construct 
a Hamiltonian for the matter field
which induces the Einstein's general relativity
at long distances in the large $N$ limit.
Our strategy is to start with a regularized Wheeler-DeWitt Hamiltonian
%%\cite{PhysRev.116.1322,PhysRev.160.1113,NICOLAI199315}
for the collective variables, 
and reverse engineer to find the corresponding Hamiltonian for the matter field.
%%similar to \eq{mCR}.
We look for a Hamiltonian density 
whose action on $\cb E, \sigma \rb$
leads to 
%a regularized Wheeler-DeWitt differential operator
%for the collective variables, 
\bqa
\hat \cH(x;E, \sigma) \cb E, \sigma  \rb = 
\tilde h^{E, \sigma}(x) \cb E, \sigma  \rb,
\label{WDE}
\eqa
where 
$\tilde h^{E, \sigma}(x)$ 
is a regularized Wheeler-DeWitt differential operator\cite{PhysRev.116.1322,PhysRev.160.1113,NICOLAI199315}
for the collective variables,
\bqa
\tilde h^{E, \sigma}(x)  &=&
- \tilde \kappa^2
\left(
:\frac{ G_{ijkl} }{|E|}
 E_{\mu}^{ j} E_{\nu}^{ l} 
\frac{ \delta}{\delta \emi(x) }  
\frac{ \delta}{\delta E_{\nu k}(x) } :
 + 
 : \frac{1}{2 |E| \Fs}
 \frac{ \delta}{\delta \sigma(x) }  \frac{ \delta}{\delta \sigma(x) } :
\right)  \nn
&&
 + \frac{|E|}{\tilde \kappa^2} 
 \left(  
 - R +  \frac{\Fs}{2}  g_E^{\mu \nu}(x) \nabla_\mu \sigma(x) \nabla_\nu \sigma(x)
+  V(\sigma) 
+ U_3( g_E, \sigma)
\right). 
%%+ \frac{O( \tilde \kappa^2/\kappa^2)}{\lc^4}.
%+ O( N^0 ).
\label{HWDW}
\eqa
Here 
$ G_{ijkl} = \frac{1}{4} \left( \delta_{ik} \delta_{jl} - \frac{1}{2} \delta_{ij} \delta_{kl} \right) $
is the supermetric for the kinetic term of the triad.
$\Fs$ represents a nonlinear term in the kinetic energy of the scalar.
$R$ is the curvature scalar for the three-dimensional metric $g_{E, \mu \nu}$.
$V(\sigma)$ is a potential for the scalar.
$U_3( g_E, \sigma)$ represents terms that involve 
more than two derivatives for $g_{E,\mu \nu}$ and $\sigma$,
where the higher-derivative terms are suppressed by $\left( \lc \nabla \right)$
compared to the two-derivative terms.
$F(\sigma)$, $V(\sigma)$ and $U_3( g_E, \sigma)$
are included for generality, 
but we do not need to specify their forms for our purpose.
%%Here, $\Lc$ is a cut-off length scale
%%that will be determined later.
%
It is important to note that the second order functional derivatives in \eq{HWDW} 
needs to be regularized as the derivatives acting on one point in space
is ill-defined.
Here the derivatives are regularized 
through a point-splitting scheme based on the heat-Kernel regularization, 
\bqa
: \frac{ G_{ijkl} }{|E| } 
 E_{\mu}^{ j} E_{\nu}^{ l} 
\frac{ \delta}{\delta \emi(x) } 
  \frac{ \delta}{\delta E_{\nu k}(x) } :
&\equiv& \int dy dz ~  
 K_{\mu i \nu k }(y,z; x,  \lc^2 ) 
\frac{ \delta}{\delta \emi(y) }  \frac{ \delta}{\delta E_{\nu k}(z) }, \nn
:\frac{1}{|E| \Fs}  \frac{ \delta}{\delta \sigma(x) }  \frac{ \delta}{\delta \sigma(x) } :
&\equiv& \frac{1}{\Fs} \int dy dz~  K(y,z;x,  \lc^2 ) 
 \frac{ \delta}{\delta \sigma(y) }  \frac{ \delta}{\delta \sigma(z) }.
\label{smearing}
\eqa
$K_{\mu i \nu k }(y,z; x,t)$ and $K(y,z; x,t)$ 
%%are bi-local tensor and scalar functions, respectively,
spread the two differential operators 
over the cut-off length scale $\lc$ centered at $x$.
In the heat kernel regularization scheme\cite{MANSFIELD1994113,PhysRevD.54.1500,KOWALSKIGLIKMAN199648},
the kernels satisfy the diffusion equation,
\bqa
\frac{\partial}{\partial t} K_{\mu i \nu k }(y,z; x,t) & = & 
\left[ \nabla_y^2 + \nabla_z^2 \right]  K_{\mu i \nu k }(y,z;x, t), 
\label{Df1} \\
\frac{\partial}{\partial t} K(y,z;x, t) & = & 
\left[ \nabla_y^2 + \nabla_z^2 \right]  K(y,z;x, t)
\label{Df2}
\eqa
with the boundary condition,
%\bqa
 $K_{\mu i \nu k }(y,z;x,0) =
%\frac{1}{4 |E|} \left( \delta_{ik} \delta_{jl} - \frac{1}{2} \delta_{ij} \delta_{kl} \right) 
\frac{ G_{ijkl} }{ |E(x)|} 
 E_{\mu}^{ j}(x) E_{\nu}^{ l}(x)  \delta(y-x) \delta(z-x)$, 
$ K(y,z;x,0) = \frac{1}{|E(x)|} \delta(y-x) \delta(z-x)$.
%\eqa
In \eq{Df1} and \eq{Df2},
$\nabla_y$ ($\nabla_z$) represents the covariant derivative 
acting on coordinate $y$ ($z$).
In the Euclidean space, the regulators become
$ K_{\mu i \nu k }(y,z;x,\lc^2)  =   \frac{1}{(2 \sqrt{\pi}  \lc)^6  |E|  } 
G_{ijkl}
 E_{\mu}^{ j} E_{\nu}^{ l} 
e^{- \frac{d_{y,x}^2 + d_{z,x}^2}{4 \lc^2} }$, 
$ K(y,z;x,\lc^2)  =   \frac{1}{(2 \sqrt{\pi} \lc)^6 |E|  } 
e^{- \frac{d_{y,x}^2 + d_{z,x}^2}{4 \lc^2} } 
$,
where $d_{x,y}$ is the proper distance between $x$ and $y$.
In \eq{HWDW}, $\tilde \kappa$ is the Planck scale for the induced gravity,
which is a free parameter for now.
Below, we show that 
$\tilde \kappa$ should be order of $\kappa$
in the large $N$ limit
if the underlying matter Hamiltonian 
has a well-defined large $N$ limit.
%%
%%

%%Now we use \eq{WDE}
%%to find $\hat \cH(x;E, \sigma) \cb E, \sigma  \rb$
%%through a reverse engineering.
%%
The Hamiltonian density for the matter field
that satisfies \eq{WDE} 
%%for a given $\cb E, \sigma \rb$ 
is given by
\bqa
%%\lb \Phi \cb 
\hat \cH(x;E, \sigma) 
&=&
%%\cb E, \sigma \rb = 
%% \nn && ~~~~~
\frac{1}{\tilde \kappa^2} 
\Biggl[
-
\frac{\tilde \kappa^4}{\kappa^4} 
: \frac{1}{|E|} \left(  
G_{ijkl} E_{\mu}^{ j} E_{\nu}^{ l} \hat T^{\mu i} \hat T^{\nu k} +   \frac{1}{2 \Fs} ( \hat O_\sigma )^2   
\right) :
\nn && 
+ |E| \left(  
- R +   \frac{\Fs}{2} g_E^{\mu \nu} \nabla_\mu \sigma \nabla_\nu \sigma
+  V(\sigma) 
+ U_3( g_E, \sigma)
\right)
\Biggr]
 %%\Psi( \Phi; E, \sigma ),
\label{WDE3}
\eqa
to the leading order in the large $N$ limit,
where 
\bqa
\hat T^{\mu i}(x)  & \equiv &  \kappa^2 \frac{\delta}{\delta \emi(x)} S[ \hat  \Phi; E, \sigma], \nn
\hat O_\sigma(x) & \equiv &  \kappa^2 \frac{\delta}{\delta \sigma(x)} S[ \hat  \Phi; E, \sigma]
\label{TO}
\eqa
with
%%\bqa
$S[\Phi; E, \sigma] = 
 \int dx ~ |E(x)| ~ 
{\cal L}[\Phi(x); \emi(x),\sigma(x)]
+  \frac{1}{2} S_0[E,\sigma]$.
%%\eqa
$\hat T^{\mu i}(x)$ and $\hat O_\sigma(x)$ in \eq{TO} 
scale as $O\left( N^0 \right)$ in the large $N$ limit.
In \eq{WDE3}, the double-trace operators\cite{Heemskerk:2010hk,2011JHEP...08..051F, 2013JHEP...01..030K,Lee2012,2016arXiv161103470M}
are responsible for generating
the kinetic terms in $\tilde h^{E,\sigma}$.
In order for the leading kinetic term and the potential term 
in \eq{WDE3}
to scale uniformly in the large $N$ limit, 
one needs $\frac{\tilde \kappa}{\kappa} \sim O(N^0)$.
This implies that a matter Hamiltonian which scales as $O(N^2)$ 
in the large $N$ limit induces a gravity with the Planck scale  $\kappa \sim \frac{\lc}{N}$,
which also controls the color entanglement entropy through \eq{kappa}.
From now on, 
we focus on such Hamiltonians, and set $\tilde \kappa = \kappa$.

\begin{figure}[ht]
\includegraphics[scale=0.40]{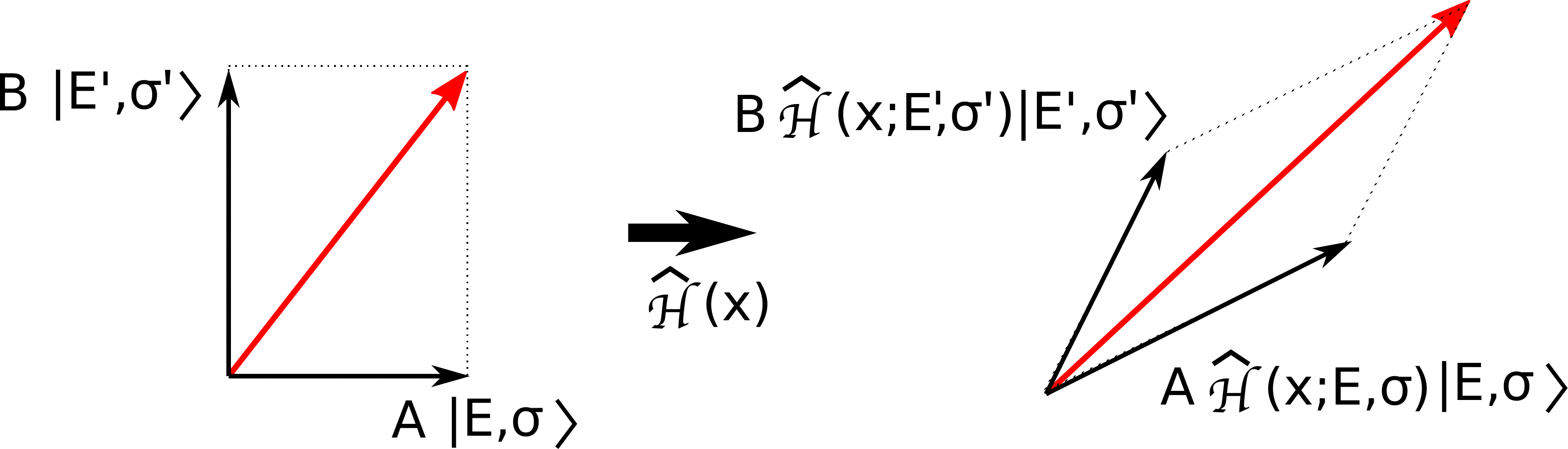}
\caption{
When $\hat {\cal H}(x)$ is applied to a state made of a linear superposition
of multiple basis states with different collective variables,
each basis state $\cb E,\sigma \rb$ is applied by 
$\hat {\cal H}(x; E,\sigma)$
which is local with respect to the distance 
measured with the metric $g_{E,\mu \nu}$.
}
\label{fig:H}
\end{figure}

\eq{WDE} implies that the evolution of $\cb E, \sigma  \rb$ generated by $\hat \cH(x;E, \sigma)$
is reproduced by the differential operator $\tilde h^{E, \sigma}(x)$
acting on the collective variables.
However, $\hat \cH(x;E, \sigma)$ is defined 
with reference to $(E,\sigma)$, 
and it does not act on states with different collective variables 
in the same way.
We encountered the same issue in Sec. \ref{sec:mini}.
In order to induce a background independent Hamiltonian 
for the collective variables, 
we use the strategy introduced in the minisuperspace cosmology.
Namely, we make a Hamiltonian to 
effectively depend on the collective variables
so that $\hat \cH(x;E, \sigma)$ associated with a specific collective variable
%, $(E,\sigma)$
is only applied to the corresponding state, $\cb E, \sigma \rb$.
This can be implemented by the following Hamiltonian density,
%%We use \eq{eq:h0} and  \eq{hE1} in \eq{eq:cH}, and 
%%perform an integration by part to write 
%%the matter Hamiltonian density, 
\bqa
&& \hat \cH(x) 
=
\frac{1}{2}   \int DE D\sigma
~
\left[
\hat {\cal H}(x,E,\sigma)
 \hat P_{E,\sigma} 
%%+ \hat P_{E,\sigma} \hat {\cal H}(x,E,\sigma)
+ h.c.
\right].
\label{pHp}
\eqa
Here $h.c.$ represents the Hermitian conjugate of the fist term. 
$\hat {\cal H}(x,E,\sigma)$ is given by \eq{WDE3}
with $\tilde \kappa = \kappa$.
%%
%$\hat P_{E,\sigma} = \cb E, \sigma \rb {\cal S}^{-1} \lb E, \sigma \cb$,
%where
%${\cal S}^{-1}$ is the inverse of the differential operator ${\cal S}$ 
%defined by the convolution of wavefunctions with $\lb E,\sigma \cb E', \sigma' \rb $,
%\bqa
%%S\left( v, \frac{\delta }{\delta v} \right)
%{\cal S}\left( E, \sigma, \frac{\delta}{\delta E}, \frac{\delta}{\delta \sigma} \right)
% ~ \chi(E,\sigma)
% =
% \int DE' D\sigma' ~
%\lb E,\sigma \cb E', \sigma' \rb 
%\chi(E', \sigma').
%\label{conv3}
%\eqa
$\hat P_{E,\sigma}$ is an operator that satisfies
\bqa
 %\cb E, \sigma \rb {\cal S}^{-1} \lb E, \sigma \cb 
 \hat P_{E,\sigma}
 \int DE' D\sigma' \cb E', \sigma' \rb \chi(E',\sigma')
%=   \cb E, \sigma \rb {\cal S}^{-1} {\cal S} \chi(E,\sigma)
= 
  \cb E, \sigma \rb\chi(E,\sigma).
 \label{conv4}
  \eqa
%%In \eq{conv4},
%%the convolution integration
%%$ \int DE' D\sigma'    \lb E, \sigma \cb E', \sigma' \rb \chi(E',\sigma')$
%%smears $\chi(E,\sigma)$,
%%and ${\cal S}^{-1}$ undoes the smearing. 
%$\hat P_{E,\sigma}$ first projects 
In \eq{pHp}, a general state is first projected 
to the state with each collective variable $(E,\sigma)$,
and then the Hamiltonian associated with the collective variables, 
$\hat {\cal H}(x;E,\sigma)$ is applied (see \fig{fig:H}).
For any $\cb E, \sigma \rb$,
$\hat {\cal H}(x)$ satisfies 
\bqa
\hat {\cal H}(x) \cb E, \sigma \rb = 
{\cal H}^{E, \sigma}(x) 
\cb E, \sigma  \rb,
\label{eq:inducedWdW}
\eqa
where
\bqa
{\cal H}^{E, \sigma}(x) 
=
\frac{1}{2} \left[
\tilde h^{E, \sigma}(x) 
+
\tilde h^{E, \sigma^\dagger}(x) 
\right]
\left\{
1 + 
O \left( 
\lc \kappa^2
 \frac{\delta}{\delta E_{\mu i}(x)}
,
 \lc \kappa^2
  \frac{\delta}{\delta \sigma(x)} 
  \right)
\right\}.
\label{HES}
\eqa
The construction of $\hat P_{E,\sigma}$
and the derivation of \eq{HES} 
are in Appendix \ref{sec:WdW}.
In the limit that 
$
\lc \kappa^2
 \frac{\delta}{\delta E_{\mu i}(x)}
,
 \lc \kappa^2
  \frac{\delta}{\delta \sigma(x)} 
\ll 1
$,
the higher derivative terms in \eq{HES} can be ignored,
and  ${\cal H}(x)$ induces 
the Wheeler-DeWitt Hamiltonian for the collective variables at long distance scales,
\bqa
\hat {\cal H}(x) \int DE D\sigma ~ \cb E, \sigma \rb \chi(E,\sigma) = 
 \int DE D\sigma ~ \cb E, \sigma \rb ~
 {\cal H}^{E, \sigma}(x) 
 \chi(E,\sigma).
\label{WDW8}
\eqa 
Unlike the case with the minisuperspace cosmology discussed in the previous section,
it is hard to perform the functional integrations over the collective variables explicitly in \eq{pHp}.
%It is likely that the resulting Hamiltonian is a strongly interacting Hamiltonian for the matrix field
%as it is the case for the minisuperspace cosmology.
In the following, we discuss general features of the Hamiltonian,
focusing on its locality.

\begin{figure}[ht]
\includegraphics[scale=0.40]{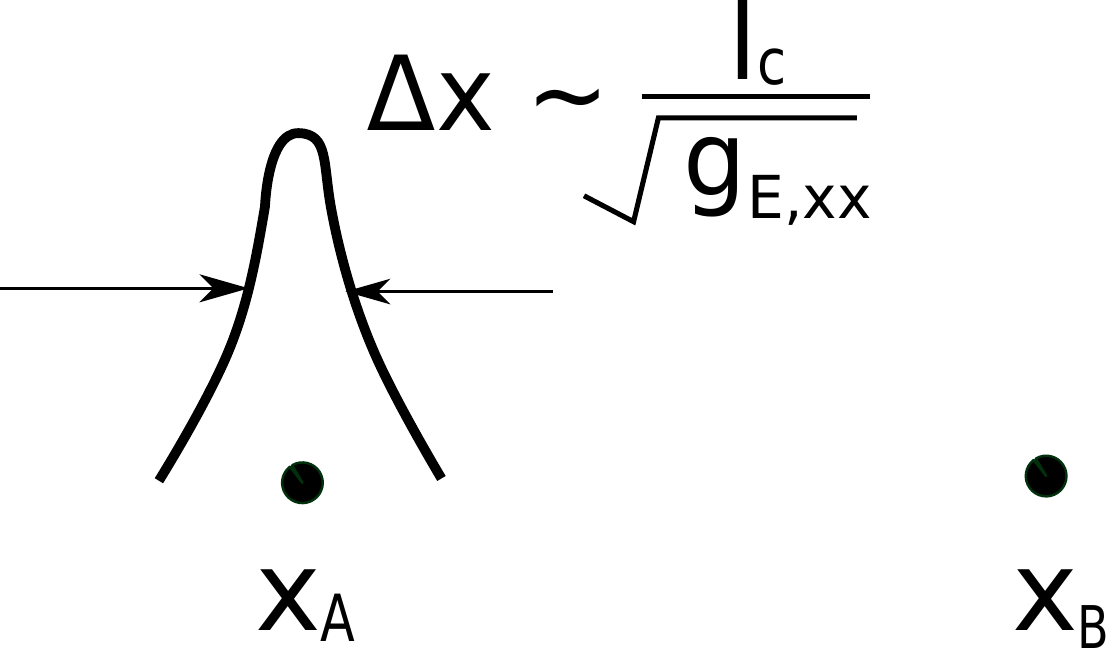}
\caption{
$\hat \cH(x_A;E,\sigma)$ has a spread over 
$\Delta x \sim \frac{\lc}{\sqrt{ g_{E,xx} } }$
along $x$ direction in the manifold.
In the presence of a large (small) mutual information between $x_A$ and $x_B$,
$g_{E,\mu \nu}$ is small (large) and the spread of the operator inserted at $x_A$ can (can not) reach $x_B$.
Therefore, the strength of coupling between $x_A$ and $x_B$
depends on entanglement of target states which determines
the metric.
}
\label{fig:relative}
\end{figure}

By choosing a space dependent speed of local time evolution,
we construct a Hamiltonian,
\bqa
\hat H_n(t) \equiv \int dx~ n(x,t) \hat \cH(x), 
\label{Hnt}
\eqa
where $n(x,t)$ in general depends both on space and time.
As a Hamiltonian for matter fields, 
one can ask how local the Hamiltonian is in space. 
In order to answer this question,
we first focus on $\hat \cH(x;E,\sigma)$ 
which is a part of $\hat \cH(x)$.
In \eq{WDE3},
the point-splitting of the kinetic terms
and the higher-derivative terms are controlled by $\lc$.
Therefore, $\hat \cH(x;E,\sigma)$ 
is local at length scales larger than $\Lc$,
if distances are measured 
with the metric $g_{E, \mu \nu}$.
However, there is no absolute sense of locality in $\hat H_n(t)$
because 
the metric that determines proper distances is 
not fixed.
Instead, $\hat \cH(x)$ is given by 
the sum of $\hat \cH(x;E,\sigma) \hat P_{E,\sigma}$
over different metrics.
To understand this point, 
let us consider two points, 
say $x_A$ and $x_B$ in the manifold.
Because the metric in $\hat \cH(x;E,\sigma) $
is determined by the state to which 
a target state is first projected by $\hat P_{E,\sigma}$,
the strength of the coupling between $x_A$ and $x_B$ 
in the Hamiltonian
is determined by the entanglement present in the target state.
A state which supports small mutual information between the two points
is projected to a state in which the proper distance between the points is large.
Accordingly, the operators in \eq{WDE3}
are spread over a small {\it coordiniate} distance,
and the coupling between the points is weak.
Conversely, for a state which supports large mutual information between the two points,
the metric in  $\cb E, \sigma \rb$  gives a small proper distance and a large coupling in $\hat H_n(t)$.
This is illustrated in \fig{fig:relative}.
There is no absolute locality 
because the metric with which locality is defined varies with states. 
Since the coupling between any two points can be large for long-range entangled states,
$\hat H_n(t)$ is not a local Hamiltonian as an operator. 
This is expected because there is no fixed notion of distance 
in any theory of background independent gravity\cite{PhysRevLett.114.031104}.
%%Nonetheless, the range of interactions is finite 
%%in the proper distance measured with the metric
%%set by a state on which the Hamiltonian acts.
Since locality of the Hamiltonian is determined relative to target states,
we call $\hat H_n(t)$ {\it relatively local}.
We emphasize that this conclusion on relative locality holds generally
for Hamiltonians which induce background independent gravity 
irrespective of specific choice of $\Psi(\Phi;E,\sigma)$. 

Ideally, one would hope to fix the regularization scheme 
and the higher-derivative terms in \eq{HWDW} such that 
$\hat \cH(x)$ and $\hat \cH_\mu (x)$ satisfy the hypersurface embedding 
algebra at the quantum level\cite{TEITELBOIM1973542}.
The commutators that involve the momentum constraint satisfy the algebra easily.
However, it is not clear whether there exists a regularized matter Hamiltonian 
which obeys the closed algebra at the quantum level.
What is guaranteed in this construction is that 
$\hat \cH(x)$ and $\hat \cH_\mu (x)$ satisfy 
the closed algebra to the leading order in the large $N$ limit
within states with slowly varying collective variables in space.
%%(For those states in which the collective variables 
%%vary little within the cut-off length scale $\Lc$ in space,
%%the specific point-splitting scheme and the higher-derivative terms are not important.)

%\subsection{Time evolution}

Now we view $\hat \cH(x)$ and $\hat \cH_\mu (x)$ as generators of symmetry.
States that are invariant under the symmetry, if exist, satisfy
\bqa
\cH^{E, \sigma}_\mu(x) \chi(E, \sigma) & = & 0,
\label{mC2} \\
%%\int dx ~n(x) ~
\cH^{E, \sigma}(x) \chi(E, \sigma) & = & 0.
\label{hC2} 
\eqa
\eq{mC2} combined with \eq{mCR} implies that the quantum state of the matter fields is 
invariant under diffeomorphism.
States that are invariant under diffeomorphism are topological 
because all physical properties are also invariant under diffeomorphism.
Such states either have no entanglement at all,
or must have infinitely long-range entanglement. 
Similarly, \eq{hC2} is a condition that a state is invariant 
under `time' translation.
States that satisfy Eqs. (\ref{mC2}) - (\ref{hC2})
generally have divergent norms
because the amplitude of the wavefunction is conserved
under the symmetry transformations that are non-compact\cite{1992grra.conf..211K,1992gr.qc....10011I}.

Therefore we consider normalizable states that break the symmetry spontaneously.
In particular, we consider normalizable semi-classical states
which satisfy the constraints to the leading order in $N$
but break the symmetry only to the sub-leading order,
\bqa
 \chi(E, \sigma) =  
 \chi_n ~
e^{
 -\int dx 
\left[
\sqrt{| \bar g|}~  
\frac{ 
\bar g^{\mu \nu} 
\bar g^{\alpha \beta} 
\left( g_{E,\mu\alpha} - \bar g_{\mu\alpha} \right)
e^{ - \frac{\lc^2 \bar \nabla^2}{2} }
\left( g_{E,\nu\beta} - \bar g_{\nu\beta} \right)
+ F(\bar \sigma)  \left( \sigma- \bar \sigma \right)
e^{ - \frac{\lc^2 \bar \nabla^2}{2} }
\left( \sigma- \bar \sigma \right)
}{2 \Delta^3}
- \frac{i}{\kappa^2} \left( \bar \pi^{\mu \nu} g_{E,\mu \nu}  +  \bar \pi_\sigma \sigma \right) 
\right]
}. \nn
\label{eq:semi}
\eqa
Here $\bar g_{\mu \nu}(x)$, $\bar \sigma(x)$, $\bar \pi^{\mu \nu}(x)$ and $\bar \pi_\sigma(x)$
are classical collective variables and their conjugate momenta,
and $\chi_n$ is a normalization constant.
$e^{ - \frac{\lc^2 \bar \nabla^2}{2} }$
suppresses fluctuations of the collective variables
with momenta larger than $\lc^{-1}$,
where $\bar \nabla_\mu$ is the covariant derivative associated with the metric $\bar g_{\mu \nu}$.
The wavefunction is manifestly invariant under $SO(3)$ gauge transformations.
Both collective variables and their conjugate momenta are well defined if
%\bqa
$\frac{\lc}{\Delta} \gg 1$,
$\frac{\lc \bar \pi^{\mu \nu}}{\sqrt{\bar g}}, \frac{\lc \bar \pi_\sigma }{\sqrt{\bar g}}
\gg \frac{1}{N^2} \left( \frac{\lc}{\Delta} \right)^{3/2}$. 
In the large $N$ and the long-wavelength limits,
Eqs.(\ref{mC2}) and (\ref{hC2}) become
the classical constraint equations\cite{PhysRev.116.1322},
\bqa
&& 
  \frac{1}{\sqrt{|\bar g|}} 
\left(
\bar \pi^{\mu \nu} \bar \pi_{\mu \nu} 
-\frac{1}{2} (\bar \pi^\mu_\mu)^2
+\frac{1}{2 F(\bar \sigma)} \bar \pi_\sigma^2
\right) 
 + 
\sqrt{|\bar g|}
 \left(  
 - \bar R +  \frac{F(\bar \sigma)}{2} \bar g^{\mu \nu} \nabla_\mu \bar \sigma \nabla_\nu \bar \sigma
+  V(\bar \sigma) 
\right) 
%%\right] 
= 0, \nn
&& 
%%\int dx ~n_\mu(x) ~ \left[
 2  \nabla_\nu \bar \pi^{\mu \nu} 
- ( \nabla^\mu \bar \sigma ) \bar \pi_\sigma 
%%\right] 
=0
\label{classicalC}
\eqa
 to the leading order in 
$1/N, \Delta/\lc, ( \lc  \nabla), ( \Lc \pi) $.
If the classical collective variables and the conjugate momenta satisfy \eq{classicalC}, 
semi-classical states obey the momentum and Hamiltonian constraints approximately.

For such states with weakly broken symmetry, 
the constraints generate non-trivial evolution
by creating Goldstone modes associated 
with the spontaneously broken symmetry,
\bqa
\cb \chi(dt) \rb 
 =    
e^{-i dt \int dx ~ \left[ 
n^{(1)}(x) \hat \cH (x) 
+ n^{(1) \mu}(x) \hat \cH_\mu (x) 
\right] }
 \cb  \chi \rb,
\eqa
where $dt$ is an infinitesimal parameter.
$n^{(1)}(x)$ and $n^{(1)  \mu}(x)$ are arbitrary functions of $x$ 
which control the local speed of the unitary transformation
and the shift respectively.
From \eq{eq:Hmu} and \eq{WDW8}, we obtain
\bqa
\cb \chi( dt ) \rb & = & \int D E^{(0)} D\sigma^{(0)}~  \cb   E^{(0)},\sigma^{(0)} \rb 
e^{-i dt \int dx \left[ n^{(1)}(x) \cH^{E^{(0)},\sigma^{(0)}} (x) 
	+ n^{(1)  \mu}  \cH^{E^{(0)},\sigma^{(0)}}_{ \mu}(x)
\right]} 
\chi( E^{(0)},\sigma^{(0)}) \nn
 & = & \int D E^{(0)} D \sigma^{(0)} DE^{(1)} D\sigma^{(1)} D\pi^{(1)} D\pi_\sigma^{(1)}~ 
\cb   E^{(1)}, \sigma^{(1)} \rb  \
 e^{\frac{i}{\kappa^2} \int dx \left[ \pi^{(1)\mu i} ( E^{(1)}_{\mu i} - E^{(0)}_{\mu i} ) + \pi_\sigma^{(1)} ( \sigma^{(1)} - \sigma^{(0)} ) \right]} \times \nn
&& e^{-\frac{i}{\kappa^2}  dt \int dx \left[ 
  n^{(1)}(x) \cH[ E^{(0)}, \sigma^{(0)}, \pi^{(1)}, \pi^{(1)}_\sigma ]
+ n^{(1) \mu} \cH_\mu[E^{(0)},\sigma^{(0)},\pi^{(1)},\pi^{(1)}_\sigma]
 \right]} 
\chi( E^{(0)},\sigma^{(0)}),
\label{eq:Z1}
\eqa
where
$D \pi D\pi_\sigma \equiv \mu^{-1}(E,\sigma) \prod_x d \pi^{\mu i}(x) d \pi_\sigma(x)$,
\bqa
\cH[E,\sigma,\pi,\pi_\sigma] &=&
: \frac{1}{|E|} \left(
G_{ijkl} E_\mu^{j} E_\nu^{l}  \pi^{\mu i}  \pi^{\nu k} + \frac{1}{2 \Fs} \pi_\sigma^2 
\right) : \nn
&&
 + 
|E| \left(  
 - R +  \frac{\Fs}{2} g_E^{\mu \nu} \nabla_\mu \sigma \nabla_\nu \sigma
+  V(\sigma) 
+ U_3( g_E, \sigma)
\right) + O(N^{-2}, \Lc \pi^3 ), \nn
\cH_\mu[E,\sigma,\pi,\pi_\sigma] & = &
   E_{\mu i} ~  \nabla_{\nu} ~ \pi^{\nu i}  
  - \nabla_\mu \sigma \pi_\sigma.
\eqa
After repeating this step infinitely many times 
in the $dt \rightarrow 0$ limit, 
one obtains
\bqa
\cb \chi(t ) \rb & = & \int DE(\tau,x) D\sigma(\tau,x) D\pi(\tau,x) D\pi_\sigma(\tau,x)  ~~ 
\cb E(t), \sigma(t) \rb ~
e^{ i S }~
\chi( E(0), \sigma(0) ),
\label{eq:Zf}
\eqa
where 
\bqa
S &=&  \frac{1}{\kappa^2} \int_0^t d\tau \int dx \Biggl\{ 
\pi^{\mu i} \partial_t E_{\mu i} + \pi_\sigma \partial_t  \sigma  
- n ~\cH[E,\sigma,\pi,\pi_\sigma]
- n^\mu ~ \cH_\mu[E,\sigma,\pi,\pi_\sigma]
\Biggr\}.
\label{fQG}
\eqa
The theory describes gravity
coupled with a scalar field
in a fixed gauge. 
In the large $N$ limit,
the saddle-point configuration 
which satisfies the classical field equations 
dominates the path integration.

The theory in \eq{fQG} has three length scales.
One is the cut-off length scale, $\lc$
which controls the relative locality of the theory.
At length scales larger than $\lc$, 
where lengths are measured with a saddle-point configuration of the dynamical metric,
the local field theory is valid
for the description of fluctuations of the collective variables
above the saddle-point configuration.
At shorter length scales, 
non-local effects kick in  
through the higher-derivative terms in the Hamiltonian.
The other length scale is the Planck scale, $\kappa \sim \frac{\lc}{N}$
below which quantum fluctuations of the collective
variables become important.
Another scale is the curvature of the saddle-point geometry set by the vacuum energy. 
Here we assume that $V(\sigma)$ is chosen such that the cosmological constant is much smaller than $\lc^{-4}$,
which in general requires a fine tuning of the potential.
In the large $N$ limit, 
the semi-classical field theory approximation is valid 
for modes whose wavelengths are larger than $\lc$.

\section{Black hole evaporation}
\label{sec:Hawking}

\subsection{Entanglement neutralization}

\label{sec:HawkingA}

The discussion in the previous section shows that
a relatively local Hamiltonian for the matter field 
induces a quantum theory for the collective variables,
which reduces to Einstein's gravity coupled with a scalar field
at long distances in the large $N$ limit.
Given an initial state $\cb \chi \rb$,
the state at parameter time $t$ is given by
\bqa
\cb \chi(t) \rb = {\cal P}_T ~ e^{ -i \int_0^t d\tau \int dx ~ \left[
n(x,\tau) \hat \cH(x)
+ n^\mu(x,\tau) \hat \cH_\mu(x)
\right]
} \cb \chi \rb,
\label{tE}
\eqa 
where ${\cal P}_T$ time-orders the evolution operator.
The state at time $t$ can be written in the basis of $\cb E, \sigma \rb$ as
$
\cb \chi(t) \rb = \int DE D\sigma~
\cb E, \sigma \rb
\chi(t; E, \sigma) 
$.
If the initial state is chosen to be a semi-classical state in \eq{eq:semi},
$\chi(t; E, \sigma)$ is sharply peaked at a saddle-point path in the large $N$ limit. 
The saddle-point path, 
$\{ \bar E_{\mu i}(x,t), \bar \sigma(x,t), \bar \pi^{\mu i}(x,t), \bar \pi_\sigma(x,t) \}$
solves the classical field equation with the initial condition,
$g_{\bar E, \mu \nu}(x,0)=\bar g_{\mu \nu}(x)$, 
$\bar \sigma(x,0)=\bar \sigma(x)$, 
$\bar E^{\nu}_{ i}(x,0) \bar \pi^{\mu i}(x,0) =  \bar \pi^{\mu \nu}(x) + \bar \pi^{\nu \mu}(x)$,
$\bar \pi_\sigma(x,0)=\bar \pi_\sigma(x)$.

\begin{figure}[ht]
\begin{center}
%\subfigure[]{
\includegraphics[scale=0.3]{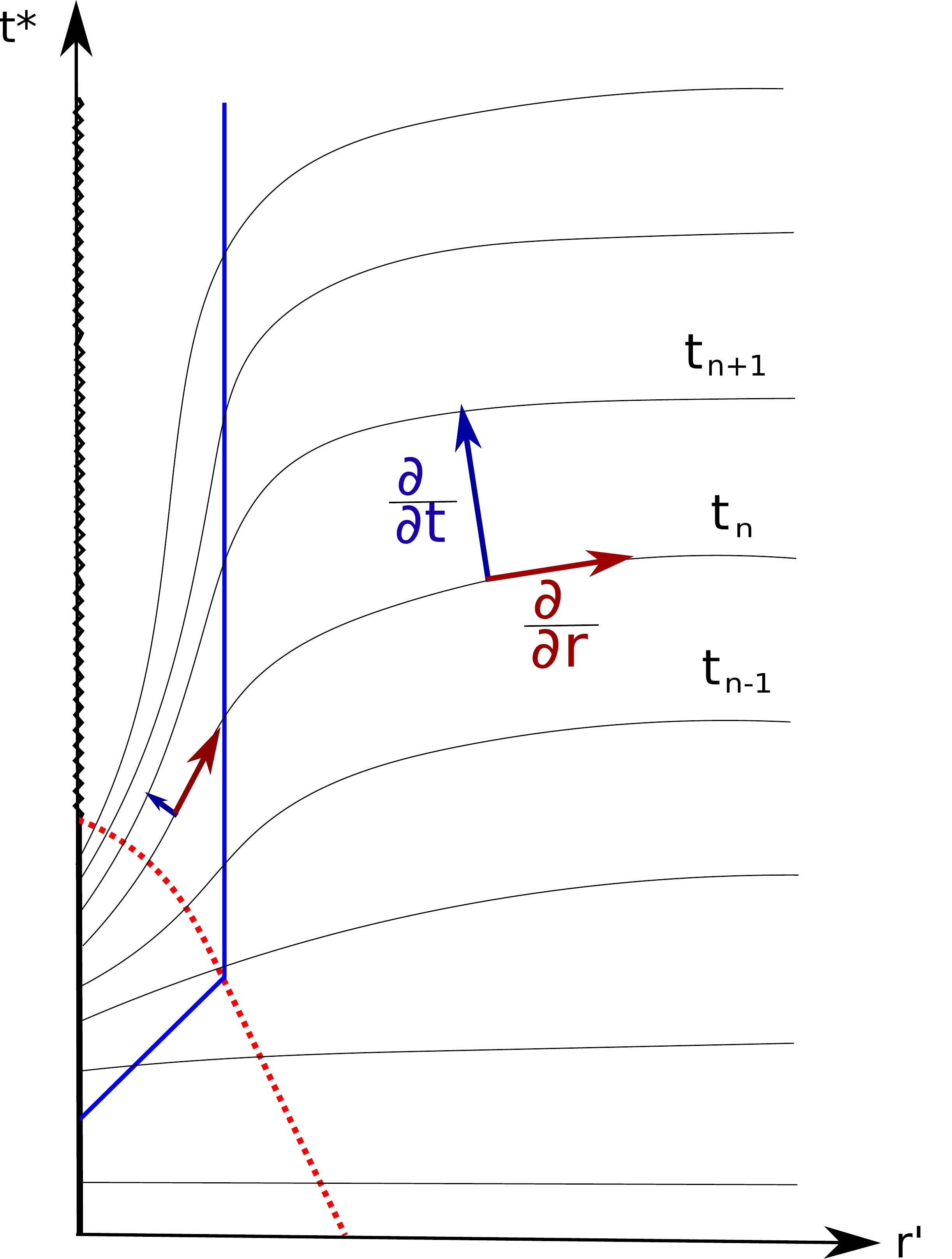} 
\label{fig:collapse}
%} ~~~~~
%\subfigure[]{\includegraphics[scale=0.5]{multiple_precess} 
%\label{fig:multiple_precess}
%}
\end{center}
\caption{
A collapsing mass shell (red dashed line) and 
the horizon (blue solid line) in the in-going Eddington-Finkelstein coordinate system,
where the transverse directions ($\theta, \phi$) are suppressed.
The metric outside the collapsing mass shell is given by the Schwarzschild metric,
$ds^2 = - \left( 1 - \frac{2M}{r'} \right) dv^2 + 2 dr' dv + r^{'2} ( d \theta^2 + \sin \theta^2 d \phi^2 )$
where $t^*=v - r'$.
The curves represent time slices 
which march forward with a constant lapse far from the black hole 
while avoiding the singularity inside the horizon.
$\frac{\partial}{\partial r}$  $\left( \frac{\partial}{\partial t} \right) $
represents the vector field that is tangential (perpendicular) to each time slice.
It is noted that 
$\frac{\partial}{\partial r}$,
which is distinct from $\frac{\partial}{\partial r'}$,
is chosen to be space-like everywhere. 
}
\label{BHcollapse}
\end{figure}

Suppose the initial condition of the semi-classical state
is chosen such that the classical solution describes a gravitational 
collapse of a spherically symmetric mass shell, 
which forms a macroscopic black hole 
with mass $M$ in the asymptotically flat Minkowski space.
We assume that the size of the horizon 
is much larger than the cut-off length scales,
$r_H  = 2 M \kappa^2 \gg \lc$
right after the black hole is formed. 
Across the horizon, 
the underlying quantum state supports color entanglement,
\bqa
S_\Phi^{i} = \frac{\pi r_H^2}{\kappa^2} \sim N^2 \left( \frac{r_H}{\lc} \right)^2.
\label{Si}
\eqa
We identify \eq{Si} as the Bekenstein-Hawking entropy\cite{PhysRevD.7.2333,Hawking1975}.
On the other hand, the singlet \ee is negligible
when the black hole is just formed.

In describing the consequent evolution of the black hole,
we choose 
$n^\mu(x,t)=0$
and 
$n(x,t) >0$ 
at all $x,t$
in \eq{tE}.
Far away from the black hole,
the lapse is chosen to approach a non-zero constant.
Inside the black hole, the lapse is chosen
such that time slices do not hit the `singularity',
and the theory at each time slice 
stays within the realm of the semi-classical field theory.
As time progresses, the space that connects
the interior of the black hole and the asymptotic region is stretched 
as is illustrated in  \fig{BHcollapse}.

For a large but finite $N$, 
one should include quantum fluctuations
of the collective variables in \eq{eq:Zf}.
Due to quantum fluctuations,
the black hole emits Hawking radiation,
and its mass decreases in time.
Let $\bar g_{\mu \nu}'(x,t)$ describe an evaporating black hole geometry 
that satisfies the classical field equation
in the presence of a time dependent black hole mass
and the energy-momentum tensor of the Hawking radiation.
The rate at which the black hole mass decreases should be self-consistently
determined by the Hawking radiation 
created by small fluctuations around the classical geometry, 
$g_{\mu \nu}(x,t) = \bar g_{\mu \nu}'(x,t) + \delta g_{\mu \nu}(x,t)$, 
$\sigma(x,t) = \bar \sigma'(x,t) + \delta \sigma(x,t)$.
For large black holes,
the Hawking radiation should be well approximated by
the adiabatic approximation because the rate at which the mass decreases 
is much smaller than $r_H^{-1}$.

As the black hole evaporates, 
the horizon shrinks and the color entanglement entropy decreases.
On the other hand, the singlet entanglement entropy increases
because Hawking radiation is emitted 
in the form of fluctuations of the collective variables.
This is easy to understand  in the weakly coupled effective theory for the collective variables.
From the perspective of the fundamental matrix field,
it is not obvious why Hawking radiation is emitted only in the singlet sector.
However, the underlying theory for the matrix field in \eq{pHp} is likely to be a strongly coupled field theory, 
and it is conceivable that there is only $O(1)$ Hawking radiation\cite{0264-9381-27-9-095015}.
If there was Hawking radiation of $O(N^2)$ color degrees of freedom, 
the induced theory for the collective variables could not be the semi-classical general relativity
which we know is the correct description of \eq{pHp}.
%%In particular, Hawking radiation of color degrees of freedom 
%%would cause the color entanglement entropy to increase across the horizon.
%%This would lead to an increasing horizon size
%%because the area of horizon is a measure of the color entanglement entropy.
Therefore the increasing entanglement between the Hawking radiation
and the degrees of freedom inside the horizon should be 
in the singlet sector.
In this regard, black hole evaporation can be viewed as
an {\it entanglement neutralization process}
in which  entanglement across horizon is transferred from
color degrees of freedom to singlet degrees of freedom.

\subsection{Late time evolution}

\label{sec:HawkingB}

The fate of $\chi(t;E, \sigma)$ in the large $t$ limit
largely depends on which of the following two possibilities is realized. 
The first possibility is that \eq{tE} 
evolves to a state which ceases to support
a well-defined horizon as early as the Page time.
The second is that $\chi(t;E, \sigma)$ remains sharply peaked around
the time dependent classical metric $\bar g_{\mu \nu}'(x,t)$ with a well defined horizon.
Resolving this issue is a complicated dynamical question.
It may well be that the answer depends on details of initial states.
In this section, we consider consequences of the second possibility,
assuming that there exists some initial states 
which continue to support a well-defined horizon throughout the evolution 
before the size of black hole reaches the cut-off length scale. 
The reason we focus on the second possibility is because
the black hole information puzzle arises in that case\cite{Hawking1975}.
Our goal here is to understand how the puzzle can be in principle
resolved in the current framework.

\begin{figure}[ht]
\begin{center}
\subfigure[]{
\includegraphics[scale=0.3]{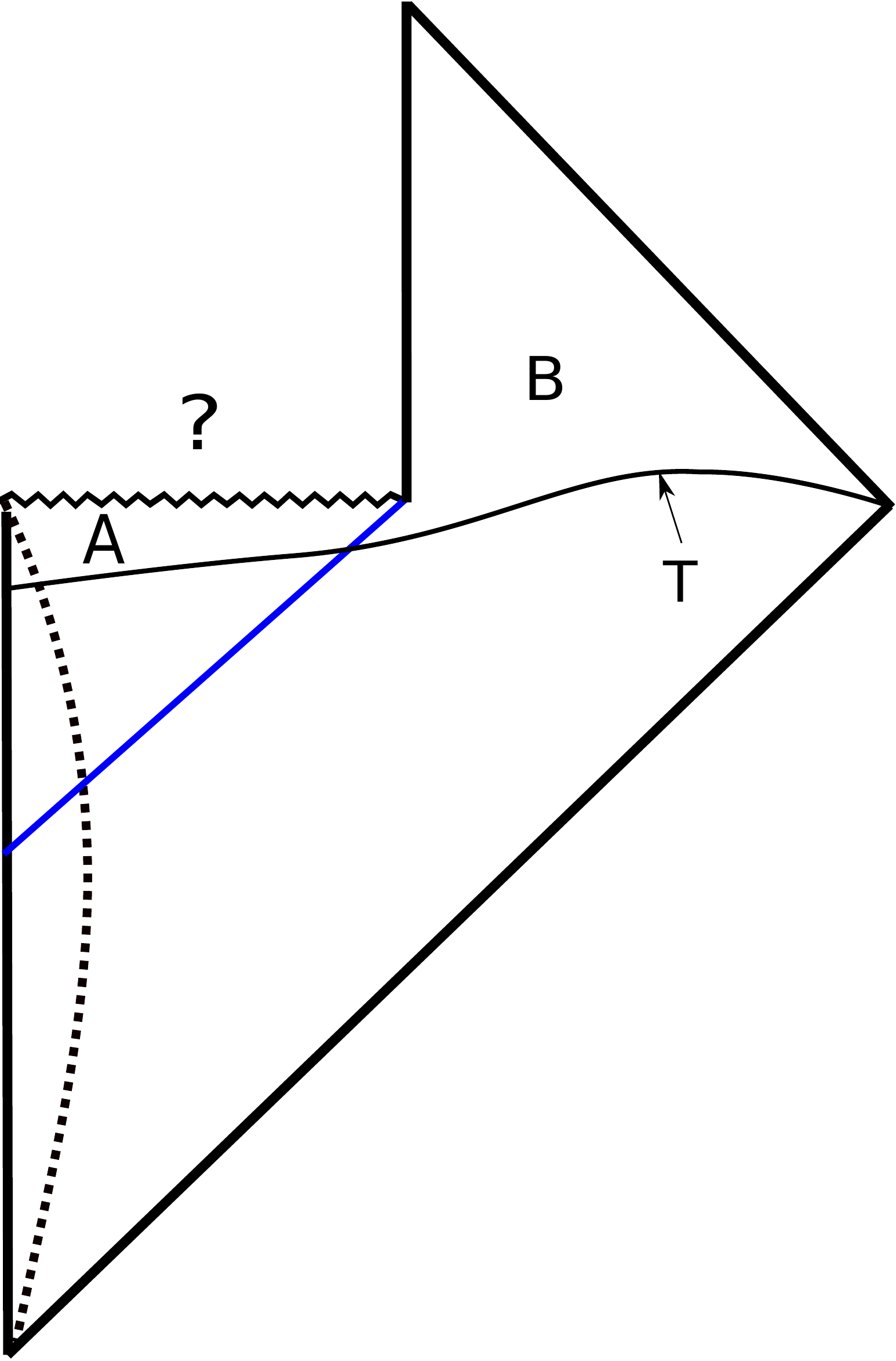} 
\label{fig:PD_evaporation}
} 
\subfigure[]{
\includegraphics[scale=0.35]{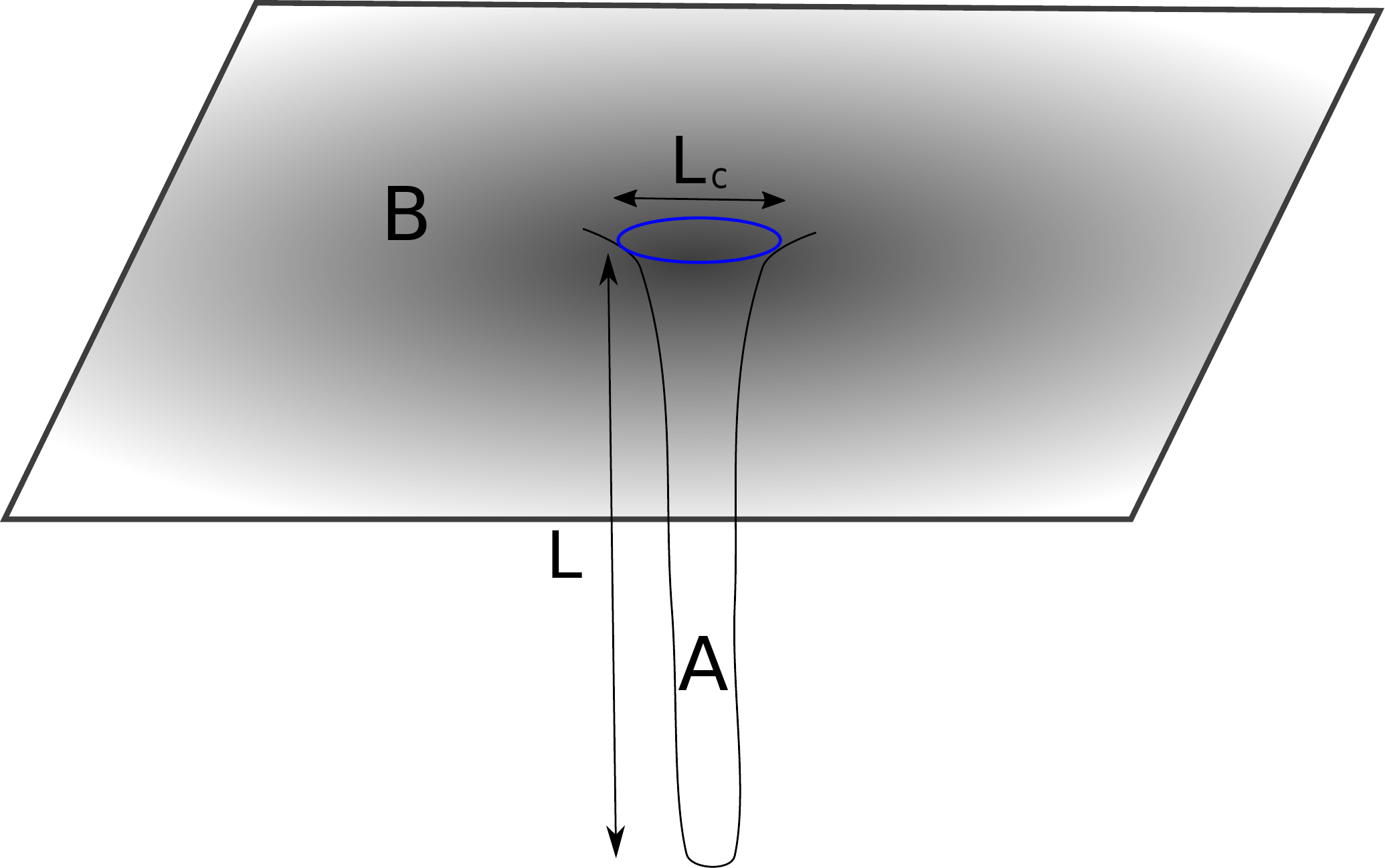} 
\label{fig:mutual_info}
} ~~~~~
\end{center}
\caption{
(a) The Penrose diagram of an evaporating black hole. 
The question mark is intended to represent the unknown 
fate of the interior of the black hole.
(b) The geometry for the quantum state at $t=T$.
When the size of the horizon is $\lc$,
the region inside the horizon is stretched to
a long funnel with size 
$L \sim M^{7/2} \kappa^5 \lc^{-1/2}$.
}
\label{evap_mutual_info}
\end{figure}

Let us choose the lapse such that at time $T$
the long throat inside the horizon as well as the horizon itself
reaches the size of $\sim \lc^2$ in the transverse direction. 
Upto this point, the local semi-classical description is still valid.
The asymptotic time that takes 
for a large black hole with mass $M$ to evolve to this state scales as $T \sim M^3 \kappa^4$.
During this time, the throat inside the horizon  stretches to the size, 
$L \sim \sqrt{g_{rr}} M^3 \kappa^4 \sim M^{7/2} \kappa^5 \lc^{-1/2}$,
where $g_{rr} \sim \frac{M \kappa^2}{\lc}$ is used for the metric inside the horizon.

Under the time evolution, 
the size of the horizon continues to shrink,
and so does the color entanglement entropy.
At $t=T$, the color \ee across the horizon becomes
$S_{\Phi}^{f} \sim \frac{\lc^2}{\kappa^2} \sim N^2$.
On the other hand, the Hawking radiation generates
a large entanglement across the horizon, 
\bqa
S_{H} \sim S_\Phi^{i}. 
\eqa
In the context of the present induced gravity, 
the `information puzzle'\cite{
Hawking1975,
THOOFT1990138,
PhysRevD.48.3743,
PhysRevLett.71.3743,
0264-9381-26-22-224001,
Almheiri2013,
2017tasi.conf..353P} 
can be phrased as the statement 
that the small color entanglement entropy,
which is identified as the Bekenstein-Hawking entropy,
can not account for
the large entanglement created by the Hawking radiation.
However, this is not necessarily paradoxical 
because the color entanglement entropy captures only a part of the full entanglement.
The other part is the singlet entanglement entropy which is supported
by correlations between fluctuations of the collective variables.

In this theory, the time evolution is unitary by construction.
%%One question that needs to be answered is 
%%whether the local semi-classical approximation remains valid
%%inside the horizon in the presence of fluctuations of the collective variables
%%that are needed to maintain purity of the quantum state
%%near the end of the evaporation process.
In order to support the entanglement 
with the Hawking radiation outside the horizon,
at least $e^{S_{H}}$ states need to be excited
inside the horizon. 
%As was shown in the previous section,
%two states with different collective variables
%but with a same local proper volume
%become more or less orthogonal if 
%\bqa
%\delta h^{\mu}_{\nu} \sim \frac{1}{N}, ~~ \delta \sigma \sim \frac{e^{-3/2 \sigma}}{N}
%\label{gauss}
%\eqa
%over any volume $\lc^{3}$.
In the large $N$ limit,
modes with wavelengths larger than $\lc$
are described by the weakly coupled field theory,
and they have Gaussian fluctuations which are order of 
$\delta h^{\mu}_{\nu}, ~ \delta \sigma \sim \frac{1}{N}$
\footnote{
For example, we can choose $\Fs \sim e^{3\sigma}$, $V(\sigma) \sim \lc^{-2} e^{4 \sigma}$
such that 
$\delta h^{\mu}_{\nu} \sim \frac{1}{N}$,
$\delta \sigma \sim \frac{e^{-3/2 \sigma}}{N}$
for $e^\sigma \ll 1$.
%$\delta \sigma \sim \frac{e^{-3/2 \sigma}}{N}$ 
}.
%The modes described by the effective field theory 
%can contribute $O(1)$ bit per volume $\lc^3$.
%%
The volume of the throat inside the horizon 
is $V \sim M^{7/2} \kappa^5 \lc^{3/2}$ at $t=T$.
If all field theory modes with wavelengths larger than $\lc$ are excited,
the total number of states that are available inside the horizon is $e^{S_{col}}$ with
\bqa
S_{col} \sim  M^{7/2} \kappa^5 \lc^{-3/2} \sim N^2 \left( \frac{r_H}{\lc} \right)^{7/2}.
\label{Scol}
\eqa 
For a macroscopic initial black hole with $r_H \gg \lc$, $S_{col} \gg S_H$.
%This means that there are more than enough singlet modes described by the semi-classical field theory
%that can support the entanglement with the Hawking radiation.
%
The number of field theory modes available inside the horizon 
is much larger than what the color \ee (Bekenstein-Hawking entropy) accounts for
near the end of evaporation process\footnote{
It is noted that a renormalization of the Newton's constant by $1/N$ corrections
is not enough to incorporate the singlet entanglement entropy within the color entanglement entropy.}.
The states counted in \eq{Scol} include highly excited states.
However, one does not need all of them.
In order to account for $S_H \sim N^2 \left( \frac{r_H}{\lc} \right)^2$,
it is enough to excite modes with wavelengths
larger than $\lambda \sim \lc  \left( \frac{r_H}{\lc} \right)^{1/2}$.
In the $r_H \gg \lc$ limit, 
only those excitations whose wavelengths are much larger than $\lc$ 
are needed to account for the entanglement with the early Hawking radiation.

%%What fails is the statistical interpretation of the Bekenstein-Hawking entropy.
The large number of singlet states that are not captured by the Bekenstein-Hawking entropy
suggests that the quantum state inside the horizon is far from equilibrium.
The lack of equilibrium can arise dynamically 
because of the relative nature of the Hamiltonian.
Since the strength of coupling between two given points in the manifold is determined by states,
points that were once connected by a strong coupling can dynamically decouple at a later time 
if the state at later time supports little color entanglement.
In this sense, not only states but also the Hamiltonian effectively flow in time
under the evolution generated by the relatively local Hamiltonian.
The growth of a long geometry inside the horizon with a small contact with the exterior
is a form of dynamical localization where the coupling across the horizon becomes weaker in time.

%%%%%%%%%%%
%%%%%%%%%%%
%%%%%%%%%%%

\section{Summary and Discussion}
\label{sec:summary}

In this paper, it is shown that 
a quantum theory of gravity can be induced from quantum matter.
Metric is introduced as a collective variable 
which controls entanglement of matter fields. 
There exists a Hamiltonian for matter fields
whose induced dynamics for metric
coincides with the general relativity
at long distances in the large $N$ limit.
The Hamiltonian that gives rise to the 
background independent gravity is non-local.
However, it has a relative locality
in that the range of interactions is controlled
by entanglement present in target states.
Within the induced theory of gravity,
a black hole evaporation can be understood as 
a unitary process where entanglement of matter 
is gradually transferred 
from color degrees of freedom
to singlet collective degrees of freedom.
We close with some remarks and speculations on open questions.

\subsection{Baby universe as a dynamical localization}

\begin{figure}[ht]
\begin{center}
%\subfigure[]{
\includegraphics[width=5cm,height=4cm]{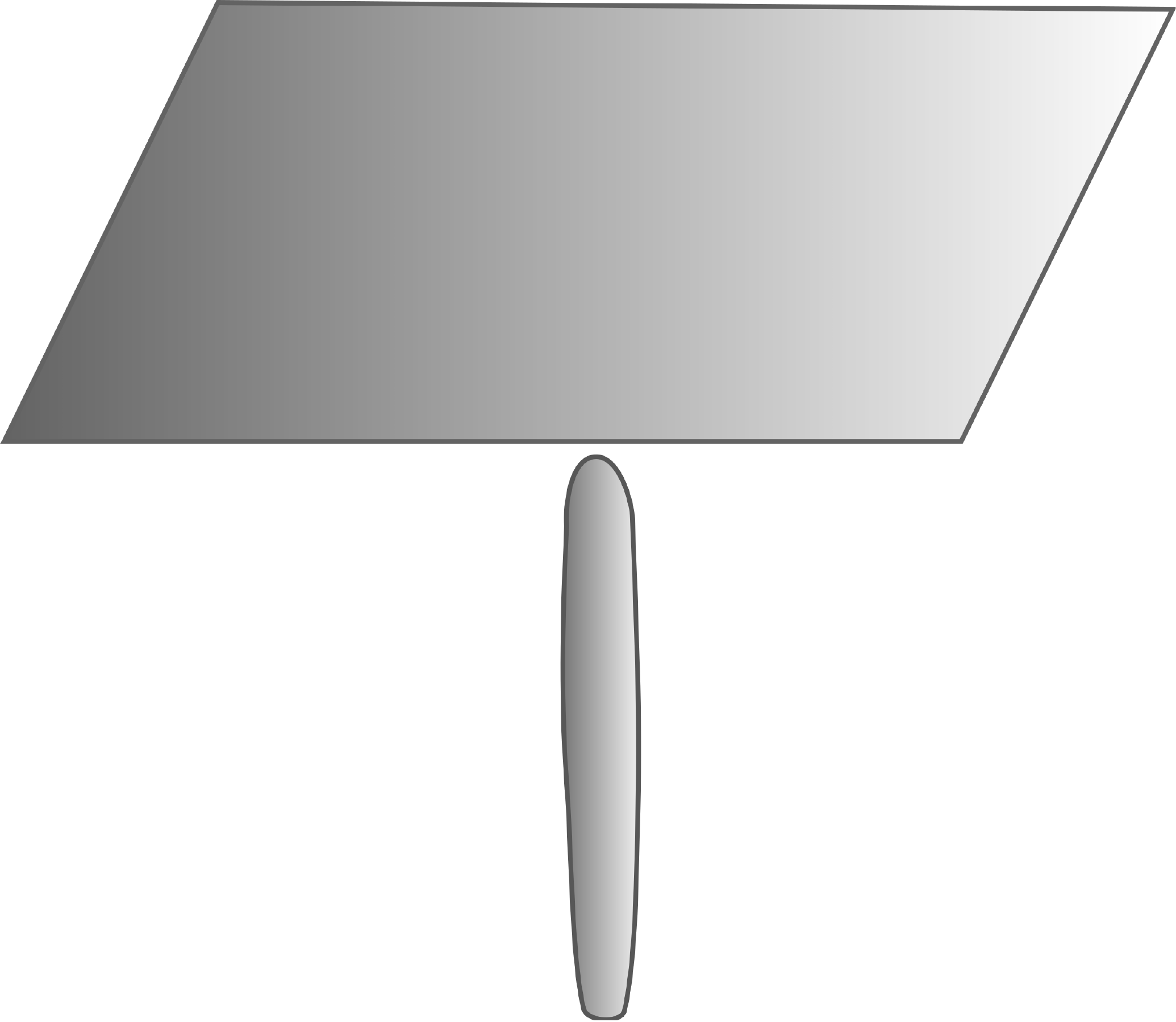} 
%\label{fig:babyU}
%} ~~~~~
%\subfigure[]{
%\includegraphics[width=7cm,height=4cm]{babyU_matrix.pdf} 
%\label{fig:babyU_matrix}
%} ~~~~~
\end{center}
\caption{
%(a) 
One possible outcome of evolving the state in \fig{fig:mutual_info} further in time
is a topological phase transition 
where the region inside the horizon is disconnected from the outer universe.
%(b) The matrix element to create a baby universe by inserting a local operator 
%(represented as $\times$ inside the ket) in a smooth geometry 
%is suppressed as $e^{-N^2 V' \lc^{-3} }$, 
%where $V'$ is the proper volume of the baby universe.
}
\label{fig:babyU}
\end{figure}

What happens if we continue to evolve the state in \eq{tE} beyond time $T$?
It is hard to answer the question without considering 
the full theory beyond the local semi-classical approximation.
Here we consider possibilities that do not modify the usual rules of quantum mechanics.
If the long throat inside the horizon remains attached to the outer space,
it gives rise to a long-lived (or stable) remnant (for a review on remnant, see Ref. \cite{CHEN20151} and references there-in).
If the region inside the horizon becomes geometrically disconnected from the exterior,
a baby universe can form as in \fig{fig:babyU}\cite{PhysRevD.37.904,PhysRevD.50.7403}.
From the perspective of the underlying matter fields, 
a baby universe corresponds to a dynamical localization 
where the region inside the horizon dynamically decouples from the exterior. 
Here localization is driven not by disorder but by the relative nature of the Hamiltonian,
where the strength of couplings between the interior and the exterior of the horizon
dynamically flows to zero at late time.

Since the proper volume inside the horizon can be arbitrarily large, 
there can be infinitely many different remnants or baby universes.
Although this seems unphysical, this is allowed within the present theory
because the number of degrees of freedom within a given region 
of the manifold is not fixed.
Any background independent quantum theory of gravity 
should include such states in the Hilbert space.
The presence of infinitely many internal states does not necessarily 
lead to an infinite production rate 
if the matrix element between a state with a smooth geometry
and a state with a remnant or a baby universe is exponentially
suppressed as the volume of the `hidden' space increases.
For example, let $\emi(x)$ be the flat Euclidean geometry, $g_{E,\mu \nu} = \delta_{\mu, \nu}$,
and $E_{\mu i}'(x)$ represent a geometry which coincides with the Euclidean metric for $|x|>R$
but has a long funnel with proper volume $V' \gg R^3$ for $|x|<R$.
Let $\hat O$ be an operator that has a support within $|x|<R$.
The matrix element is given by
\bqa
\lb E', \sigma' \cb \hat O \cb E, \sigma \rb ~\sim 
\int D^{ (\hat E, \hat \sigma) } \Phi ~ O ~ e^{- \int_{|x|<R} dx \Bigl[ |E'| {\cal L}(\Phi;E',\sigma') + |E| {\cal L}(\Phi;E,\sigma) \Bigr] }.
\eqa  
%%where the contribution from $|x|>R$ is canceled by the normalization factor.
If the matrix element is small enough,
the net production rate can be suppressed.

%Because $V' \gg R^3$, $G_{\mu \nu} \approx g_{E', \mu \nu}$
%for $|x|<R$ in \eq{GM}. 
%Therefore, $\lb E', \sigma' \cb \hat O \cb E, \sigma \rb \sim e^{-N^2 V'/\lc^3}$.
%This is illustrated in \fig{fig:babyU_matrix}.
%While the number of semi-classical states with a hidden geometry of volume $V'$ scales as $e^{ V'/\lc^3}$,
%the small matrix element in the large $N$ limit.

\subsection{dS/CFT} 

By construction, 
$\hat \cH_\mu (x)$
and $\hat \cH(x)$
satisfy the closed algebra\cite{TEITELBOIM1973542}
to the leading order in the $1/N$ and the derivative expansions.
However, the commutator between two Hamiltonian constraints
may have an anomaly that involves higher derivative terms and $1/N$ corrections.
Whether one can choose a regularization scheme for the matter Hamiltonian
such that the algebra is closed exactly
is an open question\cite{PhysRevD.36.3641,PhysRevD.37.3495,Lee:2012xba,PhysRevD.95.066003}.

\begin{figure}[ht]
\begin{center}
\includegraphics[scale=0.4]{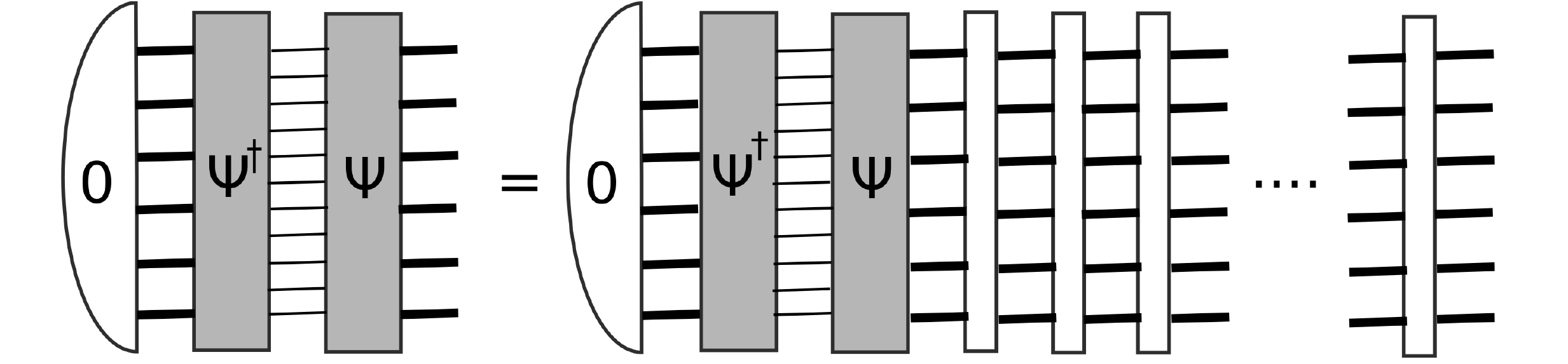} 
\label{fig:multiple_precess}
\end{center}
\caption{
A pictorial representation of the overlap 
between $\cb 0 \rb$ 
and $\cb E, \sigma \rb$.
Since $\cb 0 \rb$ is annihilated by the Hamiltonian and momentum constraints,
the overlap is invariant under the insertions of 
the evolution operator, 
which gives rise to a path integration
of the metric and the scalar field in the bulk. 
}
\label{fig:dscft}
\end{figure}

Suppose there exists a state $\cb 0 \rb$ 
which is annihilated by $\hat \cH(x)$ and $\hat \cH_\mu (x)$.
The existence of such states doesn't necessarily
require that the algebra is closed at the operator level.
Although $\cb 0 \rb$  is not normalizable in general,
the overlap with a normalizable state, $\cb E, \sigma \rb$ is well defined.
The overlap,
which can be viewed as a wavefunction of universe, 
is invariant under the insertion of  
the evolution operator generated by 
the Hamiltonian and momentum constraints\cite{Lee2016},
\bqa
\lb 0 \cb E, \sigma \rb
= \lb 0 \cb
%{\cal P}_T ~ e^{ -i \int_0^t d\tau \int dx ~ \left[
~ e^{ - i dt  \int dx ~ \left[
n(x,\tau) \hat \cH(x)
+ n^\mu(x,\tau) \hat \cH_\mu(x)
\right]
}
\cb 
E, \sigma \rb.
\eqa
Repeated insertions of the evolution operators lead to
\bqa
\lb 0 \cb E, \sigma \rb & = & 
\left.
\int DE(\tau,x) D\sigma(\tau,x) D\pi(\tau,x) D\pi_\sigma(\tau,x)  ~~ 
\lb 0 \cb E(t), \sigma(t) \rb ~
e^{ i S } \right|_{ E(0,x)=E(x), \sigma(0,x)=\sigma(x)},
\label{dSCFT}
\eqa
where the bulk action is given by \eq{fQG}.
Therefore, the overlap is given by
the $(D+1)$-dimensional path integration
with a Dirichlet boundary condition
for the collective variables
as is represented in \fig{fig:dscft}.
The bulk path integral can be viewed as 
the gravitational dual for 
the generating functional 
of the non-unitary boundary field theory 
in the dS/CFT correspondence\cite{1126-6708-2001-10-034,2011arXiv1108.5735A}.
If the lapse and the shift are integrated over,
the bulk path integration becomes a projection operator 
which imposes the Hamiltonian and momentum constraints
on the boundary state, $\cb E, \sigma \rb$.

\subsection{Non-gaussian states}

The standard lore in the AdS/CFT correspondence 
is that a bulk theory includes only a small number of 
fields if the dual boundary theory is 
in a strong coupling regime and
the majority of operators have large 
scaling dimensions\cite{Heemskerk:2009pn}.
The number of dynamical fields one has to keep in the bulk
is determined by the number of independent operators
from which all other operators can be constructed
as composite operators\cite{Lee:2013dln}.
Although there are in general infinitely many such operators,
if most of them acquire large scaling dimensions
they correspond to heavy fields in the bulk,
which can be integrated out without sacrificing locality
in the bulk.

From this perspective, 
it is rather surprising that 
a simple bulk theory 
is obtained from
the Gaussian wavefunction
defined at a temporal boundary.
The reason why we have only dynamical metric and a scalar field in the bulk
is because the initial states and the Hamiltonian for the matter field
are fine-tuned so that the time evolution generates
deformations contained 
within the sector of the energy-momentum tensor
and one scalar operator only.
From the point of view of RG flow, 
we are in the basin of attraction
toward a multi-critical point
via a fine tuning.
In general, there exist `relevant' perturbations
which take the flow away from the multi-critical point
once perturbations are turned on.
In the present work, 
we simply didn't consider such perturbations
in the initial state.
In order to suppress other operators without fine tuning,
one probably needs non-Gaussian wavefunctions
which describe strongly coupled boundary theories.

\section*{Acknowledgments}

The author thanks Bianca Dittrich for discussions and help in literature search,
and Ted Jacobson for comments on the draft
and communications on the issue of diffeomorphism invariance of the inner product.
The research was supported by
the Natural Sciences and Engineering Research Council of
Canada.
Research at the Perimeter Institute is supported
in part by the Government of Canada
through Industry Canada,
and by the Province of Ontario through the
Ministry of Research and Information.

\bibliography{references}

\newpage
\begin{appendix}

\section{Diffeomorphism invariance of the inner product}
\label{diffeo_inner}

First we note that 
\eq{Kf} is covariant under diffeomorphism, that is,
\bqa
K_{(\tilde E,\tilde \sigma)} f^{(\tilde E,\tilde \sigma)}_n (x) = 
 \lambda^{( E, \sigma)}_n  
 f^{(\tilde E,\tilde \sigma)}_n(x),
 \eqa
where $f^{(\tilde E,\tilde \sigma)}_n (\tilde x) =f^{( E, \sigma)}_n (x)$
with $\tilde x = x + \xi(x)$,
and $(\tilde E, \tilde \sigma)$
is related to  $( E,  \sigma)$
through the diffeomorphism.
This means
$ \lambda^{( \tilde E, \tilde \sigma)}_n  
= \lambda^{( E, \sigma)}_n $
and
$
  \tr{  \Gamma \left[ \lc^2 K_{(\td E,\td \sigma)} \right]    }   
= \tr{  \Gamma \left[ \lc^2 K_{(E,\sigma)} \right]    }   
$.
Similarly,
$
 \tr{     \Gamma \left[ \lc^2 K_{(\td E',\td \sigma')} \right]}
=   \tr{     \Gamma \left[ \lc^2 K_{(E',\sigma')} \right]}
$.
Furthermore, 
$
 J^{ (\td E',\td \sigma')}_{( \td E, \td \sigma) }
=
 J^{ ( E', \sigma')}_{(  E,  \sigma) }
$
because the matrix elements in \eq{matrix_element}
are invariant under diffeomorphism, 
\bqa
 \int d \td x | \td E'| ~ f^{(\td E',\td \sigma')*}_m(\td x) f^{( \td E, \td \sigma)}_n(\td x)
 =
  \int dx | E'| ~ f^{( E', \sigma')*}_m(x) f^{( E,  \sigma)}_n(x).
\label{matrix_element2}
\eqa
Finally,
\bqa
\int 
&& D^{( \td E, \td \sigma)} \td \Phi ~
~ e^{
- \frac{1}{2} \int d^3 x    ~ tr ~  \td \Phi \left( 
 |\td E'| e^{-  \Gamma \left[ \lc^2 K_{(\td E',\td \sigma')} \right] }
+ |\td E| e^{-  \Gamma \left[ \lc^2 K_{(\td E,\td \sigma)} \right] }
\right) \td \Phi
 } \nn
&& = 
 \int 
D^{( E, \sigma)} \Phi ~
~ e^{
- \frac{1}{2} \int d^3 x    ~ tr ~  \Phi \left( 
 |E'| e^{-  \Gamma \left[ \lc^2 K_{(E',\sigma')} \right] }
+ |E| e^{-  \Gamma \left[ \lc^2 K_{(E,\sigma)} \right] }
\right) \Phi
 }
\eqa
because $D^{( \td E, \td \sigma)} \td \Phi = D^{( E, \sigma)} \Phi$
and $\int d^3 x    ~ tr ~  \Phi \left( 
 |\td E'| e^{-  \Gamma \left[ \lc^2 K_{(\td E',\td \sigma')} \right] }
+ |\td E| e^{-  \Gamma \left[ \lc^2 K_{(\td E,\td \sigma)} \right] }
\right) \Phi$ is invariant under diffeomorphism.
This proves that 
$\lb E', \sigma' \cb E, \sigma \rb 
= \lb \tilde E', \tilde \sigma' \cb \tilde E, \tilde \sigma \rb $.

% %%%%%%%%%%%%%%%%%%%ZERO PROJECTION
\section{
$\lb E', \sigma' \cb E, \sigma \rb = 0$ 
unless 
$|E'(x)| = |E(x)|$
%%at all $x$
}
\label{zero_proper}

We prove the title of this appendix.
The vanishing overlap between states with different local proper volumes
is the result of mismatch in the gaussian wavefunctions
for large momentum modes.
For modes with momenta much larger than curvature scales,
we can use the semi-classical approximation
to represent eigenmodes in terms of wavepackets.
To keep track of position of wavepackets,
we divide the spatial manifold into blocks.
Suppose we choose a coordinate system with
$0 \leq x_1,x_2,x_3 < \lc$
for a spatial manifold with $T^3$ topology.
The $j$-th block is denoted as 
$S_j = \{ x | 
x_{j}^\mu \leq x^\mu < x_j^\mu + l, ~~ \mbox{with}~~  \mu = 1,2,3 \}$,
where $x_j^\mu = l~ j^\mu$  with $j^\mu$ being integers.
$l \equiv \frac{\lc}{P}$ with an integer $P$ 
is chosen to be small enough that 
$g_{E,\mu \nu}(x), \sigma(x), g_{E',\mu \nu}(x), \sigma'(x)$
can be regarded to be constants in each block.
A wavepacket is labeled by a block index $j$
and quantized momentum $k$ allowed within the block,
\bqa
f^{(E,\sigma)}_{jk}(x) 
& = & 
\sqrt{ \frac{8}{l^3 |E(x_j)| } } 
\sin \Bigl( k_1 (x^1-x_j^1) \Bigr) 
\sin \Bigl( k_2 (x^2-x_j^2) \Bigr) 
\sin \Bigl( k_3 (x^3-x_j^3) \Bigr), ~~ \mbox{ if $x \in S_j$ } \nn
    & = & 0, ~~~~~~~~~~~~~~~~~~ \mbox{if $x \notin S$}.
\eqa 
Here momentum is quantized as $k_\mu = \frac{ \pi n_\mu}{l}$ with positive integer $n_\mu$.
The normalization is chosen such that
$\int dx |E| 
f^{(E,\sigma)*}_{j'k'}(x) 
f^{(E,\sigma)}_{jk}(x) 
= \delta_{j',j} \delta_{k',k}$.
Field configurations which vanish at the boundaries of the blocks
can be decomposed in terms of  the wavepackets.
%$f^{(\hat E,\hat \sigma)}_{jk}(x) $,
%$f^{(E,\sigma)}_{jk}(x) $
%or 
%$f^{(E',\sigma')}_{jk}(x) $.
%\bqa
%\Phi(x) 
%= 
%\sum_{j,k} 
%\Phi^{(\hat E,\hat \sigma)}_{k} 
%f^{(\hat E,\hat \sigma)}_{jk}(x) 
%= 
%\sum_{j,k} 
%\Phi^{(E,\sigma)}_{k} 
%f^{(E,\sigma)}_{jk}(x) 
%=\sum_{j,k} 
%\Phi^{(E',\sigma')}_{jk} 
%f^{(E',\sigma')}_{jk}(x).
%\label{mode_exp}
%\eqa
Because 
$f^{(E,\sigma)}_{jk}(x) = 
\sqrt{ \frac{|\hat E(x_j)|}{|E(x_j)|} } 
f^{(\hat E,\hat \sigma)}_{jk}(x) $
and 
$f^{(E',\sigma')}_{jk}(x) = 
\sqrt{ \frac{|\hat E(x_j)|}{|E'(x_j)|} } 
f^{(\hat E,\hat \sigma)}_{jk}(x) $,
we have
$\Phi^{(E,\sigma)}_{jk} 
= \sqrt{ \frac{|E(x_j)|}{|\hat E(x_j)|} } 
\Phi^{(\hat E,\hat \sigma)}_{jk} $ and
$\Phi^{(E',\sigma')}_{jk} 
= \sqrt{ \frac{|E'(x_j)|}{|\hat E(x_j)|} } 
\Phi^{(\hat E,\hat \sigma)}_{jk} $.
The wave packets contribute to the overlap in \eq{innerg0} as
\bqa
&& \lb E', \sigma' \cb E, \sigma \rb \propto
\left[
  \prod_{jk}
\frac{ |E(x_j)| |E'(x_j)| }{|  \hat E(x_j)|^2 } 
\right]^{\frac{1}{4}}
e^{ -\frac{N^2}{4}  ~ \left\{ 
  \tr{ \Gamma \left[ \lc^2 K_{(E',\sigma')} \right] }
+ \tr{  \Gamma \left[ \lc^2 K_{(E,\sigma)} \right] }
\right\} } 
 \times \nn
&&
\int \prod_{jk} d \Phi^{( \hat E,  \hat \sigma)}_{jk} ~
~ e^{
- \frac{1}{2} \sum_{jk} \left\{
e^{ - \Gamma \left[ \lc^2  \lambda^{(E',\sigma')}_{jk} \right] } 
\frac{|E'(x_j)|}{|\hat E(x_j)|}  
+ e^{ - \Gamma \left[ \lc^2  \lambda^{(E,\sigma)}_{jk} \right]  } 
\frac{|E(x_j)|}{|\hat E(x_j)|}  
\right\}
\left| \Phi^{( \hat E, \hat \sigma)}_{jk} \right|^2
},
\label{innerga1}
\eqa 
where
$ \lambda^{(E,\sigma)}_{jk}  = 
g_{E}^{\mu \nu}(x_j) k_\mu k_\nu + \frac{e^{2 \sigma(x_j)}}{\lc^2}$
and
$ \lambda^{(E',\sigma')}_{jk}  = 
g_{E'}^{\mu \nu}(x_j) k_\mu k_\nu + \frac{e^{2 \sigma'(x_j)}}{\lc^2}$.
Since
$ \Gamma \left[ \lc^2  \lambda^{(E,\sigma)}_{jk} \right] 
,  \Gamma \left[ \lc^2  \lambda^{(E',\sigma')}_{jk} \right] \rightarrow 0$
in the large $k$ limit, 
the contribution from large momentum modes becomes
\bqa
&& \lb E', \sigma' \cb E, \sigma \rb \propto
\left[
 \prod_j \prod_{k}'
\frac{ 4 |E(x_j)| |E'(x_j)| }{ \left[ |E(x_j)| + |E'(x_j)|  \right]^2 } 
\right]^{\frac{1}{4}}
e^{ -\frac{N^2}{4}  ~ \left\{ 
  \tr{ \Gamma \left[ \lc^2 K_{(E',\sigma')} \right] }
+ \tr{  \Gamma \left[ \lc^2 K_{(E,\sigma)} \right] }
\right\} },
\label{innerga2}
\eqa 
where $\prod_k'$ include momenta 
with $ g_{E}^{\mu \nu}(x_j) k_\mu k_\nu, g_{E'}^{\mu \nu}(x_j) k_\mu k_\nu \gg \lc^{-2}$.
It is noted that $\frac{ 4 |E(x_j)| |E'(x_j)| }{ \left[ |E(x_j)| + |E'(x_j)| \right]^2 } < 1$ 
unless  $ |E(x_j)|  = |E'(x_j)|$.
Since there are infinitely many momentum modes 
with $ g_{E}^{\mu \nu}(x_j) k_\mu k_\nu, g_{E'}^{\mu \nu}(x_j) k_\mu k_\nu \gg \lc^{-2}$
in each block,
$ \lb E', \sigma' \cb E, \sigma \rb $ vanishes 
if there is any block in which $ |E(x_j)|  \neq |E'(x_j)|$.
Here  we ignored the modes that describe fluctuations of $\Phi(x)$ 
at the boundaries between blocks.
Those modes do not change the conclusion
because they form fluctuations of measure zero.

\section{Overlap}
\label{sec:overlap}

In this appendix, we compute the overlap in \eq{innerg01}
between states associated with metrics close to the Euclidean metric.
Since the overlap is zero for $|E(x)| \neq |E'(x)|$,
we consider the case with $|E(x)| = |E'(x)|$.
In this case, the exponent in the integrand of \eq{innerg01} can be written as
\bqa
\int dx |E| tr \left[  \Phi \frac{ e^{-  \Gamma \left[ \lc^2 K_{(E',\sigma')} \right] }
+e^{-  \Gamma \left[ \lc^2 K_{(E,\sigma)} \right] } }{2} \Phi \right]
= 
\int dx |E| tr \left[ \Phi ~ e^{-  \Gamma \left[ \lc^2 K^{''}  \right] } \Phi \right]
+ O( \lc^2 h_{\mu \nu}^2, \lc^2 \delta \sigma^2 ), \nn
\eqa
where $K^{''} \equiv \frac{1}{2} \left[ K_{(E',\sigma')} + K_{(E,\sigma)} \right]$.
To the leading order in 
$h_{\mu \nu} \equiv g_{E',\mu \nu}-g_{E,\mu \nu}$, 
$\delta \sigma \equiv \sigma' - \sigma$ and $\lc$,
\eq{innerg01} becomes
\bqa
 \lb E', \sigma' \cb E, \sigma \rb & \approx &
e^{
-\frac{N^2}{4}  
~ \left\{ \tr{ \Gamma \left[ \lc^2 K_{(E',\sigma')} \right] }
+   \tr{ \Gamma \left[ \lc^2 K_{(E,\sigma)} \right] }
\right\} }  
%\times \nn &&
\int 
D^{( E, \sigma)} \Phi ~
~ e^{
- \int d^3 x  |E|  ~ tr ~  \left[ \Phi 
e^{-  \Gamma \left[ \lc^2 K^{''} \right] }
 \Phi \right]
 },
 \nn
\label{innerg_largeN2}
\eqa 
where we use 
$J^{ (E',\sigma')}_{(E,  \sigma) } =1$
for $|E'(x)| = |E(x)|$.
%%
%
%
%\section{The quadratic expansion of the overlap }
%\label{quadratic_expansion}
%
%In this appendix, we derive  \eq{overlap_general}.
The logarithm of \eq{innerg_largeN2} can be written as
\bqa
\ln \lb E', \sigma' \cb E, \sigma \rb \approx
-\frac{N^2}{4} \int_{\lc^2}^\infty \frac{dt}{t} 
 ~ \tr{ 
   e^{- K_{(E', \sigma')} t} 
+  e^{- K_{(E, \sigma)} t} 
-2 e^{- K^{''} t} 
}.
\label{overlap3}
\eqa
%to the leading order in $h_{\mu \nu}$, $\delta \sigma$ and $\lc$ in \eq{innerg01}.
The Kernel with perturbations in the collective variables is written as
$K^{''}  = K +  \delta K$ and
$K_{(E', \sigma')}= K + 2 \delta K$,
where $K \equiv K_{(E, \sigma)}$.
The exponential of the perturbed Kernel can be expressed as
\bqa
e^{- ( K + \delta K) t} 
&=& e^{-  K t} 
- \int_0^t d \tau ~ e^{-  K (t-\tau)} \delta K e^{-  K \tau}  \nn
&& + \int_0^t d \tau_1 \int_0^{\tau_1} d \tau_2  ~
e^{-  K (t-\tau_1)} \delta K e^{-  K (\tau_1 - \tau_2)} \delta K e^{-  K \tau_2}
+ O(\delta K^3).
\label{expansion}
\eqa
The terms linear in $\delta K$ are all canceled in \eq{overlap3}
and one obtains
\bqa
\ln \lb E', \sigma' \cb E, \sigma \rb
& \approx &
-\frac{N^2}{2} \int_{\lc^2}^\infty \frac{dt}{t} 
 \int_0^t d \tau_1 \int_0^{\tau_1} d \tau_2  ~
\tr{ \delta K e^{-  K (\tau_1 - \tau_2)} \delta K e^{-  K (t-\tau_1+\tau_2) }} 
+ O(\delta K^3)
\nn
&= &
-\frac{N^2}{4} \int_{\lc^2}^\infty dt 
 \int_0^t d \tau  ~
\tr{ \delta K e^{-  K \tau} \delta K e^{-  K (t-\tau) }}
+ O(\delta K^3),
\label{overlap4}
\eqa
where the cyclic property of the trace has been used.
By inserting the complete set of basis of $K_{(E,\sigma)}$ in \eq{overlap4},
one obtains 
\bqa
\ln \lb E', \sigma' \cb E, \sigma \rb
&= &
-\frac{N^2}{4} 
\sum_{n,m}
\int_{\lc^2}^\infty dt 
 \int_0^t d \tau  ~
e^{ -(\lambda_m^{(E,\sigma)} - \lambda_n^{(E,\sigma)}) \tau - \lambda_n^{(E,\sigma)} t }
\left| \lb n \cb \delta K \cb m \rb \right|^2 \nn
&&  + O(\delta K^3),
\label{overlap5}
\eqa
where
$
 \lb n \cb \delta K \cb m \rb = \int dx ~|E| f^{(E,\sigma) *}_n(x) \delta K f^{(E,\sigma)}_m(x)
$.
The subsequent integrations over $\tau$ and $t$
gives 
\bqa
\ln \lb E', \sigma' \cb E, \sigma \rb 
\approx -\frac{N^2}{4} \sum_{n,m}
\frac{ | \lb n \cb \delta K \cb m \rb |^2  }{ \lambda_m^{(E,\sigma)} - \lambda_n^{(E,\sigma)}}
\left(
\frac{e^{-\lc^2 \lambda_n^{(E,\sigma)} }}{\lambda_n^{(E,\sigma)}}
-
\frac{e^{-\lc^2 \lambda_m^{(E,\sigma)} }}{\lambda_m^{(E,\sigma)}}
\right).
% + O( \delta K^3 ).
\label{overlap_general}
\eqa
%where 
%%%$\delta K \equiv K^{''} - K_{(E, \sigma) }$ and
%$
% \lb n \cb \delta K \cb m \rb = \int dx ~|E| f^{(E,\sigma) *}_n(x) \delta K f^{(E,\sigma)}_m(x)
%$.
%%See Appendix \ref{quadratic_expansion} for the derivation of \eq{overlap_general}.

Now, we compute \eq{overlap_general} between 
a state with the Euclidean metric 
and a state with small perturbations.  
%%in Eqs. (\ref{Euc_ovr}) - (\ref{Gammas}).
Suppose the space manifold has $T^3$ topology,
and $\emi$ describes a flat Euclidean metric, 
\bqa
\gemn = a^2 \delta_{\mu \nu}
\label{EM}
\eqa
in a coordinate system with $0 \leq x_1,x_2,x_3 < \lc$,
where $a$ is a scale factor that determines the proper size of the manifold
in the unit of the short-distance cut-off.
Let $\emi$ be the triad for the flat Euclidean metric, $\gemn = a^2 \delta_{\mu \nu}$,
and $\sigma$ is a constant.
The eigenvectors of $K_{(E,\sigma)}$ are the plane waves,
$f_p(x) = \frac{1}{(a\lc)^{3/2}} e^{i p x}$
with discrete momentum,
$p_\mu =\frac{2\pi}{\lc} n_\mu$ with integer $n_\mu$
and eigenvalue,
$\lambda_p = p^2 + \frac{e^{2 \sigma}}{\lc^2}$.
States with perturbed collective variables
are parameterized by the deformed metric
and scalar field,
$g_{E',\mu \nu}(x) = g_{E,\mu \nu} + h_{\mu \nu}(x)$,
$\sigma'(x)  = \sigma  + \delta \sigma(x)$,
where $h_{\mu \nu}(x)$ and $\delta \sigma(x)$ are small perturbations.

In the presence of the perturbation in the collective variables,
the kernel is modified to be
$K_{(E', \sigma')}= K + 2 \delta K$
with
\bqa
\delta K = \frac{1}{2} \left[
\frac{1}{\de} \partial_\mu \left( \de h^{\mu \nu} \partial_\nu - \de \frac{h}{2} g_E^{\mu \nu} \partial_\nu \right)
+ \frac{e^{2 \sigma}}{\lc^2} \left( \frac{h}{2} + 2 \delta \sigma \right)
\right].
\label{deltaK}
\eqa
From \eq{overlap_general},
we write the overlap between the state with the Euclidean metric and 
a state with the perturbation as 
\bqa
\ln \lb E', \sigma' \cb E, \sigma \rb
= -\frac{N^2
e^{-e^{2 \sigma} }
}{4} 
%%\frac{e^{-e^{2 \sigma} }}{(a \lc)^3} 
\sum_{p,q}
\frac{ 
\left| \lb q + p \cb \delta K \cb q \rb \right|^2
  }{ (p+q)^2 - q^2 }
\left(
\frac{e^{-\lc^2 q^2 }}{ q^2 + \lc^{-2} e^{2 \sigma} }
-
\frac{e^{-\lc^2 (q+p)^2 }}{ (q+p)^2 + \lc^{-2} e^{2 \sigma} }
\right)
\label{over3}
\eqa
to the quadratic order in $\delta K$,
where
\bqa
\lb q + p \cb \delta K \cb q \rb  =
\frac{1}{2  (a \lc)^{3/2}}  \left\{
\left[ -q^\mu ( q^\nu + p^\nu ) + \frac{1}{2}( q(q+p) + \lc^{-2} e^{2\sigma} ) g_E^{\mu \nu} \right] h_{\mu \nu}(p)
+ 2 e^{2\sigma} \lc^{-2} 
\delta \sigma(p) 
\right\}. \nn
\eqa
In the large $a$ limit,
the summation over the momenta 
can be done through integration,
and we obtain 
%%\eq{Gammas0} with \eq{Gammas}.
%%
%%
%%
%%
%%The overlap between the state with the flat metric in \eq{EM} and a constant $\sigma$,
%%and a state with 
%$g_{E',\mu \nu} = g_{E,\mu \nu}+h_{\mu \nu}$,
%$\sigma' = \sigma + \delta \sigma$
%is given by
\bqa
\ln \lb E', \sigma' \cb E, \sigma \rb 
& \approx & 
-\frac{N^2}{4^2 \lc^3} e^{- e^{2 \sigma}  } \sum_p
\Bigl[
I^{\mu \nu \alpha \beta}(p) 
h_{\mu \nu}(p) h_{\alpha \beta}(-p) \nn
&&
+ 2 I^{\mu \nu }(p) h_{\mu \nu}(p) \delta \sigma(-p)
+ I(p) \delta \sigma(p) \delta \sigma(-p)
\Bigr]
\label{Euc_ovr}
\eqa
to the leading order in $\lc^{-1}$, $h_{\mu \nu}$ and $\delta \sigma$.
Here 
$h_{\mu \nu}(p) = \int dx |E| f_p^*(x) h_{\mu \nu}(x)$ and
$\delta \sigma(p) = \int dx |E| f_p^*(x) \delta \sigma(x)$
with $f_p(x) = \frac{1}{(a \lc)^{3/2}} e^{i px}$.
%%The momentum is quantized as 
%%$p_\mu = \frac{2 \pi}{\lc} n_\mu$ with integer $n_\mu$.
In the large $a$ limit, 
$I^{\mu \nu \alpha \beta}(p)$,
$I^{\mu \nu }(p)$,
$I(p)$
are given by
\bqa
I^{\mu \nu \alpha \beta}(p) 
&=&
I^{4,1}(p)
~ g_E^{\mu \nu} g_E^{\alpha \beta}
+
I^{4,2}(p) 
~\left( 
  g_E^{\mu \alpha} g_E^{\nu \beta}
+ g_E^{\mu \beta} g_E^{\nu \alpha}
\right) 
+
I^{4,3}(p)
~ \left( 
 \frac{ p^\mu p^\nu}{p^2} g_E^{\alpha \beta} 
+\frac{  p^\alpha p^\beta}{p^2} g_E^{\mu \nu} 
\right)
\nn
&&
+
I^{4,4}(p)
~ \left( 
  \frac{p^\nu p^\beta}{p^2} g_E^{\alpha \mu} 
+ \frac{ p^\mu p^\beta}{p^2} g_E^{\alpha \nu} 
+ \frac{ p^\nu p^\alpha}{p^2} g_E^{\mu \beta} 
+ \frac{ p^\mu p^\alpha}{p^2} g_E^{\beta \nu} 
\right)
+I^{4,5}(p) \frac{p^\mu p^\nu p^\alpha p^\beta}{p^4}, \nn
I^{\mu \nu}(p) &=& 
I^{2,1}(p)
~ g_E^{\mu \nu} + I^{2,2}(p) \frac{p^\mu p^\nu}{p^2}, \nn
I(p) &=& 
I^{0,1}(p),
\label{Gammas0}
\eqa
where
\bqa
I^{4,1}(p) & = &
-\frac{\pi ^{3/2} \left(8 \sqrt{\pi } e^{3 \sigma} C_\sigma -14 e^{2 \sigma}+1\right)}{24} 
-\frac{\pi ^{3/2} \left(8 \sqrt{\pi } e^{\sigma} C_\sigma +2 e^{2 \sigma}-1\right)}{48} ( \lc p)^2 +O\left( ( \lc p)^4 \right), \nn 
I^{4,2}(p) & = &
\frac{\pi ^{3/2} \left(2 \sqrt{\pi } e^{3 \sigma} C_\sigma - 2 e^{2 \sigma}  + 1 \right)}{6}
+\frac{\pi ^{3/2} \left(\sqrt{\pi } e^{\sigma} C_\sigma -1\right)}{12} ( \lc p)^2 +O\left( (\lc p)^4 \right), \nn
I^{4,3}(p) & = & -\frac{\pi ^{3/2}-2 \pi ^2 e^{\sigma} C_\sigma }{12} ( \lc p)^2 +O\left( (\lc p)^4 \right), \nn
I^{4,4}(p) & = & \frac{\pi^{3/2}-\pi ^2 e^{\sigma} C_\sigma }{12} ( \lc p)^2 +O\left( (\lc p)^4 \right), \nn
I^{4,5}(p) & = & O\left( (\lc p)^4 \right), \nn
I^{2,1}(p) & = & 
\pi^{3/2} e^{2\sigma} 
-\frac{1}{6} \pi^{3/2}  e^{\sigma} \left(2 \sqrt{\pi } C_\sigma + e^{\sigma} \right)  (\lc p)^2
+O\left( (\lc p)^4 \right), \nn
I^{2,2}(p) & = & \frac{1}{3} \pi^2 e^{\sigma} C_\sigma ( \lc p)^2  +O\left( (\lc p)^4 \right), \nn
I^{0,1}(p) & = & 
4 \pi^{2}  e^{3\sigma} C_\sigma 
- \frac{1}{3} \pi^{3/2} e^{\sigma} 
\left( \sqrt{\pi } C_\sigma +2  e^{\sigma} \right) ( \lc p)^2
+O\left( (\lc p)^4 \right).
\label{Gammas}
\eqa
Here $ C_\sigma   \equiv    e^{e^{ 2 \sigma}   } \text{erfc}( e^\sigma )$,
where $\text{erfc}(x)$ is the complimentary error function. 
It has the asymptotic behavior,
$\lim_{\sigma \rightarrow -\infty} C_\sigma = 1$,
$\lim_{\sigma \rightarrow \infty} C_\sigma = \frac{1}{\sqrt{\pi}} e^{-\sigma}$.
%%See Appendix \ref{computation_overlap} for details.
%%
For $|E'(x)| = |E(x)|$, $g_E^{\mu \nu} h_{\mu \nu}=0$ to the leading order in $h_{\mu \nu}$,
and only $I^{4,2}$ and $I^{0,1}$ are important 
in \eq{Gammas0} in the small momentum limit.
The overlap decays exponentially in $h_{\mu \nu}$ and $\delta \sigma$
with the width which is controlled by the eigenvalues of the matrix,
\bqa
{\cal M}(p) =
\frac{N^2}{4^2 \lc^3} e^{- e^{2 \sigma}  } 
\left(
\begin{array}{cc}
I^{ \mu \nu \alpha \beta}(p) &   I^{\mu \nu }(p) \\
 I^{\alpha \beta }(p)  	& I(p)
\end{array}
\right).
\label{Mp}
\eqa
%%For $a \gg 1$ and $e^{\sigma} \ll 1$,
It is noted that two states with 
$|h_{\mu}^{\nu}(p)|,  ~e^{3/2 \sigma} |\delta \sigma(p)| >  \frac{\lc^{3/2}}{N}$
for any wavevector $p \ll \lc^{-1}$ are almost orthogonal.
In real space, this implies that two states whose collective variables differ by 
$
h_{\mu}^{\nu}(x) \sim \frac{1}{N}$, 
$|\delta \sigma(x)| \sim \frac{e^{-3/2 \sigma} }{N}$
over a proper volume $\lc^3$ 
are nearly orthogonal. 
This follows from the fact that
if $h_{\mu }^{\nu}(x) \sim \frac{1}{N}$ over a proper volume $\lc^3$,
$h_{\mu}^{\nu}(p) \sim \frac{ \lc^3 }{N \sqrt{(a \lc)^{3/2}} }$
for $p^2 < \lc^{-2}$.
Therefore, 
$\ln \lb E', \sigma' \cb E, \sigma \rb 
\sim
-\frac{N^2}{\lc^3} \sum_{p^2 < \lc^{-2} } |h_{\mu}^{\nu}(p) |^2
\sim -1$.
In the large $N$ limit,
the overlap becomes proportional to the delta function,
\bqa
\lim_{N \rightarrow \infty}
\lb E', \sigma' \cb E, \sigma \rb 
%& = &  e^{ \cJ }
& \propto &
\prod_{p} \left[
%\left[ \det {\cal M}(p) \right]^{-1/2} ~
\delta \Bigl( \delta \sigma(p) \Bigr)
\prod_{(\mu,\nu)}  
\delta \Bigl( h_{\mu \nu}(p) \Bigr) 
\right].
\label{innerg1}
\eqa

\section{Decomposition of the entanglement entropy for semi-classical states}
\label{EE_decomposition}

In this appendix, we derive Eqs. (\ref{eq:decom}) - (\ref{Zn22})
from \eq{eq:SA}.
In manipulating \eq{eq:SA},
it is convenient to rewrite \eq{eq:psi} as
\bqa
  \Psi(\Phi; E, \sigma) = 
    e^{
- tr \int dx dy ~ 
\Phi(x) 
t_{E,\sigma}(x,y) 
\Phi(y)
-  \frac{1}{2} S_0[E,\sigma]
%     S[ \Phi; E, \sigma] 
},
\label{eq:psi_app}
\eqa 
where
$t(x,y)$ can be written as
\bqa
t(x,y) = 
\frac{1}{2} \int dz |E(z)| 
\left[
 \frac{e^{2\sigma}}{\lc^2} \delta(z-x) \delta(z-y)
 +
g_E^{\mu \nu}(z) 
\partial_\mu \delta(z-x)
\partial_\nu \delta(z-y)
+ O(\lc^2 \nabla^4)
\right]
\label{texp}
\eqa
in the small $\lc$ limit
with fixed $\frac{e^{2\sigma}}{\lc^2} $.
In \eq{eq:psi_app},
the derivative terms in $\Psi(\Phi;E,\sigma)$ are
represented as bi-local couplings.
%%where $t_{E,\sigma}(x,y)$ decays exponentially
%%in the distance between $x$ and $y$.
%over the proper length scale $\lc$.
%%
%%
%%
%%
%%
%%
%%
%%
In terms of the bi-local representation of the wavefunction,
$Z_n$ in \eq{eq:SA} is written as
\bqa
Z_n & = &  \int \prod_{j=1}^n \left[ D^{(\hat E, \hat \sigma)} \Phi^j DE_1^j D\sigma_1^j DE_2^j D\sigma_2^j    
 \right]  \nn
&&
 e^{
- \sum_j 
tr \int_{x,y \in \bar A} ~ dx dy ~
\Phi^j(x)
 \Bigl(
  t_{E_2^j,\sigma_2^j}(x,y)
+ t_{E_1^j,\sigma_1^j}(x,y)
 \Bigr)   
\Phi^j(y)
} \times \nn
&& e^{
- \sum_j 
tr \int_{x,y \in  A}~  dx dy ~
\Phi^j(x)
 \Bigl(
t_{ E_2^{j+1},\sigma_2^{j+1}}(x,y)
+ t_{E_1^j,\sigma_1^j}(x,y)
 \Bigr)  
\Phi^j(y)
 } \times \nn
&&
 e^{
- \sum_j 
tr \int_{x \in \bar A, y \in A}~  dx dy ~
\Phi^j(x)
 \Bigl(
   t_{E_1^j,\sigma_1^j}(x,y)
+  t_{E_1^j,\sigma_1^j}(y,x)
 \Bigr)   
\Phi^j(y)
} \times \nn
&& e^{
- \sum_j 
tr \int_{x \in \bar A, y \in A}~  dx dy ~
\Phi^j(x)
 \Bigl(
  t_{ E_2^{j},\sigma_2^{j}}(x,y)
+ t_{ E_2^{j},\sigma_2^{j}}(y,x)
 \Bigr)  
\Phi^{j-1}(y)
 } \times \nn
&&
\Bigl(
\prod_{j=1}^n
\tilde \chi^*(E_1^j,\sigma_1^j)
\tilde \chi(E_2^j,\sigma_2^j)
 \Bigr).
\label{Zn0}
\eqa
Here
$\tilde \chi(E,\sigma) = e^{- \frac{1}{2}S_0[E,\sigma]} ~ \chi(E,\sigma)$.
The variables with replica indices outside the range of $[1,n]$ 
are identified cyclically with period $n$ :
$E^{n+j}_{\mu i}=E^j_{\mu i}$, $\sigma^{n+j}=\sigma^j$.
Each term in \eq{Zn0} has obvious meaning.
The second and third lines represent
the bi-local couplings within region $\bar A$ and $A$, respectively.
The fourth and fifth lines represent
the bi-local couplings between $\bar A$ and $A$ 
in the bra and ket of the wavefunction, respectively.
The last line is the contribution from the collective variables.
It is the last three terms that generate non-trivial entanglement.

Now we insert the following expression for the identity 
inside the integration of \eq{Zn0},
\bqa
1 &  = & 
\prod_{j=1}^n \Bigg\{
\tilde \mu^{-1} ( E_1^{[j]}, \sigma_1^{[j]})
~ e^{S_0[E_1^{[j]},\sigma_1^{[j]}] } ~
\prod_{x} 
\left[
\delta \Bigl(  \sigma^j_2(x) - \sigma^{[j]}_1(x) \Bigr)
\prod_{(\mu,\nu)}  
\delta \Bigl( g_{ E_2^j,\mu\nu}(x) - g_{E_1^{[j]},\mu\nu}(x) \Bigr) 
\right] 
\times \nn
&&
\left[
\int
% \prod_{j=1}^n  
D^{(\hat E, \hat \sigma)} \Phi^j ~
 e^{
- \sum_j 
tr \int ~ dx dy ~
\Phi^j(x)
% \Bigl(
 t_{E_1^{[j]},\sigma_1^{[j]}}(x,y)
% \Bigr)   
\Phi^j(y)
- \sum_j 
tr \int ~ dx dy ~
\Phi^j(x)
% \Bigl(
 t_{ E_2^j, \sigma_2^j}(x,y)
% \Bigr)   
\Phi^j(y)
} \right]^{-1}
\Bigg\},
\label{eq1}
\eqa
where $x,y$ run over the entire space, 
and
\bqa
\{ E^{[j]}(x) ,\sigma^{[j]}(x) \} 
& = & \{ E^j(x) , \sigma^j(x) \} ~~~~~~~~~~~ \mbox{for $x \in \bar A$}, \nn
& = & \{ E^{j-1}(x) , \sigma^{j-1}(x) \} ~~~~ \mbox{for $x \in  A$}.
\eqa
The functional integration of $\Phi$ in the last line is unconstrained,
and \eq{innerg} has been used.
The integrations over $E_2^j, \sigma_2^j$ result in 
\bqa
&&
 Z_n =   \int \prod_{j=1}^n \left[  DE^j D\sigma^j   \right]  
\prod_{j=1}^n
 \Biggl\{
 e^{S_0[E^{[j]},\sigma^{[j]}] } ~
\tilde \chi^* \Bigl( E^{j}(x_{\bar A}),\sigma^j(x_{\bar A}); E^{j}(x_{A}), \sigma^{j}(x_A) \Bigr) \times \nn
&&
\left.
\tilde \chi \Bigl( E^{j}(x_{\bar A}),\sigma^j(x_{\bar A}); E^{j-1}(x_A), \sigma^{j-1}(x_A) \Bigr)
 \right|_{
% \begin{array}{l}
{\scriptstyle \sigma^{2,..,n}(x) = \sigma^1(x)} ,   
{\scriptstyle g_{E^{2,..,n},\mu\nu}(x) = g_{E^1,\mu \nu}(x)}, 
{\scriptstyle  x \in \partial A }
%\end{array}
 }
 \Biggr\} \times \nn
&&
\int \prod_{j=1}^n  D^{(\hat E, \hat \sigma)} \Phi^j ~
\Biggl[
 e^{
- 2 \sum_j 
tr \int_{x,y \in \bar A} ~ dx dy ~
\Phi^j(x)
  t_{E^j,\sigma^j}(x,y)
\Phi^j(y)} \times \nn
&&
~~~~~~~~~~
~~~~~~~~~~~~~~~~~~~~
e^{
- 2 \sum_j 
tr \int_{x,y \in  A}~  dx dy ~
\Phi^j(x)
 t_{E^j,\sigma^j}(x,y)
\Phi^j(y)
 } 
\times \nn
&&
~~~~~~~~~~~~~~~~~~~~~~~~~~~~~~
 e^{
- \sum_j 
tr \int_{x \in \bar A, y \in A}~  dx dy ~
\Phi^j(x)
 \Bigl(
   t_{E^j,\sigma^j}(x,y)
+  t_{E^j,\sigma^j}(y,x)
 \Bigr)   
\Phi^j(y)
} \times \nn
&& 
~~~~~~~~~~~~~~~~~~~~~~~~~~~~~~
e^{
- \sum_j 
tr \int_{x \in \bar A, y \in A}~  dx dy ~
\Phi^j(x)
 \Bigl(
  t_{ E^{[j]},\sigma^{[j]}}(x,y)
+ t_{ E^{[j]},\sigma^{[j]}}(y,x)
 \Bigr)  
\Phi^{j-1}(y)
 } 
\Biggr] \times \nn
&&
\left[
\int \prod_{j=1}^n D^{(\hat E, \hat \sigma)} \Phi^j 
 e^{
- 2 \sum_j 
tr \int_{x,y} ~ dx dy ~
\Phi^j(x)
  t_{E^{[j]},\sigma^{[j]}}(x,y)
\Phi^j(y)
 } 
\right]^{-1}.
\label{Zn}
\eqa
It is noted that the delta functions in \eq{eq1}
twists the boundary condition for the collective variables in \eq{Zn},
and force $\sigma^j(x)$ and $g_{E^j,\mu \nu}$ to be independent of $j$
in $\partial A$.

If $\chi(E,\sigma)$ is sharply peaked at a classical configuration, 
$\bar E(x), \bar \sigma(x)$
(and its SO(3) gauge orbits),
\eq{Zn} can be approximately factorized 
into the contribution from the matter fields 
and the contribution from the collective variables as
\bqa
Z_n \approx 
Z_n^\Phi( \bar E, \bar \sigma) 
Z_n^{E, \sigma},
\eqa
where
\bqa
 Z_n^\Phi( \bar E, \bar \sigma) 
&=& 
\int \prod_{j=1}^n  D^{(\bar E, \bar \sigma)} \Phi^j ~
\Biggl[
 e^{
- 2 \sum_j 
tr \int_{x,y \in \bar A} ~ dx dy ~
\Phi^j(x)
  t_{\bar E, \bar \sigma}(x,y)
\Phi^j(y)} \times \nn
&&
~~~~~~~~~~
e^{
- 2 \sum_j 
tr \int_{x,y \in  A}~  dx dy ~
\Phi^j(x)
 t_{\bar E,\bar \sigma}(x,y)
\Phi^j(y)
 } 
\times \nn
&&
~~~~~~~~~~
 e^{
- \sum_j 
tr \int_{x \in \bar A, y \in A}~  dx dy ~
\Phi^j(x)
 \Bigl(
   t_{\bar E,\bar \sigma}(x,y)
+  t_{\bar E,\bar \sigma}(y,x)
 \Bigr)   
\Phi^j(y)
} \times \nn
&& 
~~~~~~~~~~
e^{
- \sum_j 
tr \int_{x \in \bar A, y \in A}~  dx dy ~
\Phi^j(x)
 \Bigl(
  t_{ \bar E,\bar \sigma}(x,y)
+ t_{ \bar E,\bar \sigma}(y,x)
 \Bigr)  
\Phi^{j-1}(y)
 } 
\Biggr] \times \nn
&&
~~~~~~~~~~
\left[
\int \prod_{j=1}^n D^{(\bar E, \bar \sigma)} \Phi^j 
 e^{
- 2 \sum_j 
tr \int_{x,y} ~ dx dy ~
\Phi^j(x)
  t_{\bar E,\bar \sigma}(x,y)
\Phi^j(y)
 } 
\right]^{-1}, \nn
\label{Zn21_ap} \\
 Z_n^{E, \sigma}
&=& 
\int 
\prod_{j=1}^n 
DE^j D\sigma^j  ~
\Bigg\{
~ e^{S_0[E^{[j]},\sigma^{[j]}] } ~
\tilde \chi^* \Bigl( E^{j}(x_{\bar A}),\sigma^j(x_{\bar A}); E^{j}(x_{A}), \sigma^{j}(x_A) \Bigr) \times \nn
&& 
\left.
\tilde \chi \Bigl( E^{j}(x_{\bar A}),\sigma^j(x_{\bar A}); E^{j-1}(x_A), \sigma^{j-1}(x_A) \Bigr)
 \right|_{
 \begin{array}{l}
{\scriptstyle \sigma^{2,..,n}(x_{\partial A}) = \sigma^1(x_{\partial A})} ,   \\
{\scriptstyle g_{E^{2,..,n},\mu\nu}(x_{\partial A}) = g_{E^1,\mu \nu}(x_{\partial A})}
\end{array}
 }
\Bigg\}
.
 \label{Zn22_ap}
\eqa
As a result, the entanglement entropy is given by the sum of 
two contributions,
\bqa
S(A) \approx S_\Phi(A) + S_{E, \sigma}(A),
\eqa
where 
$S_\Phi(A) = 
 -\lim_{n \rightarrow 1} \frac{1}{n-1}
 \left(
 Z_n^\Phi
 -1
  \right)$
  and 
  $S_{E, \sigma}(A) = 
 -\lim_{n \rightarrow 1} \frac{1}{n-1}
 \left(
 Z_n^{E, \sigma}
 -1
  \right)$.

$S_\Phi(A)$ is the entanglement  
supported by the matter field
defined in the classical geometry set by the collective variables, 
$\{ \bar E_{\mu i}, \bar \sigma \}$.
The third and the fourth lines in \eq{Zn21_ap}
describe the couplings between $A$ and $\bar A$.
It penalizes configurations of the matter fields 
which have discontinuities across the boundary.
In the small $\lc$ limit, 
the cross couplings between $A$ and $\bar A$ are localized near the boundary, 
and it can be replaced with a function that depends
on $ \Phi^{j}(x) - \Phi^{j-1}(x)$ at the boundary,
\bqa
&& e^{
- \sum_j 
tr \int_{x \in \bar A, y \in A}~  dx dy ~
\left[
\Phi^j(x)
 \Bigl(
   t_{\bar E,\bar \sigma}(x,y)
+  t_{\bar E,\bar \sigma}(y,x)
 \Bigr)   
\Phi^j(y)
+
\Phi^j(x)
 \Bigl(
  t_{ \bar E,\bar \sigma}(x,y)
+ t_{ \bar E,\bar \sigma}(y,x)
 \Bigr)  
\Phi^{j-1}(y)
\right]
 }  \nn
%% &= &  C(\bar E, \bar \sigma) 
 & \approx &
 \prod_{r=2}^N 
  \prod_{x \in \partial A} F \Bigl( \Phi^{r}(x) - \Phi^{r-1}(x); \bar E(x)    \Bigr). 
 \label{eq:df}
\eqa
It is noted that
$F \Bigl( \Phi^{r}(x) - \Phi^{r-1}(x); \bar E(x) \Bigr) $
does not depend on $\bar \sigma$ 
in the small $\lc$ limit
because the term associated with $\sigma$ 
in \eq{texp}
is ultra-local.
The integration measure in
the numerator of \eq{Zn21_ap} 
can be decomposed into
modes with a Dirichlet boundary condition
and  modes localized at the boundary as
\bqa
\int \prod_{j=1}^n D^{(\bar E, \bar \sigma)} \Phi^j 
=
\left. 
\int \prod_{j=1}^n D^{(\bar E, \bar \sigma)} \Phi^j 
\right|_{ \Phi^{2,..,n}(x_{\partial A}) = \Phi^1(x_{\partial A}) } 
\times
\prod_{r=2}^n \prod_{x \in \partial A} D^{(\bar E, \bar \sigma)} \Phi^r(x).
\label{decompose}
\eqa
%%The first term on the right hand side of \eq{decompose} denotes
Here
$\left. 
\prod_{j=1}^n D^{(\bar E, \bar \sigma)} \Phi^j 
\right|_{ \Phi^{2,..,n}(x_{\partial A}) = \Phi^1(x_{\partial A}) } 
\equiv 
\prod_{j=1}^n 
\prod_m
\prod_{a,b}
d \Phi_{ab,m}^{0(\bar E, \bar \sigma) j} 
$
with $ \Phi_{ab,m}^{0 (\bar E, \bar \sigma) j} $ representing
the $m$-th normal mode which satisfies the boundary condition,
$ \Phi^{2,..,n}(x_{\partial A}) = \Phi^1(x_{\partial A})$.
\eq{decompose} should be viewed as the defining expression for 
$\prod_{r=2}^n \prod_{x \in \partial A} D^{(\bar E, \bar \sigma)} \Phi^r(x)$.
Since
$\prod_r \prod_{x \in \partial A} F \Bigl( \Phi^{r}(x) - \Phi^{r-1}(x); \bar E(x)    \Bigr) $
is sharply peaked at $\Phi^{2,..,n}(x_{\partial A}) = \Phi^1(x_{\partial A})$
in the small $\lc$ limit,
the integrations in \eq{Zn21_ap} can be factorized as
\bqa
 Z_n^\Phi( \bar E, \bar \sigma) 
& \approx & 
C(\bar E, \bar \sigma) 
\int \prod_{j=1}^n  D^{(\bar E, \bar \sigma)} \Phi^j ~
\Biggl[
 e^{
- 2 \sum_j 
tr \int_{x,y \in \bar A} ~ dx dy ~
\Phi^j(x)
  t_{\bar E, \bar \sigma}(x,y)
\Phi^j(y)} \times \nn
&&
~~~~~~~~~~
e^{
- 2 \sum_j 
tr \int_{x,y \in  A}~  dx dy ~
\Phi^j(x)
 t_{\bar E,\bar \sigma}(x,y)
\Phi^j(y)
 } 
\Biggr]_{
 \Phi^{2,..,n}(x_{\partial A}) = \Phi^1(x_{\partial A}) }
\times \nn
&&
\left[
\int \prod_{j=1}^n D^{(\bar E, \bar \sigma)} \Phi^j 
 e^{
- 2 \sum_j 
tr \int_{x,y} ~ dx dy ~
\Phi^j(x)
  t_{\bar E,\bar \sigma}(x,y)
\Phi^j(y)
 } 
\right]^{-1},
\label{Zn23}
\eqa
where 
\bqa
 C(\bar E, \bar \sigma)  \equiv
\int 
\prod_{x \in \partial A} 
\left[
\prod_{r=2}^n D^{(\bar E, \bar \sigma)} \Phi^r(x) ~
\prod_{r=2}^n F \Bigl( \Phi^{r}(x) - \Phi^{r-1}(x); \bar E(x) \Bigr) 
\right].
\eqa
In order to determine $ C(\bar E, \bar \sigma)$,
we first note that
$ Z_n^\Phi( \bar E, \bar \sigma) = C(\bar E, \bar \sigma)$ 
in the $\sigma \rightarrow \infty$ limit.
This follows from the fact that 
$ 
\lim_{\sigma \rightarrow \infty}
\int \prod_{j=1}^n 
D^{(\bar E, \bar \sigma)} \Phi^j ~
\left.
e^{ - 2   
\int dx  |\bar E|  {\cal L}[\Phi^j; \bar E, \bar \sigma] }
\right|_{ \Phi^{2,..,n}(x_{\partial A}) = \Phi^1(x_{\partial A}) } 
=
\lim_{\sigma \rightarrow \infty}
\int \prod_{j=1}^n D^{(\bar E, \bar \sigma)} \Phi^j ~
e^{ - 2   \int dx |\bar E|  {\cal L}[\Phi^j; \bar E, \bar \sigma] }   
=1$.
Because  $S_\Phi(A)$ should vanish in the large $\sigma$ limit,
we have $ \lim_{n \rightarrow 1} \frac{ C(\bar E, \infty) -1}{n-1} = 0$.
For a finite $\bar \sigma$, we have
$ C(\bar E, \bar \sigma)  
= \left[ 
\tilde J^{ (\bar E,\bar \sigma)}_{( \bar E,  \infty) } 
\right]^{n-1} 
C(\bar E, \infty )  $,
where
$\tilde J^{ (\bar E,\bar \sigma)}_{( \bar E,  \infty) } $
is the Jacobian for the change of basis at the boundary.
The Jacobian is finite and independent of $\lc$.
For example, for a flat metric with a constant $\bar \sigma$,
$J^{ (\bar E,\bar \sigma)}_{( \bar E,  \infty) } =1$
irrespective of the value of $\bar \sigma$.
As a result, 
$ C(\bar E, \bar \sigma)  $ 
does not give rise to
a singular contribution to the entanglement entropy
in the small $\lc$ limit.
Therefore, $S_\Phi(A)$ can be written as \eq{FF}
to the leading order in $\lc$\cite{PhysRevB.79.115421}.
On the other hand, $S_{E,\sigma}(A)$ is the entanglement
generated by fluctuations of the collective variables.

\section{Computation of the entanglement entropy}
\label{computation_EE}

Here we derive \eq{EE} 
for the state 
with the Euclidean metric in \eq{EM}
and a constant $\sigma$.
In the three torus,
we compute the entanglement entropy
for region $A$ 
which is parameterized by the coordinate,
$A = \{(
x_1,x_2,x_3) \cb 
0 \leq x_1 < l, 
0 \leq x_2 < l_c, 
0 \leq x_2 < l_c 
\}$.
In the presence of the Dirichlet boundary condition, 
$\Phi(0,x^2,x^3) = \Phi(l,x^2,x^3) = 0$,
the eigenmodes in region $A$ are modified such that
\bqa
f_{k} = \sqrt{ \frac{2}{a^3 \lc^2 l} } \sin (k_1 x^1) e^{i (k_2 x^2 + k_3 x^3)},
\eqa
where $(k_1,k_2,k_3) = \left( \frac{n_1 \pi}{l}, \frac{2 n_2 \pi}{\lc}, \frac{2 n_3 \pi}{\lc} \right)$
with eigenvalue $\lambda_k = k^2 + \lc^{-2} e^{2 \sigma}$.
Here $n_1$ represents positive integers while $n_2, n_3$ are general integers.
The free energy from region $A$ is given by
\bqa
F_A = -\frac{N^2}{2} \int_{\lc^2}^\infty \frac{dt}{t} ~
e^{- e^{2\sigma} \lc^{-2} t}
\left[ \sum_{n_1=1}^\infty e^{ -\left( \frac{ \pi n_1}{a l} \right)^2 t } \right]
\left[ \sum_{n=-\infty}^\infty e^{ -\left( \frac{ 2 \pi n}{a \lc} \right)^2 t } \right]^2.
\eqa
Similarly, the free energy from the complement of $A$ is 
\bqa
F_{\bar A} = -\frac{N^2}{2} \int_{\lc^2}^\infty \frac{dt}{t} ~
e^{- e^{2\sigma} \lc^{-2} t}
\left[ \sum_{n_1=1}^\infty e^{ -\left( \frac{ \pi n_1}{a (\lc-l)} \right)^2 t } \right]
\left[ \sum_{n=-\infty}^\infty e^{ -\left( \frac{ 2 \pi n}{a \lc} \right)^2 t } \right]^2.
\eqa
Subtracting the free energy without the Dirichlet boundary condition,
\bqa
F = -\frac{N^2}{2} \int_{\lc^2}^\infty \frac{dt}{t} ~
e^{- e^{2\sigma} \lc^{-2}  t}
\left[ \sum_{n=-\infty}^\infty e^{ -\left( \frac{ 2 \pi n}{a \lc} \right)^2 t } \right]^3,
\eqa
one obtains
\bqa
F_A + F_{\bar A} - F = 
\frac{N^2}{2} \int_{\lc^2}^\infty \frac{dt}{t} 
~ e^{- e^{2\sigma} \lc^{-2} t}
\frac{ a^2 \lc^2}{ 4 \pi t}
%\left[ 
%1 + O( e^{ }  )
%\right].
\eqa
in the limit that $a \gg 1$. 
Integration over $t$ gives \eq{EE}
in the $e^{\sigma} \rightarrow 0$ limit.

\section{Proof of \eq{mCR}}
\label{diffeomorphism}

We consider the unitary operator
that generates diffeomorphism for the matter field,
\bqa
\hat T(n) = e^{     
- i \int dx ~ n^\mu(x) \hat \cH_\mu(x)
},
\eqa
where $n^\mu(x)$ is an infinitesimal translation.
$\hat T(n)$ shifts $\Phi(x)$ by $n^\mu(x)$ as
\bqa
\hat T(n) \cb \Phi \rb = A(n)  \cb \tilde \Phi \rb,
\eqa
where $\tilde \Phi( \tilde x) = \Phi(x)$
with $\tilde x^\mu = x^\mu + n^\mu(x)$.
$A(n)$ is a constant which is determined from the normalization condition, 
\bqa
&& \delta^{(\hat E, \hat \sigma)}( \Phi' - \Phi ) 
 =   \lb \Phi' \cb \hat T(n)^{\dagger} \hat T(n) \cb \Phi \rb  \nn
&&  = | A(n) |^2 \lb \tilde \Phi' \cb \tilde \Phi \rb
= | A(n) |^2 \delta^{(\hat E, \hat \sigma)}( \tilde \Phi' - \tilde \Phi ),
\label{deldel} 
\eqa
where
$\delta^{(\hat E, \hat \sigma)}( \Phi' - \Phi )  = \prod_{a,b} \prod_n 
\delta \Bigl( \Phi^{' (\hat E, \hat \sigma)}_{ab,n} - \Phi^{(\hat E, \hat \sigma)}_{ab,n} \Bigr)$.
Let $\tilde {\hat \sigma}(  x) = \hat \sigma(x) - n^\mu \nabla_\mu \hat \sigma$ and 
$\tilde {\hat E}_{\mu i}(x) = \hat E_{\mu i}(x)  - \nabla_\mu n^\nu \hat E_{\nu i}$.
From $\int D^{(\hat E, \hat \sigma)}\Phi ~\delta^{(\hat E, \hat \sigma)}( \Phi' - \Phi )
= \int D^{(\tilde {\hat E}, \tilde {\hat \sigma})}\Phi ~\delta^{(\tilde {\hat E}, \tilde {\hat \sigma})}( \Phi' - \Phi )$
and 
$D^{(\tilde {\hat E}, \tilde {\hat \sigma} )}\Phi 
= J^{(\tilde {\hat E}, \tilde {\hat \sigma} )}_{(\hat E, \hat \sigma)}
D^{(\hat E, \hat \sigma)}\Phi $,
we have
%\bqa
$
\delta^{(\tilde {\hat E}, \tilde {\hat \sigma})}( \Phi' - \Phi )
=
J^{(\hat E, \hat \sigma)}_{(\tilde {\hat E}, \tilde {\hat \sigma} )}
\delta^{(\hat E, \hat \sigma)}( \Phi' - \Phi ).
%\label{deldel2}
%\eqa
$
On the other hand,
\bqa
\delta^{(\hat E, \hat \sigma)}( \Phi' - \Phi )
= 
\delta^{(\tilde {\hat E}, \tilde {\hat \sigma})}( \tilde \Phi' - \tilde \Phi )
=
J^{(\hat E, \hat \sigma)}_{(\tilde {\hat E}, \tilde {\hat \sigma} )}
\delta^{(\hat E, \hat \sigma)}( \tilde \Phi' - \tilde \Phi ).
\label{deldel3}
\eqa
\eq{deldel} and \eq{deldel3} leads to
\bqa
A(n) = 
\left[ J^{(\hat E, \hat \sigma)}_{(\tilde {\hat E}, \tilde {\hat \sigma} )} \right]^{1/2}.
\eqa
Now we examine how $\hat T(n)$ acts on $\cb E, \sigma \rb$ :
\bqa
\hat T(n) \cb E, \sigma \rb &  = & \int D^{ (\hat E, \hat \sigma) } \Phi ~   
\left[ J^{(\hat E, \hat \sigma)}_{(\tilde {\hat E}, \tilde {\hat \sigma} )} \right]^{1/2}    
\cb \tilde \Phi \rb 
\Psi( \Phi; E, \sigma) \nn
& = & \int D^{ (\tilde {\hat E}, \tilde {\hat \sigma} )  } \tilde \Phi ~ 
\left[ J^{(\hat E, \hat \sigma)}_{(\tilde {\hat E}, \tilde {\hat \sigma} )}
J^{( E, \sigma)}_{(\tilde {E}, \tilde {\sigma} )} \right]^{1/2}
~
\cb \tilde \Phi \rb 
\Psi( \tilde \Phi; \tilde  E, \tilde \sigma) \nn
& = & \int 
D^{ (\hat E, \hat \sigma) } \tilde \Phi ~ 
\left[ J^{(\tilde {\hat E}, \tilde {\hat \sigma} )}_{(\hat E, \hat \sigma)}
J^{( E, \sigma)}_{(\tilde {E}, \tilde {\sigma} )} \right]^{1/2}
~
\cb \tilde \Phi \rb 
\Psi( \tilde \Phi; \tilde  E, \tilde \sigma) \nn
&=& \cb \tilde E, \tilde \sigma \rb.
\eqa
From the first to the second lines,
we use 
$D^{ (\hat E, \hat \sigma) } \Phi = D^{ (\tilde {\hat E}, \tilde {\hat \sigma} )  } \tilde \Phi$
and \eq{diffeo}.
For the next equality, 
we use
$D^{ (\tilde {\hat E}, \tilde {\hat \sigma} )  } \tilde \Phi =
J^{ (\tilde {\hat E}, \tilde {\hat \sigma} )}_{ (\hat E, \hat \sigma) }   
 D^{ (\hat E, \hat \sigma) } \tilde \Phi $.
In the last equality, we use
the fact that 
\eq{matrix_element} is invariant under diffeomorphism :
$
J^{
(\tilde {\hat E}, \tilde {\hat \sigma} )}_{
(\hat E, \hat \sigma)
}
J^{
( E, \sigma)}_{
(\tilde {E}, \tilde {\sigma} )
}
= 
\frac{
J^{
(\tilde {\hat E}, \tilde {\hat \sigma} )}_{
(\tilde {E}, \tilde {\sigma} )
}
}
{
J^{
(\hat E, \hat \sigma)}_{
( E, \sigma)
}
}
= 
1$.
The linear terms in $n^\mu$ gives \eq{mCR}.

\section{Induced Wheeler-DeWitt Hamiltonian}
\label{sec:WdW}

In this appendix, we prove \eq{eq:inducedWdW}.
As a first step, 
we construct the projection operator $\hat P_{E,\sigma}$ 
in \eq{pHp} that satisfies
\bqa
 \hat P_{E,\sigma} ~ \int DE' D\sigma' \cb E', \sigma' \rb \chi(E',\sigma')
~ = ~
  \cb E, \sigma \rb\chi(E,\sigma).
  \label{proa}
  \eqa
We first try $\cb E, \sigma \rb \lb E, \sigma \cb$ 
as a candidate for the projection operator. 
However, it does not satisfy \eq{proa} 
because the convolution integration in 
$ \int DE' D\sigma' \lb E, \sigma \cb E', \sigma' \rb \chi(E',\sigma')$
generates a significant smearing of the wavefunction.
In the present case, the smearing modifies $\chi(E,\sigma)$
beyond the wavefunction normalization unlike the case in Sec. \ref{sec:miniB}.
The difference is caused by the fact that 
in the Gaussian model $\kappa^{-1}$
is proportional to the number of matter fields
unlike the non-Gaussian model 
considered for the minisuperspace cosmology.
The smearing can be represented as a differential operator,
\bqa
%%\cb E, \sigma \rb
\int DE' D\sigma' ~
\lb E,\sigma \cb E', \sigma' \rb 
\chi(E', \sigma') = 
%S\left( v, \frac{\delta }{\delta v} \right)
%%\cb E, \sigma \rb
{\cal S}\left( E, \sigma, \frac{\delta}{\delta E}, \frac{\delta}{\delta \sigma} \right)
 ~ \chi(E,\sigma),
\label{conv3a}
\eqa
where
\bqa
{\cal S}\left( E, \sigma, \frac{\delta}{\delta E}, \frac{\delta}{\delta \sigma} \right)
= 
\left[ 
e^{  \int dy_1 dy_2 |E(y_1)|  |E(y_2)|
%\frac{1}{|E|}
{\cal M}^{-1}_{ab}(y_1,y_2) 
 \frac{\delta}{\delta v_a(y_1)}
 \frac{\delta}{\delta v_b(y_2)}
 }
 \right]
\label{S}
\eqa
in the large $N$ limit with
$v_a(x) = \left( g_{E,\mu \nu}(x), \sigma(x) \right)$.
${\cal M}^{-1}(y_1,y_2)  \sim \frac{1}{|E|^2 N^2} e^{-d_{y_1,y_2}/\lc}$
is given by the inverse of \eq{Mp},
which decays exponentially in the proper distance between $y_1$ and $y_2$.
%%(See Appendix \ref{sec:overlap}).
%
Inside $[...]$ of \eq{S}, it is understood that the functional differentiations are
ordered to the right 
so that they act only on $\chi(E,\sigma)$.
However, to the leading order in $1/N^2$, 
the normal ordering can be ignored because
${\cal S}
%%\left( E, \sigma, \frac{\delta}{\delta E}, \frac{\delta}{\delta \sigma} \right)
=
e^{  \int dy_1 dy_2 |E(y_1)|  |E(y_2)|
%\frac{1}{|E|}
{\cal M}^{-1}_{ab}(y_1,y_2) 
 \frac{\delta}{\delta v_a(y_1)}
 \frac{\delta}{\delta v_b(y_2)} 
+O(N^{-4})
} $.
The metric differentiation 
is defined through the chain rule,
$\frac{\delta}{\delta g_{E,\mu\nu}(y)} = 
\frac{ \partial E_{\rho j}(y) }{ \partial g_{E,\mu \nu}(y) }
\frac{\delta}{\delta E_{\rho j}(y) }$,
where $\frac{ \partial E_{\rho j}(y) }{ \partial g_{E,\mu \nu}(y) }$
is evaluated in a fixed gauge for the local $SO(3)$ symmetry.
Since $\chi(E,\sigma)$ depends on the triad only through the metric, 
$\frac{\delta}{\delta g_{E,\mu\nu}(y)}$ is independent of the gauge choice.
In order to undo the smearing
introduced in \eq{conv3a},
we insert the inverse of ${\cal S}$ as
\bqa
\hat P_{E,\sigma} = \cb E, \sigma \rb {\cal S}^{-1}   \lb E, \sigma \cb, 
\label{PESa}
\eqa
where ${\cal S}^{-1}$ is the inverse of ${\cal S}$,
which can be computed order by order in $1/N^2$.
To the leading order, it is given by
${\cal S}^{-1}
=
e^{  -\int dy_1 dy_2 |E(y_1)|  |E(y_2)|
%\frac{1}{|E|}
{\cal M}^{-1}_{ab}(y_1,y_2) 
 \frac{\delta}{\delta v_a(y_1)}
 \frac{\delta}{\delta v_b(y_2)} 
+O(N^{-4})
} $.
The existence of ${\cal S}^{-1}$ relies on the
linear independence of $\cb E,\sigma \rb$.
This follows from the fact that states for
a large number ($N^2$) of matter fields are
parameterized by $O(1)$ collective fields. 
Therefore two states with different collective variables
should be linearly independent
for a sufficiently large $N$ 
while they are not necessarily orthogonal.
\eq{PESa} satisfies \eq{conv4} because
\bqa
\cb E, \sigma \rb {\cal S}^{-1} \lb E, \sigma \cb 
%% \hat P_{E,\sigma}
 \int DE' D\sigma' \cb E', \sigma' \rb \chi(E',\sigma')
=   \cb E, \sigma \rb {\cal S}^{-1} {\cal S} \chi(E,\sigma)
= 
  \cb E, \sigma \rb\chi(E,\sigma).
 \label{conv4a}
  \eqa

Now it is straightforward to check \eq{eq:inducedWdW}.
From \eq{WDE} and \eq{pHp},
we obtain 
\bqa
&&  \hat \cH(x) ~ \int DE' D\sigma'  \cb E', \sigma' \rb \chi(E',\sigma') \nn
&=&
\frac{1}{2} \int DE D\sigma   DE' D\sigma'
~
\left\{
\tilde h^{E,\sigma}(x) \cb E, \sigma \rb {\cal S}^{-1} 
\lb E, \sigma \cb
+ 
\cb E, \sigma \rb {\cal S}^{-1}
\tilde h^{E,\sigma }(x)
\lb E, \sigma \cb
\right\}
\cb E', \sigma' \rb  \chi(E',\sigma')    \nn
&=&
\frac{1}{2}   \int DE D\sigma
~
\left\{
\tilde h^{E,\sigma}(x) \cb E, \sigma \rb
{\cal S}^{-1} {\cal S}
+ 
\cb E, \sigma \rb
{\cal S}^{-1}
\tilde h^{E,\sigma }(x)
{\cal S}
\right\}
~
\chi( E, \sigma ) \nn
&=&
\frac{1}{2}   \int DE D\sigma
~
\left\{
\tilde h^{E,\sigma}(x) \cb E, \sigma \rb
+ 
{\cal S}
\tilde h^{E,\sigma \dagger }(x)
{\cal S}^{-1}\cb E, \sigma \rb
\right\}
~
\chi( E, \sigma ),
\label{pHpa1}
\eqa
where ${\cal S}^\dagger = {\cal S}$ has been used.
The hermiticity of ${\cal S}$ follows from the fact that
$\int 
DE D\sigma 
DE' D\sigma' ~
\chi^*(E,\sigma)
\lb E,\sigma \cb E', \sigma' \rb 
\chi'(E', \sigma')$
can be either written as
$\int 
DE D\sigma ~
\chi^*(E,\sigma)
{\cal S}
\chi'(E, \sigma)$
or
$\int 
DE D\sigma ~
\left[ {\cal S} \chi(E,\sigma) \right]^*
\chi'(E, \sigma)$.

Since ${\cal S}$ does not commute with $\tilde h^{E, \sigma \dagger}(x)$,
the unsmearing can not be done perfectly.
The residual effect of smearing gives rise to higher derivative terms 
in the collective variable in \eq{pHpa1}.
Using the Hausdorff-Campbell formula,
one can isolate the residual term as
\bqa
{\cal S} 
\tilde h^{E, \sigma \dagger}(x) 
{\cal S}^{-1}
= 
\left[
1 + 
O \left( \frac{\lc^3}{|E| N^2} 
 \frac{\delta}{\delta v_a(x)} 
\right)
\right]
\tilde h^{E, \sigma \dagger}(x),
\label{HausdorffCampbell} 
\eqa
where it is used that 
operators in 
$\tilde h^{E, \sigma  \dagger}(x)$  in \eq{HWDW}
are centered at $x$ with a spread $\lc$ in space,
and  ${\cal M}^{-1}(x,y)$ decays exponentially 
beyond the length scale $\lc$.
In terms of the conjugate momenta,
\bqa
\pi^{\mu i}(x) = i \kappa^2 \frac{ \delta}{\delta \emi(x)}, ~~~~
\pi_\sigma(x) = i \kappa^2 \frac{ \delta}{\delta \sigma(x)},
\eqa
%%which control the time derivative of the collective fields,
the higher derivative terms in 
\eq{HausdorffCampbell} 
are of the order of
$
O \left( \lc   \frac{\pi}{|E|}  \right)
$
where $\pi$ denotes either
$\pi^{\mu i}$
or 
$\pi$.
%%Therefore higher order terms in momenta are also controlled by $\lc$.
%%
%%
%%
%%
Therefore, the action of $\hat {\cal H}(x)$
on $\cb E, \sigma \rb$ results in  
\bqa
 \hat \cH(x) 
  \cb E, \sigma \rb 
 & = &
\frac{1}{2} 
\left[
\tilde h^{E,\sigma}(x) + 
\tilde h^{E,\sigma \dagger }(x)
\right] 
\left\{
1 + 
O \left( 
\lc \kappa^2
 \frac{\delta}{\delta E_{\mu i}(x)}
,
 \lc \kappa^2
  \frac{\delta}{\delta \sigma(x)} 
  \right)
\right\}
 \cb E, \sigma \rb. 
\label{pHpa2}
\eqa
The higher derivative correction from the smearing is small
if the conjugate momentum is small compared to the $\lc^{-1}$.

\end{appendix}

\end{document}